\documentclass[11pt,twoside]{report}
\usepackage{fleqn,espcrc1}
\usepackage{amssymb}
\usepackage{graphicx}
\def\openone{\leavevmode\hbox{\small1\kern-3.8pt\normalsize1}}

\newcommand{\AmS}{{\protect\the\textfont2
  A\kern-.1667em\lower.5ex\hbox{M}\kern-.125emS}}
\title{\bf \LARGE Wannier-Stark resonances in 
        optical and semiconductor superlattices}
\author{Markus Gl\"uck\address[kl]{FB Physik, Universit\"at Kaiserslautern, 
        D-67653 Kaiserslautern, Germany}, 
        Andrey R. Kolovsky\addressmark[kl]\address{L. V. Kirensky Institute of Physics,
        660036 Krasnoyarsk, Russia}, and	
        Hans J\"urgen Korsch\addressmark[kl]}
\begin{document}
\maketitle
\begin{abstract}
\vspace*{2.0cm}
\noindent
{\bf Abstract:}\\
In this work, we discuss the resonance states of a quantum particle
in a periodic potential plus a static force. Originally this problem was
formulated for a crystal electron subject to a static electric
field and it is nowadays known as the Wannier-Stark problem.
We describe a novel approach to the Wannier-Stark problem developed in recent 
years. This approach allows to compute the complex energy spectrum of a
Wannier-Stark system as the poles of a rigorously constructed scattering matrix
and solves the Wannier-Stark problem without any
approximation. The suggested method is very efficient from the numerical point 
of view and has proven to be a powerful analytic tool for Wannier-Stark resonances
appearing in different physical systems such as optical lattices or 
semiconductor superlattices.
\end{abstract}

\vspace*{5mm}
\tableofcontents

\chapter{Introduction}
\label{sec1}

The problem of a Bloch particle in the presence of additional external 
fields is as old as the quantum theory of solids. 
Nevertheless, the topics introduced in the early studies of the system, 
Bloch oscillations \cite{Bloc28}, Zener tunneling \cite{Zene34} and the 
Wannier-Stark ladder \cite{Wann60}, are still the subject of current 
research. The literature on the field is vast and manifold, 
with different, sometimes unconnected lines of evolution. 
In this introduction we try to give a survey of the field, 
summarize the different theoretical approaches and discuss 
the experimental realizations of the system. It should be noted 
from the very beginning that most of the literature deals with 
one-dimensional single-particle descriptions of the system, 
which, however, capture the essential physics of real systems. 
Indeed, we will also work in this context. 

\section{Wannier-Stark problem}
\label{sec1a}

In the one-dimensional case the Hamiltonian of a Bloch particle in an additional 
external field, in the following referred to as the Wannier-Stark Hamiltonian, 
has the form
\begin{equation}
\label{1a0}
H_W = \frac{p^2}{2m} + V(x) + F x, \quad V(x+d)= V(x), 
\end{equation} 
where $F$ stands for the static force induced by the external field.
Clearly, the external field destroys the translational symmetry of the
field-free Hamiltonian $H_0 = p^2/2m + V(x)$. Instead, from an arbitrary 
eigenstate with $H_W \Psi = E_0 \Psi$, one can by a translation over $l$ 
periods $d$ construct a whole ladder of eigenstates with energies
$E_l = E_0 + l dF$, the so-called Wannier-Stark ladder. Any superposition
of these states has an oscillatory evolution with the time period
\begin{equation}
\label{1a0a}
T_B=\frac{2\pi\hbar}{dF} \;,
\end{equation} 
known as the Bloch period. There has been a long-standing controversy about 
the existence of the Wannier-Stark ladder and Bloch oscillations
\cite{Zak68,Wann69,Zak69,Rabi71,Shoc72,Rabi72,Chur81,Chur83,Krie86,Emin87,Hart88,Klei90,Zak91,Page91,Leo91,Zhao92b}, 
and only recently agreement about the nature
of the Wannier-Stark ladder was reached. The history of this
discussion is carefully summarized in \cite{Krie86,Nenc91,Bouc95,Ross98a}.
\begin{figure}[t]
\begin{center}
\includegraphics[width=10cm]{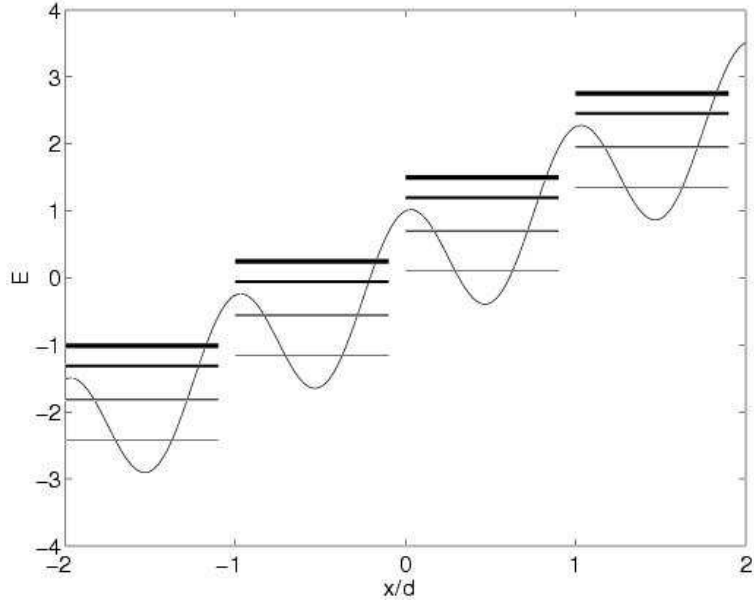}
\caption{\it Schematic illustration of the Wannier-Stark ladder 
of resonances. The width of the levels is symbolized by the
different strength of the lines.}
\label{fig1b}
\end{center}
\end{figure}

From today's point of view the discussion mainly dealt with the effect 
of the single band approximation (effectively a projection on a 
subspace of the Hilbert space) on the spectral properties of the 
Wannier-Stark Hamiltonian.
Within the single band approximation, the $\alpha$'th 
band of the field-free Hamiltonian $H_0$ forms, 
if the field is applied, the Wannier-Stark ladder with the quantized energies
\begin{equation}
\label{1a1}
E_{\alpha,l} = \bar{\epsilon}_\alpha + dFl \;, \quad l=0\pm1,\ldots \;,
\end{equation}
where $\bar{\epsilon}_\alpha$ is the mean energy of the
$\alpha$-th band (see Sec.~\ref{sec1b}). This Wannier-Stark quantization was 
the main point to be disputed in the discussions mentioned above.
The process, which is neglected in the single band approximation and
which couples the bands, is Zener tunneling \cite{Zene34}. For smooth potentials
$V(x)$, the band gap decreases with increasing band index. Hence, as  
the tunneling rate increases with decreasing band gap, the Bloch particles 
asymmetrically tend to tunnel to higher bands and the band population 
depletes with time (see Sec.~\ref{sec1c}). This already gives a hint
that Eq.~(\ref{1a1}) can be only an approximation to the actual spectrum of
the sytem. Indeed, it has been proven that the spectrum of the 
Hamiltonian (\ref{1a0}) is continuous \cite{Avro77,Bent83b}. 
Thus the discrete spectrum (\ref{1a1}) can refer only to resonances
\cite{Herb81,Agle85,Comb91,Bent91,Mois98}, and Eq.~(\ref{1a1}) should be
corrected as
\begin{equation}
\label{1a2}
\mathcal{E}_{\alpha,l} = E_\alpha + dFl 
-{\rm i}\,\frac{\Gamma_\alpha}{2} \;,
\end{equation}
(see Fig.~\ref{fig1b}).
The eigenstates of the Hamiltonian (\ref{1a0}) corresponding to these
complex energies, referred in what follows as the Wannier-Stark 
states $\Psi_{\alpha,l}(x)$, are metastable states with 
the lifetime given by $\tau=\hbar/\Gamma_\alpha$. To find the complex
spectrum (\ref{1a2}) (and corresponding eigenstates) is an
ultimate aim of the Wannier-Stark problem.

Several attempts have been made to calculate the Wannier-Stark
ladder of resonances. Some analytical results have been obtained
for nonlocal potentials \cite{Avro76,Avro82} and for potentials
with a finite number of gaps \cite{Grec93,Grec94,Grec95,Grec97,Grec97b,Grec98,Busl98}. 
(We note, however, that almost all periodic potentials have an infinite 
number of gaps.) A common numerical approach is the formalism of a
transfer matrix to potentials which consist of piecewise constant
or linear parts, eventually separated by delta function barriers 
\cite{Bana78,Bent82,Bent83a,Ritz93,Chan93}. Other methods approximate
the periodic system by a finite one \cite{Roy82,Souc91,Mend94,Glut99}. 
Most of the results concerning Wannier-Stark systems, however,
have been deduced from single- or finite-band approximations and 
strongly related tight-binding models. The main
advantage of these models is that they, as well in the case of static (dc) 
field \cite{Fuku73} as in the cases of oscillatory (ac) and dc-ac fields 
\cite{Dunl86,Zhao91,Zhao92a,Hhong92,Zhao94,Zhao95,Dres97,Hang98},
allow analytical solutions. Tight-binding models have been 
additionally used to investigate the effect of disorder 
\cite{Hone93,Yama93,Holt95,Holt95a,Dres96,Yama98}, 
noise \cite{Suqi00} or alternating site energies 
\cite{Kova88,Zhao91a,Zhao97,Bao98,Rive00} on the dynamics
of Bloch particles in external fields. In two-band descriptions
Zener tunneling has been studied \cite{Rotv95,Rotv96,Hone96,Zhao98,Yan98}, 
which leads to Rabi oscillations between Bloch bands \cite{Zhao96}.
Because of the importance of tight-binding and single-band
models for understanding the properties of Wannier-Stark resonances
we shall discuss them in some more detail.

\section{Tight-binding model}
\label{sec1b}

In a simple way, the tight-binding model can be introduced
by using the so-called Wannier states (not to be confused with
Wannier-Stark states), which are defined as follows.
In the absence of a static field, the eigenstates of the field-free
Hamiltonian,
\begin{equation}
\label{1b0}
H_0=\frac{p^2}{2m}+V(x) \;,
\end{equation}
are known to be the Bloch waves
\begin{equation}
\label{1b1}
\phi_{\alpha,\kappa}(x)=\exp({\rm i}\kappa x)\chi_{\alpha,\kappa}(x) \;,\quad
\chi_{\alpha,\kappa}(x+d)=\chi_{\alpha,\kappa}(x) \;,
\end{equation}
with the quasimomentum $\kappa$ defined in the first Brillouin zone 
$-\pi/d\le \kappa<\pi/d$. The functions (\ref{1b1}) solve the eigenvalue
equation
\begin{equation}
\label{1b2}
H_0\phi_{\alpha,\kappa}(x)=\epsilon_\alpha(\kappa)\phi_{\alpha,\kappa}(x) \;,\quad
\epsilon_\alpha(\kappa+2\pi/d)=\epsilon_\alpha(\kappa) \;,
\end{equation}
where $\epsilon_\alpha(\kappa)$ are the Bloch bands. Without affecting the energy
spectrum, the free phase of the Bloch function $\phi_{\alpha,\kappa}(x)$
can be chosen such that it is an analytic and periodic function of the
quasimomentum $\kappa$ \cite{Kohn72}. Then we can expand it in a Fourier 
series in $\kappa$, where the expansion coefficients
\begin{equation}
\label{1b3}
\psi_{\alpha,l}(x)=\int_{-\pi/d}^{\pi/d} {\rm d}\kappa
\exp(-{\rm i}\kappa ld)\,\phi_{\alpha,\kappa}(x)
\end{equation}
are the Wannier functions. 

\begin{figure}[t]
\begin{center}
\includegraphics[width=13cm,height=9cm]{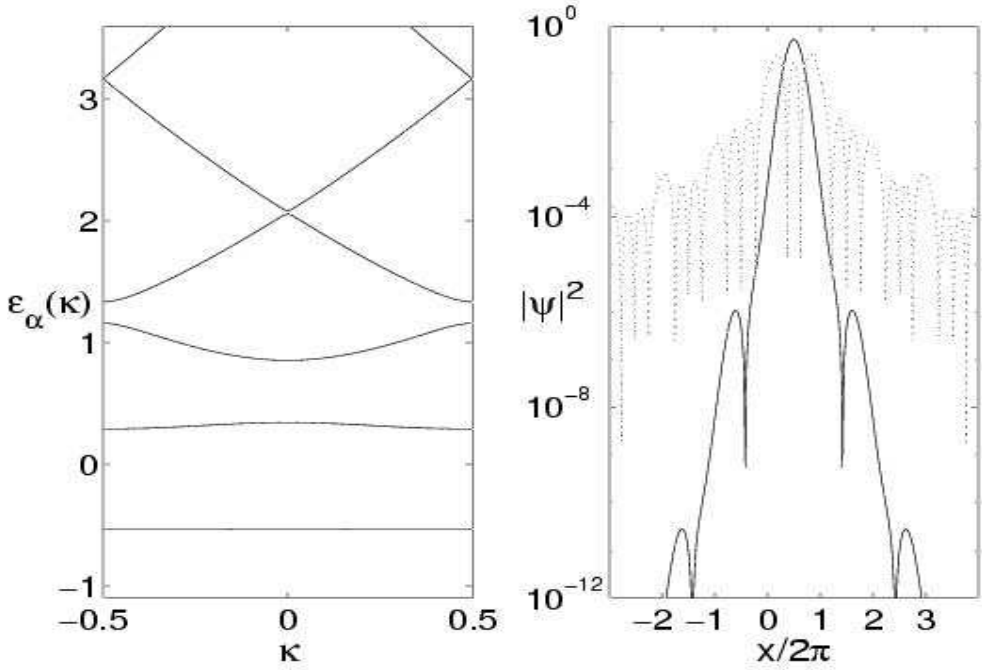}
\caption{\it Left panel -- lowest energy bands $\epsilon_\alpha(\kappa)$ for the 
potential $V(x) = \cos(x)$ with parameters $\hbar = 1$ and $m=1$. 
Right panel -- associated Wannier states $\psi_{0,0}$ 
(solid line) and $\psi_{1,0}$ (dotted line).}
\label{fig1a}
\end{center}
\end{figure}

Let us briefly recall the main properties
of the Wannier and Bloch states. Both form orthogonal sets with
respect to both indices. The Bloch functions are, in general, complex
while the Wannier functions can be chosen to be real. While the Bloch
states are extended over the whole coordinate space, the Wannier
states are exponentially localized \cite{Kohn59,Nenc83}, essentially
within the $l$-th cell of the potential. Furthermore, the Bloch functions
are the eigenstates of the translation (over a lattice period) operator 
while the Wannier states satisfy the relation
\begin{equation}
\label{1b4}
\psi_{\alpha,l+1}(x)=\psi_{\alpha,l}(x-d) \;,
\end{equation}
which directly follows from Eq.~(\ref{1b3}). Finally, the Bloch states
are eigenstates of $H_0$ but the Wannier states are not. As an
example, Fig.~\ref{fig1a} shows the Bloch band spectrum $\epsilon_\alpha(\kappa)$
and two Wannier functions $\psi_{\alpha,0}(x)$ of the system (\ref{1b0}) 
with $V(x)=\cos x$, $m=1$ and $\hbar=1$. The exponential decrease of the 
ground state is very fast, i.e. the relative occupancy of the adjacent 
wells is less than $10^{-5}$. For the second excited Wannier state
it is a few percent.

The localization property of the Wannier states suggests to use them
as a basis for calculating the matrix elements of the Wannier-Stark
Hamiltonian (\ref{1a0}). (Note that the field-free Hamiltonian (\ref{1b0}) is
diagonal in the band index $\alpha$.) The tight-binding Hamiltonian is deduced 
in the following way. Considering a particular band 
$\alpha$, one takes into account only the main and the first diagonals
of the Hamiltonian $H_0$. From the field term $x$ only the diagonal
part is taken into account. Then, denoting the Wannier states resulting from 
the $\alpha$-th band by $|l\rangle$, the tight-binding Hamiltonian reads
\begin{equation}
\label{1b5}
H_{TB}=\sum_l(\bar{\epsilon}_\alpha+dFl)\,|l\rangle\langle l|
+\frac{\Delta_\alpha}{4}
\,\big(\,|l+1\rangle\langle l|+|l\rangle\langle l+1|\,\big) \;.
\end{equation}
The Hamiltonian (\ref{1b5}) can be easily diagonalized which yields
the spectrum $E_{\alpha,l}=\bar{\epsilon}_\alpha+dFl$ with the eigenstates
\begin{equation}
\label{1b6}
|\Psi_{\alpha,l}\rangle
=\sum_m J_{m-l}\Big(\frac{\Delta_\alpha}{2dF}\Big)|m\rangle \;.
\end{equation}
Thus, all states are localized and the spectrum is the discrete 
Wannier-Stark ladder (\ref{1a1}).

The obtained result has a transparent physical meaning. 
When $F=0$ the energy levels of Wannier states $|l\rangle$ coincide
and the tunneling couples them into Bloch waves 
$|\kappa\rangle=\sum_l \exp({\rm i}\kappa l)|l\rangle$. Correspondingly, the 
infinite degeneracy of the level $\bar{\epsilon}_\alpha$ is removed, producing the 
Bloch band \,$\epsilon_\alpha(\kappa)=\bar{\epsilon}_\alpha
+(\Delta_\alpha/2)\cos(d\kappa).$\footnote{Because only the nearest off-diagonal elements are taken into account in 
Eq.~(\ref{1b5}), the Bloch bands are always approximated by a cosine dispersion
relation.}
When $F\ne0$ the Wannier levels are misaligned and the
tunneling is suppressed. As a consequence, the Wannier-Stark state
involves (effectively) a finite number of Wannier states, as indicated
by Eq.~(\ref{1b6}). It will be demonstrated later on that for the low-lying 
bands Eq.~(\ref{1a1}) and Eq.~(\ref{1b6}) approximate quite well the 
real part of the complex Wannier-Stark spectrum and
the resonance Wannier-Stark functions $\Psi_{\alpha,l}(x)$, respectively.
The main drawback of the model, however, 
is its inability to predict the imaginary
part of the spectrum (i.e.~the lifetime of the Wannier-Stark states),
which one has to estimate from an independent calculation.
Usually this is done with the help of Landau-Zener theory.

\section{Landau-Zener tunneling}
\label{sec1c}

Let us address the following question: if we take an initial state
in the form of a Bloch wave with quasimomentum $\kappa$, what will be 
the time evolution of this state when the external static field is
switched on?

The common approach to this problem is to look for the solution
as the superposition of Houston functions \cite{Hous40}
\begin{equation}
\label{1c0}
\psi(x,t)=\sum_\alpha c_\alpha(t) \psi_\alpha(x,t) \;,
\end{equation}
\begin{equation}
\label{1c0a}
\psi_\alpha(x,t)
=\exp\left(-\frac{\rm i}{\hbar}\int_0^{\rm t}{\rm d}t'\epsilon_\alpha(\kappa')\right)
\phi_{\alpha,\kappa'}(x) \;,
\end{equation}
where $\phi_{\alpha,\kappa'}(x)$ is the Bloch function with the quasimomentum
$\kappa'$ evolving according to the classical equation of motion $\dot{p}=-F$, i.e
$\kappa'=\kappa-Ft/\hbar$.
Substituting Eq.~(\ref{1c0}) into the time-dependent Schr\"odinger
equation with the Hamiltonian (\ref{1a0}), we obtain
\begin{equation}
\label{1c2}
{\rm i}\hbar \,\dot{c}_\alpha=F \sum_\beta
X_{\alpha,\beta}(\kappa')\exp\left(-\frac{\rm i}{\hbar}
\int_0^{\rm t}{\rm d}t'[\epsilon_\alpha(\kappa')-\epsilon_\beta(\kappa')]\right) c_\beta \;,
\end{equation}
where
$X_{\alpha,\beta}(\kappa)={\rm i}\int{\rm d}x\, 
\chi^*_{\alpha,\kappa}(x)\,\partial/\partial\kappa \,X_{\beta,\kappa}(x)$.
Neglecting the interband coupling, i.e.~$X_{\alpha,\beta}=0$
for $\alpha \ne \beta$, we have
\begin{equation}
\label{1c4}
c_\beta(t)\approx 0 \quad {\rm for} \quad \alpha \ne \beta 
\qquad {\rm and} \qquad {\rm i}\hbar \,\dot{c}_\alpha=F\,
X_{\alpha,\alpha}(\kappa')\,c_\alpha\;.
\end{equation}
This solution is the essence of the so-called single-band approximation.
We note that within this approximation one can use the Houston functions
(\ref{1c0a}) to construct the localized Wannier-Stark states similar to those
obtained with the help of the tight-binding model.

The correction to the solution (\ref{1c4}) is obtained by using the
formalism of Landau-Zener tunneling. In fact, when the quasimomentum $\kappa'$
explores the Brillouin zone, the adiabatic transition occurs at
the points of ``avoided'' crossings between the adjacent Bloch bands 
[see, for example, the avoided crossing between the 4-th and 5-th bands in 
Fig.~\ref{fig1a}(a) at $\kappa=0$]. Semiclassically, the probability of this 
transition is given by
\begin{equation}
\label{1c5}
P\approx\exp\left(-\frac{\pi\Delta^2_{\alpha,\beta}}
{8\hbar(|\epsilon'_\alpha|+|\epsilon'_\beta|)F}\right) \;,
\end{equation}
where $\Delta_{\alpha,\beta}$ is the energy gap between the bands
and $\epsilon'_\alpha$, $\epsilon'_\beta$ stand for the 
slope of the bands at the point
of avoided crossing in the limit $\Delta_{\alpha,\beta}\rightarrow0$
\cite{Landau}. In a first approximation, one can assume that
 the adiabatic transition occurs once for each Bloch
cycle $T_B=2\pi\hbar/dF$. Then the population of the $\alpha$-th
band decreases exponentially with the decay time
\begin{equation}
\label{1c6}
\tau=\hbar/\Gamma_\alpha \;,\quad
\Gamma_\alpha=a_\alpha F\exp(-b_\alpha/F) \;,
\end{equation}
where $a_\alpha$ and $b_\alpha$ are band-dependent constants.

In conclusion, within the approach described above one obtains from each 
Bloch band a set of localized states with energies given by Eq.~(\ref{1a1}).
However, these states have a finite lifetime given by Eq.~(\ref{1c6}).
It will be shown in Sec.~\ref{sec3a} that the estimate (\ref{1c6})
is, in fact, a good ``first order'' approximation for the
lifetime of the metastable Wannier-Stark states.

\section{Experimental realizations}
\label{sec1d}

We proceed with experimental realizations of the Wannier-Stark 
Hamiltonian (\ref{1a0}). Originally, the problem was formulated 
for a solid state electron system with an applied external electric field, and in 
fact, the first measurements concerning the existence of the Wannier-Stark
ladder dealt with photo-absorption in crystals \cite{Koss72}.
Although this system seems convenient at first glance, 
it meets several difficulties because of the intrinsic
multi-particle character of the system. Namely, the dynamics of an
electron in a solid is additionally influenced by electron-phonon
and electron-electron interactions. In addition, scattering by
impurities has to be taken into account. In fact, for all reasonable 
values of the field, the Bloch time (\ref{1a0a})  
is longer than the relaxation time, and therefore neither Bloch oscillations 
nor Wannier-Stark ladders have been observed in solids yet. 

One possibility to overcome these problems is provided by semiconductor superlattices
\cite{Esak93}, which consists of alternating layers of different semiconductors, as for
example, $GaAs$ and $Al_{x}Ga_{1-x}As$. In the most simple approach, the wave
function of a carrier (electron or hole) in the transverse direction
of the semiconductor superlattice is approximated by a plane wave for a particle
of mass $m^*$ (the effective mass of the electron in the conductance
or valence bands, respectively). In the direction perpendicular to the
semiconductor layers (let it be $x$-axis) the carrier ``sees'' a periodic 
sequence of potential barriers
\begin{equation}
\label{1d0a}
V(x)=\left\{\begin{array}{ll}
V_0& \quad {\rm if}\quad \exists\  l \in \mathbb{Z}
\quad {\rm with}\quad |x- l d|<a/2\\
\ 0\;,\quad &\quad {\rm else}
\end{array}\right. \;,
\end{equation}
where the height of the barrier $V_0$ is of the order of 100 meV and the
period $d\sim100$ \AA. Because the period of this potential is two orders
of magnitude larger than the lattice period in bulk semiconductor, the Bloch time 
is reduced by this factor and may be smaller than the relaxation time.
Indeed, semiconductor superlattices were the first systems where 
Wannier-Stark ladders were observed \cite{Mend88,Vois88,Mend93}
and Bloch oscillations have been measured in four-wave-mixing 
experiments \cite{Feld92,Leo92} as proposed in \cite{Ples92}.
In the following years, many facets of the topics
have been investigated. Different
methods for the observation of Bloch oscillation have been
applied \cite{Wasc93,Deko94,Cho96,Leo98}, and nowadays it is possible
to detect Bloch oscillations at room temperature \cite{Deko95}, 
to directly measure \cite{Lyss97} or even control \cite{Sudz98}
their amplitude. Wannier-Stark ladders have been found in 
a variety of superlattice structures 
\cite{Souc90,Whit90,Schn91a,Yu94,Pegg95}, 
with different methods \cite{Gibb93,Hama94}. 
The coupling between different Wannier-Stark ladders 
\cite{Schn90,Naka91,Tana92,Bast94,Kuem99}, the
influence of scattering \cite{Ples94,Xia94,Loes00}, the relation to the 
Franz-Keldysh effect \cite{Ribe92,Schm94,Lind95}, the influence
of excitonic interactions \cite{Dign91,Fox92,Leis94,Dign94,Lind97}
and the role of Zener tunneling \cite{Carl94,Sibi98,Helm99,Leo00} have been 
investigated. Altogether, there is a large variety of interactions 
which affect the dynamics of the electrons in semiconductor superlattices, 
and it is still quite complicated to assign which effect is due to which origin. 

A second experimental realization of the Wannier-Stark Hamiltonian
is provided by cold atoms in optical lattices. The majority of 
experiments with optical lattices deals with neutral alkali atoms 
such as lithium \cite{Ande96}, sodium \cite{Wilk96,Bhar97,Madi99}, 
rubidium \cite{Ande98,Frie98,Dutt99} or cesium \cite{Daha96,Guid98,Vant99}, 
but also optical lattices for argon have been realized \cite{Muel97}. 
The description of the atoms in an optical lattice is rather simple.
One approximately treats the atom as a two-state system 
which is exposed to a strongly detuned standing laser wave. Then 
the light-induced force on the atom is described by the potential 
\cite{Adam94,Wall95a}
\begin{equation}
\label{1d0}
V(x) = \frac{\hbar \Omega_R^2}{4 \delta} \cos^2(k_L x)\, ,
\end{equation}
where $\hbar \Omega_R$ is the Rabi frequency (which is proportional to
the product of the dipole matrix elements of the optical transition and
the amplitude of the electric component of the laser field),
$k_L$ is the wave number of the laser, and $\delta$ is the detuning of the 
laser frequency from the frequency of the atomic transition.\footnote{ 
The atoms are additionally exposed to dissipative forces, which 
may have substantial effects on the dynamics \cite{Bern00}. However,
since these forces are proportional to $\delta^{-2}$ while the 
dipole force (\ref{1d0}) is proportional to $\delta^{-1}$, 
for sufficiently large detuning one can reach the 
limit of non-dissipative optical lattices.}

In addition to the optical forces,
the gravitational force acts on the atoms. Therefore, a laser 
aligned in vertical direction yields the Wannier-Stark Hamiltonian
\begin{equation}
\label{1d1}
H=\frac{p^2}{2m}+\frac{\hbar\Omega_R^2}{8\delta}\,\cos(2k_L x)+mgx \;,
\end{equation}
where $m$ is the mass of the atom and $g$ the gravitational constant.
An approach where one can additionally vary the strength of the constant
force is realized by introducing a tunable frequency difference 
between the two counter-propagating waves which form
the standing laser wave. If this difference $\delta \omega$ increases 
linearly in time, $\delta \omega(t) = 2 k_L a t$, the two laser waves 
gain a phase difference which increases quadratically in time according to
$\delta \phi(t) = k_L a t^2$. The superposition of both waves 
then yields an effective potential 
$V(x,t)=(\hbar\Omega_R^2/4\delta)\cos^2[k_L (x-a t^2/2)]$,
which in the rest frame of the potential also yields the Hamiltonian
(\ref{1d1}) with the gravitational force $g$ substituted by $a$.
The atom-optical system provides a much cleaner realization of the single 
particle Wannier-Stark Hamiltonian (\ref{1a0}) than the solid state systems.
No scattering by phonons or lattice impurities occurs.
The atoms are neutral and therefore no excitonic effects have to be taken
into account. Finally, the interaction between the atoms can be neglected
in most cases which justifies a single particle description of the system.
Indeed, Wannier-Stark ladders, Bloch oscillations and Zener 
tunneling have been measured in several experiments in optical 
lattices \cite{Wilk96,Bhar97,Daha96,Qian96,Raiz97,Madi98}. 

Besides the semiconductor and optical lattices,
different attempts have been made to find the Wannier-Stark ladder and
Bloch oscillations in other systems like natural superlattices, optical
and acoustical waveguides, etc.
\cite{Mons90,Ster98,Pesc98,Lenz99,Kavo00,Mora99a,Pert99,Sank96,Sank97,Mate94}. 
However, here we denote them mainly for completeness. In the applications 
of the theory to real systems we confine ourselves to optical lattices and 
semiconductor superlattices. 

A final remark of this section concerns the choice of the independent
parameters of the systems. In fact, by using an appropriate scaling,
four main parameters of the physical systems -- the particle mass $m$,
the period of the lattice $d$, the amplitude of the periodic potential $V_0$
and the amplitude of the static force $F$ -- can be reduced to two
independent parameters. In what follows we use the scaling which
sets $m=1$, $V_0=1$ and $d=2\pi$. Then the independent parameters
of the system are the scaled Planck constant $\hbar'$ (entering the
momentum operator) and the scaled static force $F'$. In particular,
for the system (\ref{1d1}) the scaling $x'=2k_L x$, 
$H'=H/V_0$ ($V_0=\hbar'\Omega_R^2/4\delta$) gives
\begin{equation}
\label{1d2}
\hbar'=\left(\frac{8\omega_{\rm rec}\delta}{\Omega^2_R}\right)^{1/2} \;,\quad
\omega_{\rm rec}=\frac{\hbar k^2_L}{2m} \;,
\end{equation}
i.e. the scaled Planck constant is inversely proportional to the
intensity of the laser field.
For the semiconductor superlattice, the scaled Planck constant is
$\hbar'=2\pi\hbar/d\sqrt{m^*V_0}$.

\section{This work}
\label{sec1f}

In this work we describe a novel approach to the Wannier-Stark
problem which has been developed by the authors during the last few
years \cite{EPJD,PLA1,PRE1,JPA,PRE2,PRL1,PRL2,PLA2,PRA,EPL,JOB1,JOB2,PLA3,PE,PRB,Thesis}.
By using this approach, one finds the complex spectrum (\ref{1a1})
as the poles of a rigorously constructed scattering matrix. The suggested 
method is very efficient from the numerical points of view and 
has proven to be a powerful tool for an analysis of the Wannier-Stark states
in different physical systems.

The review consists of two parts. The first part, which includes chapters 2-3, deals
with the case of a dc field. After introducing
a scattering matrix for the Wannier-Stark system we describe the basic 
properties of the Wannier-Stark states, such as lifetime, localization of the
wave function, etc., and analyze their dependence on the magnitude of
the static field. A comparison of the theoretical predictions with some
recent experimental results is also given.

In the second part (chapters 4-7) we study the case of combined ac-dc fields:
\begin{equation}
\label{1f0}
H = \frac{p^2}{2m} + V(x) + F x +F_\omega x\cos(\omega t) \;. 
\end{equation} 
We show that the scattering matrix introduced for the case of dc field
can be extended to the latter case, provided that the period of
the driving field $T_\omega=2\pi/\omega$ and the Bloch period (\ref{1a0})
are commensurate, i.e. $qT_B=pT_\omega$ with $p,q$ being integers.  
Moreover, the integer $q$ in the last equation appears as the number
of scattering channels. The concept of the metastable quasienergy
Wannier-Bloch states is introduced and used to analyze the dynamical and 
spectral properties of the system (\ref{1f0}). Although the method
of the quasienergy Wannier-Bloch states is formally applicable only to
the case of ``rational'' values of the driving frequency 
(in the sense of equation $T_\omega/T_B=q/p$), the obtained results can be well
interpolated for arbitrary values of $\omega$.

The last chapter of the second part of the work deals with 
the same Hamiltonian (\ref{1f0}) but considers a very different topic. In chapters
2-6 the system parameters are assumed to be in the deep quantum region
(which is actually the case realized in most experiments with
semiconductors and optical lattices). In chapter 7, we turn to the
semiclassical region of the parameters, where the system  (\ref{1f0})
exhibits chaotic scattering. We perform a statistical analysis
of the complex (quasienergy) spectrum of the system and compare the results
obtained with the prediction of random matrix theory for chaotic 
scattering.

To conclude, it is worth to add few words about notations. Through the paper we
use low case $\phi$ to denote the {\em Bloch} states, which are eigenstates
of the field free Hamiltonian (\ref{1b0}). The {\em Wannier-Stark} states, 
which solve the eigenvalue problem with Hamiltonian (\ref{1a0}) and which are 
our main object of interest, are denoted by capital $\Psi$. These states should not 
be mismatched with the {\em Wannier} states (\ref{1b3}) denoted by low case $\psi$.
Besides the Bloch, Wannier, and Wannier-Stark states we shall introduce later
on the {\em Wannier-Bloch} states. These states generalize the notion of
Bloch states to the case of nonzero static field and are denoted by capital
$\Phi$. Thus we always use capital letters ($\Psi$ or $\Phi$) to refer to
the eigenfunctions for $F\ne0$ and low case letter ($\psi$ or $\phi$) in the
case of zero static field, as summarized in the table below.
\begin{table}[hb]
\begin{tabular}{c|c|c|c}\hline
\rule[-4mm]{0mm}{10mm}  function & name  &             &   dc-field\\ \hline
\rule[-4mm]{0mm}{10mm}$\phi_{\alpha,\kappa}(x)$ &Bloch&  
    delocalized eigenfunctions of the Hamiltonian $H_0$  & $F=0$\\ 
\rule[-4mm]{0mm}{10mm}$\psi_{\alpha,l}(x)$ &Wannier&  
    dual localized basis functions  & $F=0$\\ 
\rule[-4mm]{0mm}{10mm}$\Psi_{\alpha,l}(x)$ & Wannier-Stark&  
    resonance eigenfunctions of the Hamiltonian $H_W$  & $F\ne 0$\\  
\rule[-4mm]{0mm}{10mm}$\Phi_{\alpha,\kappa}(x)$ &Wannier-Bloch&  
    res.~eigenfunctions of the evolution operator $U(T_B)$  & $F\ne 0$\\ \hline
\end{tabular}
\end{table}

\chapter{Scattering theory for Wannier-Stark systems}
\label{sec2} 

In this work we reverse the traditional view in treating the two 
contributions of the potential to the Wannier-Stark Hamiltonian 
\begin{equation}
\label{2a0a}
H_W = \frac{p^2}{2} + V(x) + Fx, \quad V(x+2\pi)= V(x) \;. 
\end{equation} 
Namely, we will now consider the external field $Fx$ as part of the 
unperturbed Hamiltonian and the periodic potential as a perturbation, 
i.e. $H_W=H_0+V(x)$, where $H_0 = p^2/2+Fx$.
The combined potential $V(x) + Fx$ cannot support bound states, 
because any state can tunnel through a finite number of barriers
and finally decay in the negative $x$-direction ($F>0$). 
Therefore we treat this system using scattering theory. 
We then have two sets of eigenstates, namely the continuous 
set of scattering states, whose asymptotics define the
S-matrix $S(E)$, and the discrete set of 
metastable resonance states, whose complex energies 
$\mathcal{E} = E - {\rm i}\Gamma/2$ are given by the poles
of the S-matrix. Due to the periodicity of the potential $V(x)$, 
the resonances are arranged in Wannier-Stark ladders of 
resonances. The existence of the Wannier-Stark ladders of
resonances in different parameter regimes has been proven, e.g., 
in \cite{Herb81,Agle85,Comb91,Bent91}. 
  
\section{S-matrix and Floquet-Bloch operator}
\label{sec2a}

The scattering matrix $S(E)$ is calculated by
comparing the asymptotes of the scattering states $\Psi_S(E)$
with the asymptotes of the ``unscattered'' states $\Psi_0(E)$, which are 
the eigenstates of the ``free'' Hamiltonian
\begin{equation}
\label{2a0}
H_0 = \frac{p^2}{2} + F x\, ,\quad F > 0.
\end{equation} 
In configuration space, the $\Psi_0(E)$ are Airy functions
\begin{equation}
\label{2a1} 
\Psi_0(x;E) \sim {\rm Ai}\left( \xi -\xi_0 \right)
\longrightarrow (-\pi^2\xi)^{-1/4} \sin\left( \zeta + \pi/4 \right) \;.
\end{equation}
where $\xi = a x$, $\xi_0 = a E/F$, $a= (2F/\hbar^2)^{1/3}$,
and $\zeta = \frac{2}{3}\,(-\xi)^{3/2}$ \cite{Abra72}. 
Asymptotically the scattering 
states $\Psi_S(E)$ behave in the same way, however, they have an 
additional phase shift $\varphi(E)$,
i.e.~for $x \rightarrow -\infty$ we have 
\begin{equation}
\label{2a2}
\Psi_S(x;E) \longrightarrow (-\pi^2\xi)^{-1/4} \, 
{\rm sin}\left[\, \zeta + \pi/4 + \varphi(E)\right]\, .
\end{equation}
%
Actually, in the Stark case it is more convenient to compare the momentum 
space instead of the configuration space asymptotes. 
(Indeed, it can be shown that both approaches are equivalent
\cite{JOB2,Thesis}.) In momentum space the eigenstates (\ref{2a1}) are given by
\begin{equation}
\label{2a3}
\Psi_0(k;E)  = {\rm exp} \left[ {\rm i} \left( 
\frac{\hbar^2 k^3}{6 F} - \frac{E k}{F} \right)\,\right] .
\end{equation}
For $F > 0$ the direction of decay is the negative $x$-axis, so the 
limit $k \rightarrow -\infty$ of $\Psi_0(k;E)$ 
is the outgoing part and the limit $k \rightarrow \infty$ the 
incoming part of the free solution.

The scattering states $\Psi_S(E)$ solve the Schr\"odinger equation
\begin{equation}
\label{2a4}
H_W  \, \Psi_S(E) = E \, \Psi_S(E) \, 
\end{equation}
with $H_W = H_0 + V(x)$. (By omitting the second argument of the wave
function, we stress that the equation holds both in the momentum
and coordinate representations.) Asymptotically the potential $V(x)$ can be
neglected and the scattering states are eigenstates of the free 
Hamiltonian (\ref{2a0}). In other words, we have
\begin{equation}
\label{2a5}
\lim_{k \rightarrow \pm\infty} \Psi_S(k;E)\,
 = {\rm exp} \left[ {\rm i} \left( \frac{\hbar^2 k^3}{6 F} -  
\frac{E k}{F} \pm \varphi(E)\right)\right]\, .
\end{equation}
With the help of Eqs.~(\ref{2a3}) and (\ref{2a5}) we get 
\begin{equation}
\label{2a6}
 S(E) = \lim_{k \rightarrow \infty} 
\frac{\Psi_S(-k;E)}{\Psi_0(-k;E)}\, \frac{\Psi_0(k;E)}{\Psi_S(k;E)}\, ,
\end{equation}  
which is the definition we use in the following.
In terms of the phase shifts $\varphi(E)$ the S-matrix obviously reads
$S(E) = \exp[-{\rm i}2\varphi(E)]$ and, thus, it is unitary.

To proceed further, we use a trick inspired by the existence of the space-time
translational symmetry of the system, the so-called electric translation
\cite{Ashby65}. Namely, instead of analyzing the spectral 
problem (\ref{2a4}) for the Hamiltonian, we shall
analyze the spectral properties of the evolution operator over a
Bloch period
\begin{equation}
\label{2b0}
U=\exp\left(-\frac{{\rm i}}{\hbar}H_WT_B\right) \;,\quad T_B=\frac{\hbar}{F} \;.
\end{equation}
Using the gauge transformation, which moves the static field into
the kinetic energy, the operator (\ref{2b0}) can be presented in the form
\begin{equation}
\label{2b1}
U={\rm e}^{-{\rm i}x}\,\widetilde{U} \;,
\end{equation}
\begin{equation}
\label{2b1a}
\widetilde{U}= \widehat{\rm exp} \left( -\frac{\rm i}{\hbar}
\int_0^{T_B} \left[ \frac{(p -Ft)^2}{2} + V(x)\right]\,{\rm d}t\right) \;,
\end{equation}
where the hat over the exponential function denotes time ordering 
\footnote{Indeed, substituting into the Schr\"odinger equation,
$i\hbar\partial\psi/\partial t=H_W\psi$, the wave function in the form
$\psi(x,t)=\exp(-iFtx/\hbar)\tilde{\psi}(x,t)$, we obtain
$i\hbar\partial\tilde{\psi}/\partial t=\widetilde{H}_W\tilde{\psi}$ where
$\widetilde{H}_W=(p-Ft)^2/2+V(x)$. Thus
$\tilde{\psi}(x,T_B)=\widetilde{U}\tilde{\psi}(x,0)$ or
$\psi(x,T_B)=\exp(-ix)\widetilde{U}\psi(x,0)$.}. 
The advantage of the operator $U$ over the Hamiltonian $H_W$ is that it commutes 
with the translational operator and, thus, the formalism of the quasimomentum
can be used.\footnote{The tight-binding version of the evolution
operator (\ref{2b1}) was studied in Ref.~\cite{Niu89}.} Besides this, the 
evolution operator also allows us to treat the combined case of an ac-dc field, 
which will be the topic of the second part of this work.

There is a one to one correspondence between the eigenfunctions of the
Hamiltonian and the eigenfunctions of the evolution operator. Indeed, let
$\Psi_S(x;E)$ be an eigenfunction of $H_W$ corresponding to the energy $E$.
Then the function
\begin{equation}
\label{2b2}
\Phi_S(x;\lambda,\kappa)=\sum_l\exp(+{\rm i}2\pi l\kappa)\Psi_S(x-2\pi l;E)
\end{equation}
is a Bloch-like eigenfunction of $U$ corresponding to the eigenvalue
$\lambda=\exp(-{\rm i}ET_B/\hbar)$, i.e.
\begin{equation}
\label{2b3}
U\Phi_S(\lambda,\kappa)=\lambda\Phi_S(\lambda,\kappa) \;,\quad
\lambda=\exp(-{\rm i}E/F)  \;.
\end{equation}
Equation (\ref{2b3}) simply follows from the continuous time evolution of 
the function (\ref{2b2}), which is $\Phi_S(x;\lambda,\kappa,t)
=\sum_l\exp(+{\rm i}2\pi l\kappa)\exp[-{\rm i}(E+2\pi Fl)t/\hbar]
\Psi_S(x-2\pi l;E)$, or
\begin{equation}
\label{2b4}
\Phi_S(\lambda,\kappa,t)=
\exp(-{\rm i}Et/\hbar)\Phi_S(\lambda,\kappa-Ft/\hbar) \;.
\end{equation}
Let us also note that the quasimomentum $\kappa$
does not enter into the eigenvalue $\lambda$. Thus the spectrum of the evolution
operator $U$ is degenerate along the Brillouin zone. Besides this, the
relation between energy $E$ and $\lambda$ is unique only if we restrict
the energy interval considered to the first ``energy Brillouin zone'', 
i.e. $0\le E\le2\pi F$. 

When the energy is restricted by this first Brillouin zone, the transformation
inverse to (\ref{2b2}) reads
\begin{eqnarray}
\label{2b5}
\Psi_S(E) = \int_{-1/2}^{1/2} {\rm d}\kappa \, \Phi_S(\lambda,\kappa) \;. 
\end{eqnarray}
This relation allows us to use the asymptotes of the
Floquet-Bloch solution $\Phi_S(\lambda,\kappa)$ instead 
of the asymptotes of the $\Psi_S(E)$ in the S-matrix definition (\ref{2a6}).
In fact, since the functions $\Phi_S(x;\lambda,\kappa)$ are 
Bloch-like solution, they can be expanded in the basis of plane waves:
\begin{equation}
\label{2b6}
\Phi_S(x;\lambda,\kappa)=\sum_n C_S(n;\lambda,\kappa) \langle x|n+\kappa\rangle
\;,\quad
\langle x|n+\kappa\rangle=(2\pi)^{-1/2}e^{{\rm i}(n+\kappa)x} \;.
\end{equation}
From the integral (\ref{2b5}) the relation 
$\langle n+\kappa|\Phi_S(\lambda,\kappa)\rangle=\langle n+\kappa|\Psi_S(E)\rangle$ 
follows directly, i.e. in the momentum representation
the functions $\Psi_S(k;E)$ and $\Phi_S(k;\lambda,\kappa)$ coincide at the points 
$k = n + \kappa$. Thus we can substitute the asymptotes of $\Phi_S(k;\lambda,\kappa)$ 
in Eq.~(\ref{2a6}). This gives
\begin{equation}
\label{2b7}
 S(E) = \lim_{n \rightarrow \infty} 
\frac{C_S(-n)}{C_0(-n)} \, \frac{C_0(n)}{C_S(n)} \:,
\end{equation}
where the energy on the right-hand side of the equation enters
implicitly through the eigenvalue $\lambda=\exp(-{\rm i}E/F)$. 
Let us also note that by construction $S(E)$ in Eq.~(\ref{2b7}) does 
not depend on the particular choice of the quasimomentum $\kappa$. 
In numerical calculations this provides a test for controlling the accuracy.

\section{S-matrix: basic equations}
\label{sec2c}

Using the expansion (\ref{2b6}), the eigenvalue equation (\ref{2b3})
can be presented in matrix form
\begin{equation} 
\label{2c0}
\sum_n \widetilde{U}^{(\kappa)}_{m+1,n} C_S(n) =\lambda C_S(m) \;,
\end{equation}
where
\begin{equation} 
\label{2c0a}
\widetilde{U}^{(\kappa)}_{m,n}
=\langle m+ \kappa|\,\widetilde{U}\,|n + \kappa\rangle
\end{equation}
and the unitary operator $\widetilde{U}$ is given in Eq.~(\ref{2b1a}).
[Deriving Eq.~(\ref{2c0}) from Eq.~(\ref{2b3}), we took into
account that in the plane wave basis the momentum
shift operator $\exp(-{\rm i} x)$ has the matrix elements 
$\langle m|\exp(-{\rm i} x)|n\rangle= \delta_{m+1,n}$.]
Because $\lambda$ does not depend on the 
quasimomentum $\kappa$,\footnote{This means that the operators 
$\exp(-{\rm i}x)\widetilde{U}^{(\kappa)}$ 
are unitary equivalent -- a fact, which can be directly concluded
from the explicit form of this operator.}
we can set $\kappa=0$ and shall drop this upper matrix index in what
follows.
\begin{figure}[t]
\begin{center}
\includegraphics[width=8cm]{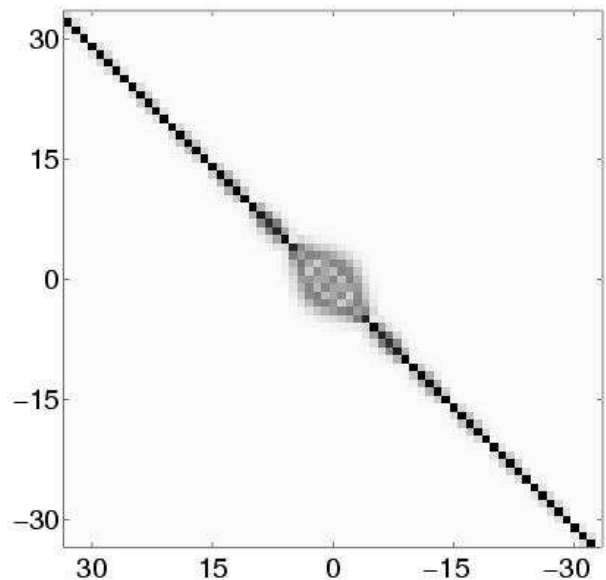}
\caption{\it Matrix of the Floquet-Bloch operator $U$ 
for $H_W = p^2/2 + \cos(x) + Fx$ with system parameters $\hbar = 0.5$ 
and $F = 0.2$: The absolute values of the elements are shown
in a grey scale plot.
With increasing indices the matrix tends to a diagonal one.}
\label{fig2a}
\end{center}
\end{figure}
For $n \rightarrow \pm \infty$, the kinetic term of
the Hamiltonian dominates the potential and the matrix 
$\widetilde{U}$ tends to a diagonal one.
This property is exemplified in Fig.~\ref{fig2a}, where we depict 
the Floquet-Bloch matrix for the potential $V(x) = \cos(x)$. 
%
Suppose the effect of the off-diagonals elements can be neglected 
for $|n| > N$. Then we have
\begin{equation}
\label{2c1}
\widetilde{U}_{m,n} \approx u_m \delta_{m,n} 
\qquad {\rm for}\,\, |m|,|n| > N\, 
\end{equation}
with 
\begin{equation}
\label{2c2}
u_m =\exp\left( - \frac{\rm i}{2\hbar} \int_0^{T_B} 
{(\hbar m -Ft)}^2 \, {\rm d}t \right) 
= \exp\left( \frac{\rm i \hbar^2}{6F}\left[ (m-1)^3 - m^3\right] 
\right) \;.
\end{equation}
For the unscattered states $\Phi_0(\lambda)$ the formulas
(\ref{2c1}) hold exactly for any $m$ and,
given a energy $E$ or $\lambda=\exp(-iE/F)$, the eigenvalue
equation can be solved to yield the discrete version of 
the Airy function in the momentum representation: 
$C_0(m)= \exp({\rm i}\hbar^2 m^3/6F -{\rm i}E m/F)$.
With the help of the last equation we have
\begin{equation}
\label{2c3}
\frac{C_0(n)}{C_0(-n)} = \exp\left[ {\rm i}\,\frac{\hbar^2 n^3}{3 F}
-{\rm i} \frac{2 E n}{F}\right] \;,
\end{equation}
which can be now substituted into the S-matrix definition (\ref{2b7}).

We proceed with the scattering states $\Phi_S(\lambda)$.  
Suppose we order the $C_S$ with indices increasing from bottom to top.
Then we can decompose the vector $C_S$ into three parts, 
\begin{equation}
\label{2c4}
C_S = \left(
\begin{array}{l}
C_S^{(+)}\\[1mm]
C_S^{(0)}\\[1mm]
C_S^{(-)}
\end{array}
\right) \;,
\end{equation}
where $C_S^{(+)}$ contains the coefficients for $n > N$, $C_S^{(-)}$ 
contains the coefficients for $n < -N\!-\!1$ and $C_S^{(0)}$ contains 
all other coefficients for $-N\!-\!1\le n \le N$. The 
coefficients of $C_S^{(+)}$ recursively depend on the coefficient 
$C_S(N)$, via 
\begin{equation}
\label{2c5}
C_S(m+1) = (\lambda/u_{m+1})\,  C_S(m) \qquad {\rm for} \quad m \ge N\, .
\end{equation}
Analogously, the coefficients of $C_S^{(-)}$ recursively depend 
on $C_S(-N-1)$, via
\begin{equation}
\label{2c6}
C_S(m) = (u_{m+1}/\lambda)\,  C_S(m+1) \qquad {\rm for} \quad m<-N-1\, .
\end{equation}
Let us define the matrix $W$ as the matrix $\widetilde{U}$, 
truncated to the size $(2N+1) \times (2N+1)$. 
Furthermore, let $B_N$ be the matrix $W$ accomplished by zero
column and row vectors:
\begin{equation}
\label{2c7}
B_N = \left( \begin{array}{cc}
        \vec{0}^{\rm t} & 0 \\
        W & \vec{0} \\
      \end{array}\right) \;.
\end{equation}
Then the resulting equation for $C_S^{(0)}$ can be written as
\begin{equation}
\label{2c8}
\left(B_N - \lambda \openone \right) C_S^{(0)}
= -u_{N+1} \, C_S(N+1) \, e^1\, ,
\end{equation}
where $e^1$ is a vector of the same length as $C_S^{(0)}$, 
with the {\it first} element equal to one and all others equal to zero. 
For a given $\lambda$, Eq.~(\ref{2c8}) matches the 
asymptotes $C_S^{(+)}$ and $C_S^{(-)}$ by linking $C_S^{(+)}$,
via $C_S(N+1)$ and Eq.~(\ref{2c5}), to $C_S^{(0)}$ and, via $C_S(-N-1)$ 
and Eq.~(\ref{2c6}), to $C_S^{(-)}$. Let us now introduce the row vector 
$e_1$ with all elements equal to zero except the {\it last} one, which 
equals one. Multiplying $e_1$ with
$C_S^{(0)}$ yields the last element of the latter one, 
i.e.~$C_S(-N-1)$. Assuming that $\lambda$ is not an eigenvalue
of the matrix $B_N$ (this case is treated in the next section) we can 
multiply Eq.~(\ref{2c8}) with the inverse of $(B_N - \lambda \openone)$,
which yields
\begin{eqnarray}
\label{2c8a}
\frac{C_S(-N-1)}{C_S(N+1)}=-u_{N+1}
\,e_1 \left[\, B_N - {\rm e}^{-{\rm i}E/F}\, 
\openone\,\right]^{-1}e^1 \;. 
\end{eqnarray}
Finally, substituting Eq.~(\ref{2c3}) and Eq.~(\ref{2c8a}) into
Eq.~(\ref{2b7}), we obtain
\begin{eqnarray}
\label{2c9}
S(E) = \lim_{N\rightarrow \infty} A(N+1)
\,e_1 \left[\, B_N - {\rm e}^{-{\rm i}E/F}\, 
\openone\,\right]^{-1}e^1 \;, 
\end{eqnarray}
with a phase factor $A(N)=-u_N C_0(N)/C_0(-N)$, which ensures
the convergence of the limit $N\rightarrow\infty$. The derived
Eq.~(\ref{2c9}) defines the scattering matrix of the 
Wannier-Stark system and is one of our basic equations.

To conclude this section, we note that Eq.~(\ref{2c9}) also
provides a direct method to calculate the so-called Wigner delay time
\begin{eqnarray}
\label{2c10}
\tau(E) = - {\rm i}\,\hbar \,\frac{\partial \ln S(E)}{\partial E} 
=-2\hbar \,\frac{\partial\varphi(E)}{\partial E} \;.
\end{eqnarray}
As shown in Ref.~\cite{PRE2},
\begin{eqnarray}
\label{2c11}
\tau(E) =   \lim_{N\rightarrow \infty} \frac{\hbar}{F}
\left[\, \left( C_S^{(0)}, C_S^{(0)}\right) - 2(N+1) \,\right] \;.
\end{eqnarray}
Thus, one can  calculate the delay time from the norm of the $C_S^{(0)}$, 
which is preferable to (\ref{2c10}) from the numerical point of view,
because it eliminates an estimation of the derivative.
In the subsequent sections, we shall use the Wigner delay time to analyze  
the complex spectrum of the Wannier-Stark system.

\section{Calculating the poles of the S-matrix}
\label{sec2d}

Let us recall the S-matrix definitions for the Stark system, 
\begin{eqnarray}
\label{2d0}
S(E) &=& \lim_{k \rightarrow \infty} 
\frac{\Psi_S(-k;E)}{\Psi_S(k;E)} \, \frac{\Psi_0(k;E)}{\Psi_0(-k;E)}
=  \lim_{n \rightarrow \infty} 
\frac{C_S(-n)}{C_S(n)} \, \frac{C_0(n)}{C_0(-n)} \;.
\end{eqnarray} 
The S-Matrix is an analytic function of the (complex) energy, 
and we call its isolated poles located in the lower half 
of the complex plane, i.e.~those which have an imaginary part 
less than zero, resonances. 
In terms of the asymptotes of the scattering states, resonances 
correspond to scattering states with purely outgoing asymptotes, 
i.e.~with no incoming wave. (These are the so-called
Siegert boundary conditions \cite{Sieg39}.) 
As one can see directly from (\ref{2c3}),
poles cannot arise from the contributions of the free solutions.
In fact, $C_0(n)/C_0(-n)$ decreases exponentially as a function of $n$
for complex energies $\mathcal{E} = E - {\rm i}\Gamma/2$. 
Therefore, poles can arise only from the scattering states $C_S$.

Actually, we already noted the condition for poles in the
previous section. In the step from equation (\ref{2c8}) 
to the S-matrix formula (\ref{2c9}) 
we needed to invert the matrix $(B_N - \lambda \openone)$.
We therefore excluded the case when $\lambda$ is an eigenvalue of $B_N$.
Let us treat it now. If $\lambda$ is an eigenvalue of $B_N$, 
the equation defining $C^{(0)}_S$ then reads
\begin{equation}
\label{2d1}
\left(B_N - \lambda \openone \right) C_S^{(0)} = 0 \;.
\end{equation}
The scattering state $C_S$ we get contains no incoming wave, 
i.e.~it fulfills the Siegert boundary condition.
In fact, the first element $C^{(0)}_S(N)$ is equal to zero, which follows 
directly from the structure of $B_N$, and consequently $C_S^{(+)}=0$.
In addition, the eigenvalues fulfill 
$|\lambda| \le 1$,\footnote{This property follows directly from 
non-unitarity of $B_N$:
$B_N^\dagger B_N = \openone - e_1^{\rm t} e_1$.}
which in terms of the energy $\mathcal{E} = E -{\rm i}\Gamma/2$ 
means $\Gamma \ge 0$. Let us also note that, according to Eq.~(\ref{2c6}),
the outgoing wave $C_S^{(-)}$ diverges exponentially as 
$C_S^{(-)}(n)\sim |\lambda|^{-n}$. 
\begin{figure}[t]
\begin{center}
\includegraphics[width=7cm]{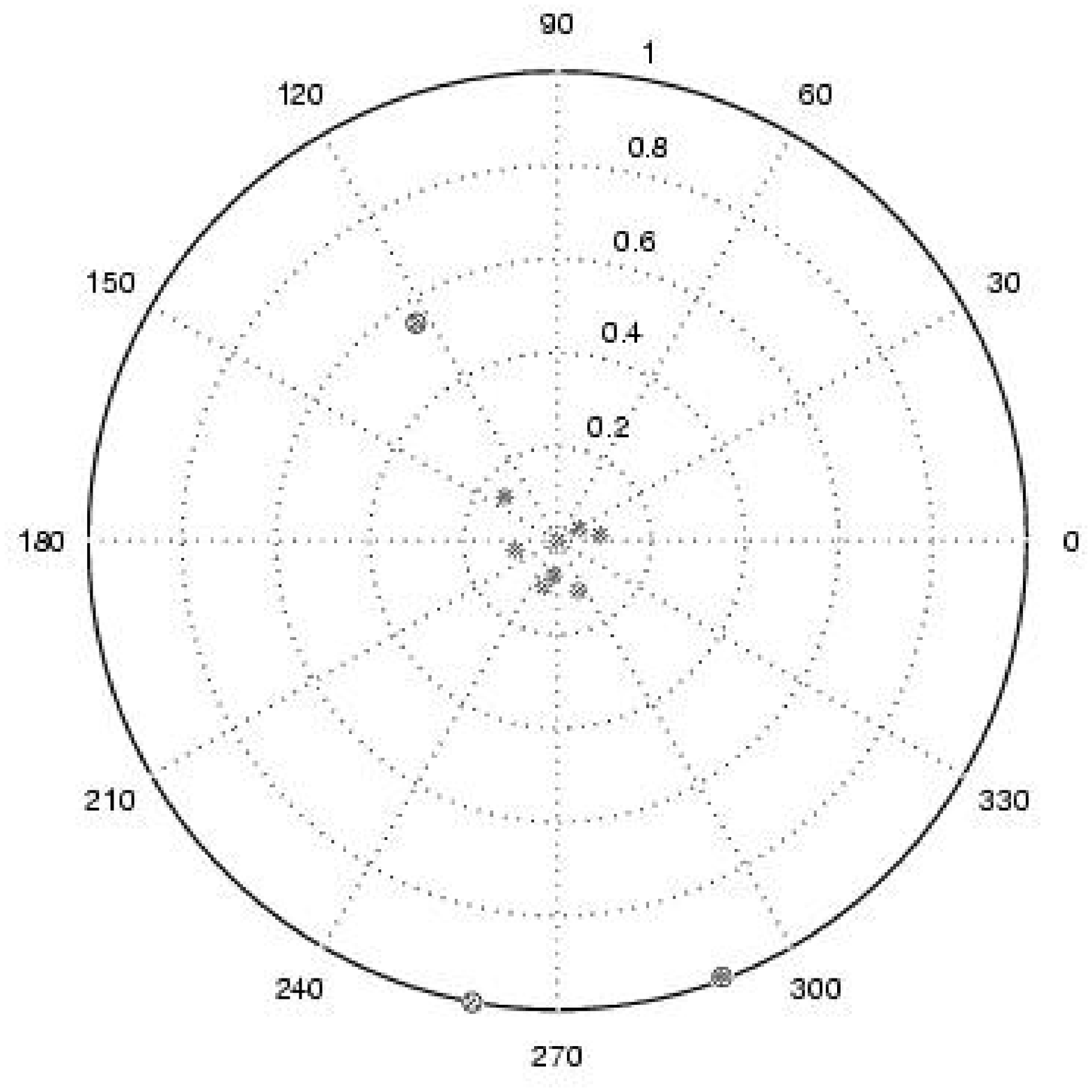}
\hspace*{5mm}
\includegraphics[width=6cm,height=7cm]{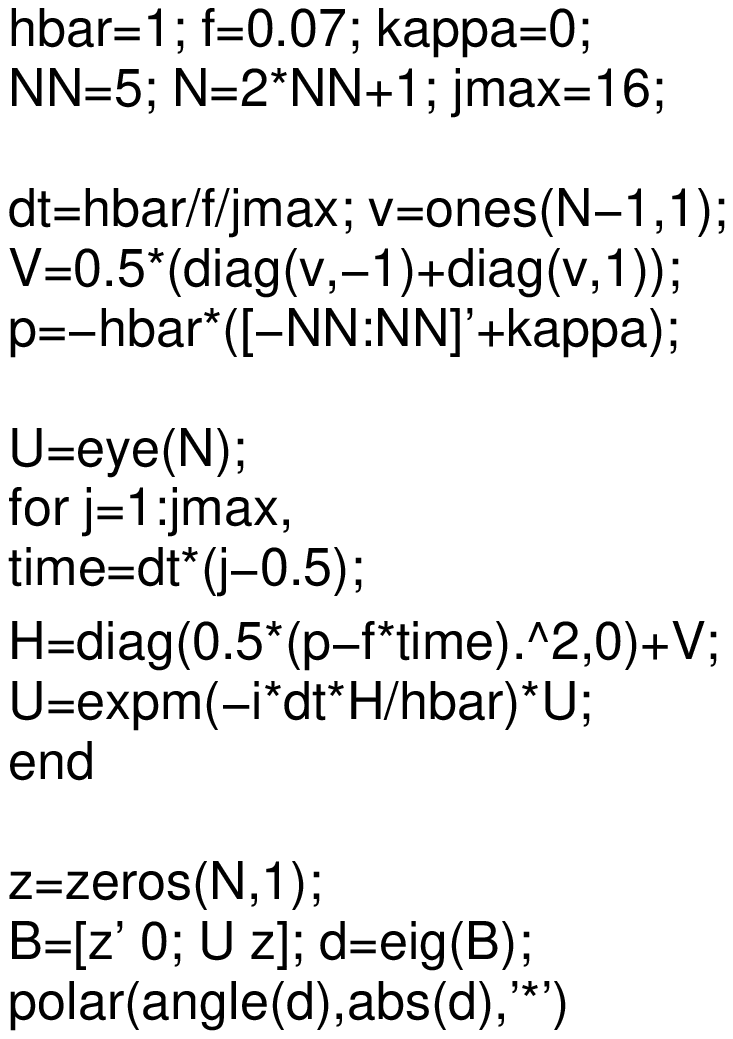}
\end{center}
\caption{\it The eigenvalues $\lambda$ of the matrix $B_N$ calculated
for system (\ref{2a0a}) with $V(x)=\cos x$, $\hbar = 1$ and $F = 0.07$.
The numerical parameters are $N=5$, $j_{\rm max}=16$ and $\kappa=0$.
The eigenvalues corresponding to the first three Wannier-Stark ladders
are marked by circles. On the right to the figure is the {\sc matlab} 
source code which generates the depicted data.}
\label{fig2b}
\end{figure}

Equation (\ref{2d1}) provides the basis for a
numerical calculation of the Wannier-Stark resonances. 
A few words should be said about the numerical algorithm.
The time evolution matrix (\ref{2b1a}) can be calculated by using $2N+1$ 
plane wave basis states 
$\langle x|n\rangle=(2\pi)^{-1/2}\exp({\rm i}nx)$ via
\begin{equation}
\label{2d2}
\widetilde{U}^{(\kappa)} \approx \prod_{j=1}^{j_{\rm max}} \exp \left( - 
\frac{{\rm i}}{\hbar} \widetilde{H}^{(\kappa)}(t_j) \Delta t \right)
\end{equation} 
where $t_j = (j\!-\!1/2)\Delta t$, $\Delta t = T_B/j_{\rm max}$
and $\widetilde{H}^{(\kappa)}(t_j)$ is the truncated matrix of the
operator $\widetilde{H}^{(\kappa)}(t)=(p-Ft+\hbar\kappa)^2/2+V(x)$.
Then, by adding zero elements,  we obtain the matrix $B_N$ and calculate 
its eigenvalues $\lambda$. The resonance energies are
given by $\mathcal{E} = {\rm i} F \ln \lambda$.
As an example, Fig~\ref{fig2b} shows the eigenvalues $\lambda_\alpha$
in the polar representation for the system (\ref{2a0a}) with $V(x)=\cos x$.
Because of the numerical error (introduced by truncation
procedure and round error) not all eigenvalues correspond to the S-matrix
poles. The ``true'' $\lambda$ can be distinguished from the ``false'' $\lambda$
by varying the numerical parameters $N$, $j_{\rm max}$ and the quasimomentum 
$\kappa$ (we recall that in the case of dc field $\lambda$ is independent 
of $\kappa$). The true $\lambda$ are stable against variation of the parameters, 
but the false $\lambda$ are not. In Fig~\ref{fig2b},
the stable $\lambda$ are marked by circles and can be shown (see next
section) to correspond to Wannier-Stark ladders originating from the first
three Bloch bands. By increasing the accuracy, more true $\lambda$ 
(corresponding to higher bands) can be detected. 

\section{Resonance eigenfunctions}
\label{sec2f}

\begin{figure}[t]
\begin{center}
\includegraphics[width=7.5cm,height=6.5cm]{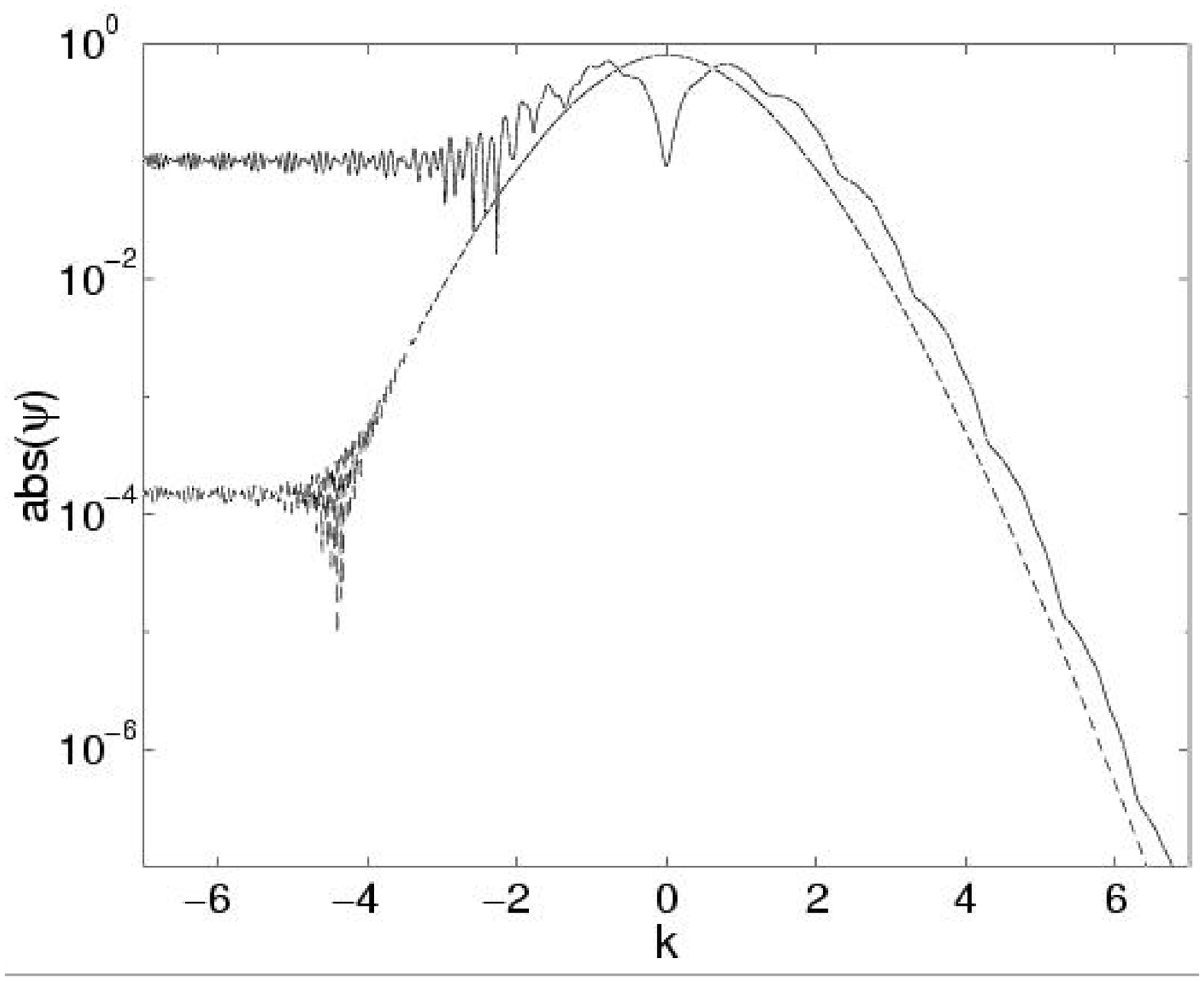}
\hspace{0.3cm}
\includegraphics[width=7.5cm,height=6.5cm]{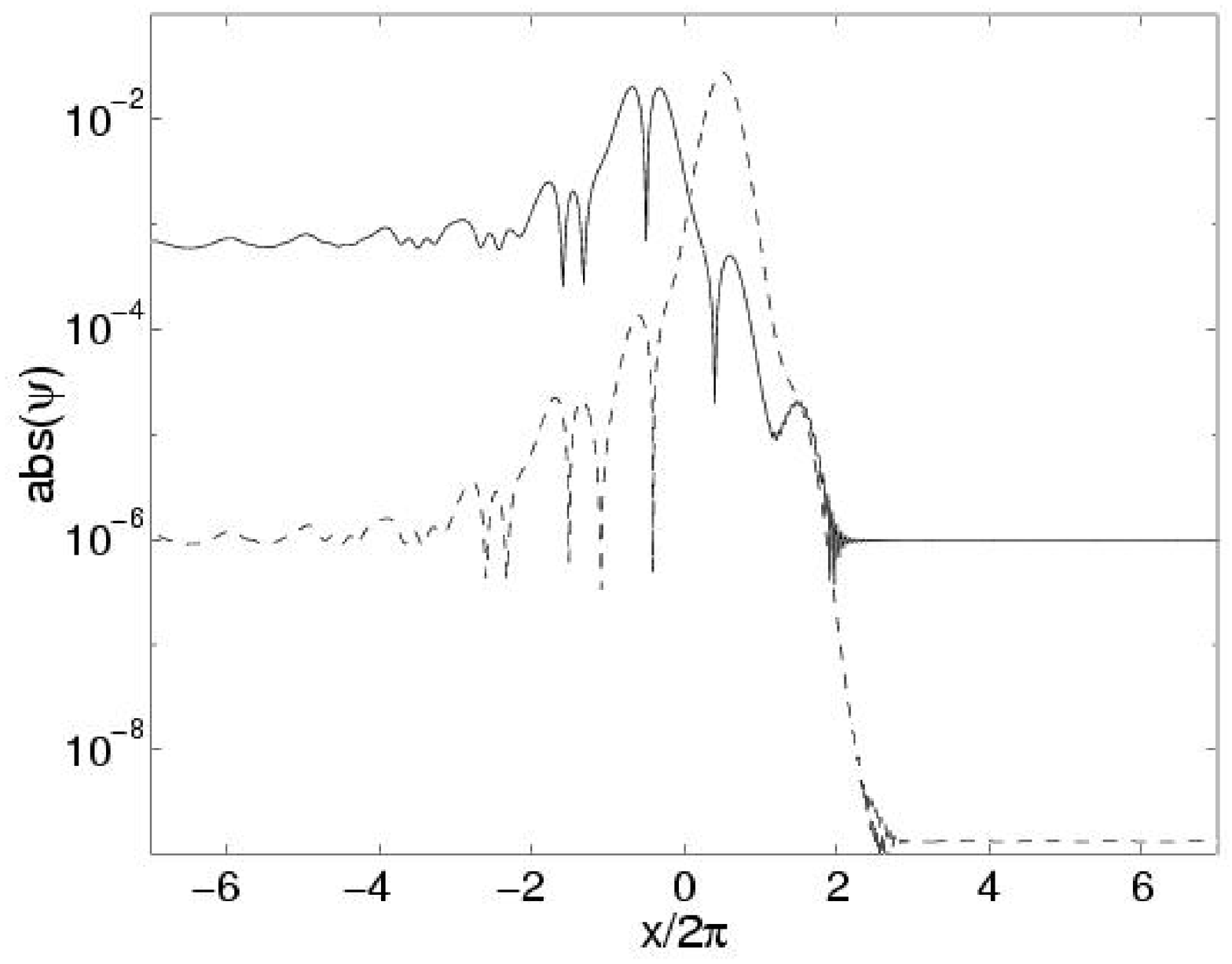}
\caption{\it Resonance wave functions of the two most stable resonances
of system (\ref{2a0a}) with parameters $\hbar = 1$ and $F = 0.07$
in momentum and in configuration space. The ground state is plotted 
as a dashed, the first excited state as a solid line. In the second
figure the first excited state is shifted by one space period to enhance
the visibility.}
\label{fig2c}
\end{center}
\end{figure}

According to the results of preceding section, the resonance Bloch-like
functions $\Phi_{\alpha,\kappa}$, referred to in what follows as the 
Wannier-Bloch functions, are given (in the momentum representation) by
\begin{equation}
\label{2f0}
\Phi_{\alpha,\kappa}(k)=\sum_n C_\alpha(n)\,\delta(n+\kappa-k) \;.
\end{equation} 
where $C_\alpha(n)$ are the elements of the eigenvector of Eq.~(\ref{2d1})
in the limit $N\rightarrow\infty$. The change of the notation
$\Phi_S(\lambda,\kappa)\rightarrow\Phi_{\alpha,\kappa}$ indicates
that from now on we deal with the resonance eigenfunctions
corresponding to the discrete (complex) spectrum $\mathcal{E}_\alpha$.
The Wannier-Stark states $\Psi_{\alpha,l}$, which are the resonance
eigenfunction of the Wannier-Stark Hamiltonian $H_W$, 
are calculated by using  Eq.~(\ref{2b4})
and  Eq.~(\ref{2b5}). In fact, according to  Eq.~(\ref{2b4}), the 
quasimomentum $\kappa$ of the Wannier-Bloch function 
changes linearly with time and explores the whole Brillouin zone during one
Bloch period. Thus, one can obtain the Wannier-Stark states 
$\Psi_{\alpha,l}$ by calculating the eigenfunction $\Phi_{\alpha,\kappa}$
of the evolution operator $U$ for, say, $\kappa=0$ and propagating it over
the Bloch period. (Additionally, the factor $\exp(-{\rm i}\mathcal{E}_\alpha t/\hbar)$
should be compensated.) We used the discrete version of the
continuous evolution operator, given by (\ref{2d2}) with the upper limit 
$j_{\rm max}$ substituted by the actual number of timesteps. 
Resonance Wannier-Stark functions corresponding 
to two most stable resonances are shown in Fig.~\ref{fig2c}.
\begin{figure}[t]
\begin{center}
\includegraphics[width=15.5cm,height=5.5cm]{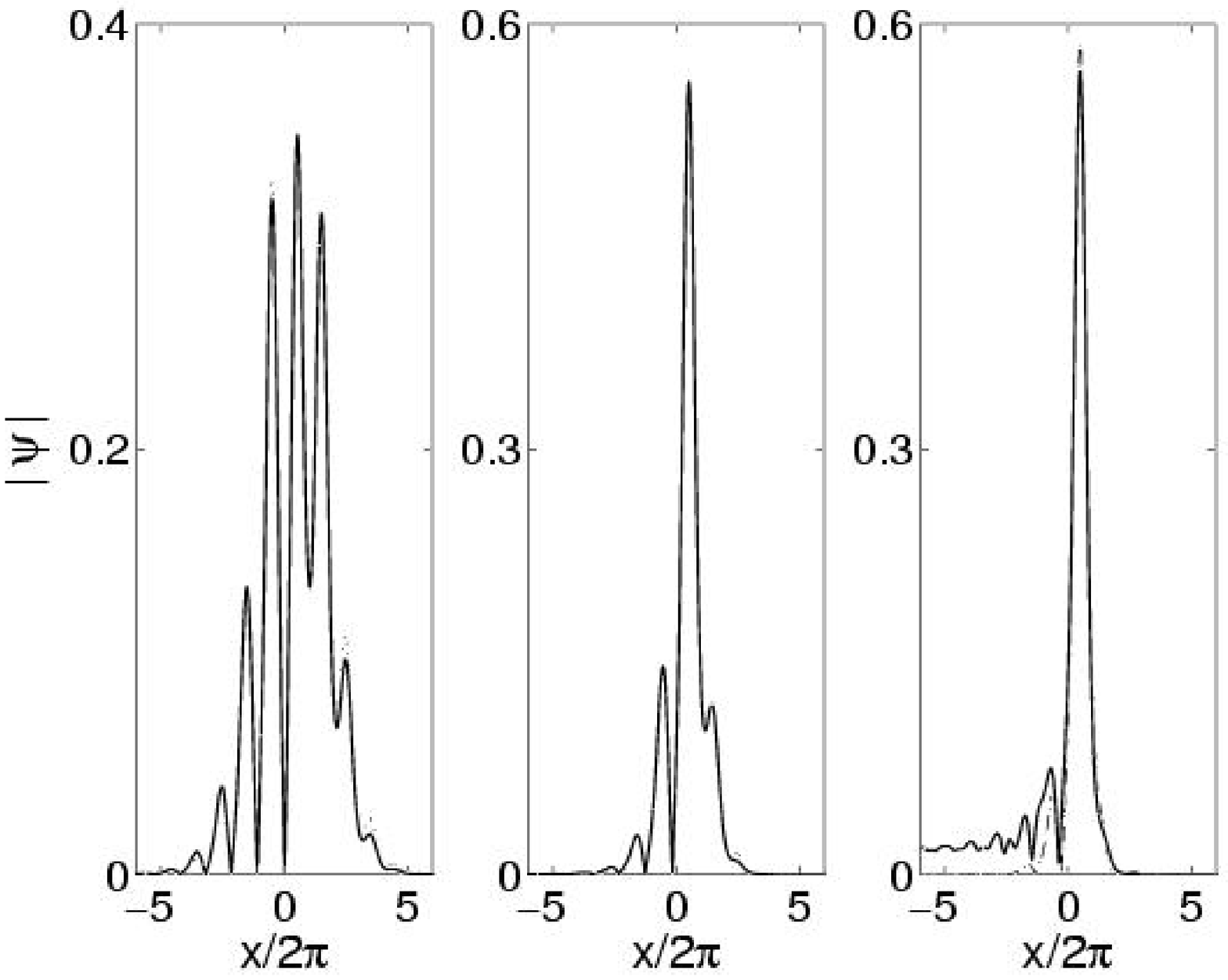}
\includegraphics[width=15.5cm,height=5.5cm]{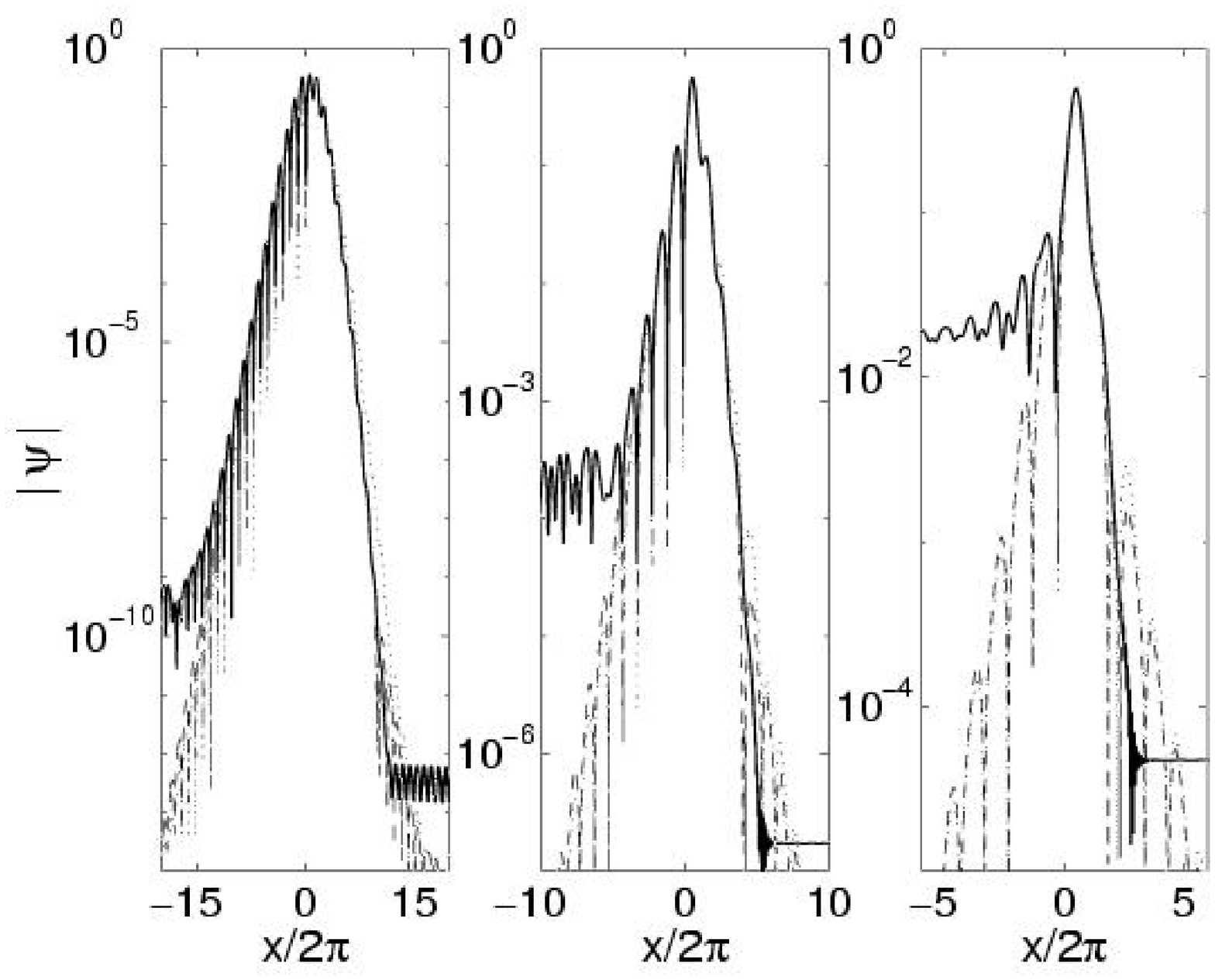}
\caption{\it Comparison of the wave functions calculated within the
different approaches for $\hbar = 2$ and $F = 0.01, 0.03, 0.1$, 
shown on a linear (top) and on a logarithmic scale (bottom).
The dotted line is the tight-binding, the dashed line
the single-band and the solid line is the scattering result.}
\label{fig2d}
\end{center}
\end{figure}

The left panel in Fig.~\ref{fig2c} shows the wave functions in the momentum
representation, where the considered interval of $k=p/\hbar$ is defined
by the dimension of the matrix $B_N$, i.e. $|k|\le N$. The (faster than
exponential) decrease in the positive direction is clearly visible.
The tail in the negative direction reflects the decay of resonances.
Although it looks to be constant in the figure, its magnitude actually
increases exponentially (linearly in the logarithmic scale of the figure)
as $k\rightarrow -\infty$. 
The wave functions in the coordinate representation (right panel)
are obtained by a Fourier transform. Similar to the momentum space
the resonance wave functions decrease in positive $x$-direction
and have a tail in the negative one. Obviously, a finite momentum
basis implies a restriction to a domain in space, who's size can be
estimated from energy conservation as $|x|\le\hbar^2N^2/2F$.
Additionally the Fourier transformation introduces
numerical errors due to which the wave functions decay
only to some finite value in positive direction. We note, however, that for 
most practical purposes it is enough to know the Wannier-Stark states in the 
momentum representation.

Now we discuss the normalization of the Wannier-Stark states.
Indeed, because of the presence of the exponentially diverging tail, the
wave functions $\Psi_{\alpha,l}(k)$ or $\Psi_{\alpha,l}(x)$
can not be normalized in the usual sense. This problem is easily resolved
by noting that for the non-hermitian eigenfunctions (i.e. in the case
considered here) the notion of scalar product is modified as
\begin{equation}
\label{2f1}
\int{\rm d}x \Psi^*_{\alpha,l}(x)\Psi_{\alpha,l}(x)\rightarrow
\int{\rm d}x \Psi^L_{\alpha,l}(x)\Psi^R_{\alpha,l}(x) \;,
\end{equation}
where $\Psi^L_{\alpha,l}(x)$ and $\Psi^R_{\alpha,l}(x)$ are the left
and right eigenfunctions, respectively. In Fig.~\ref{fig2c} the right
eigenfunctions are depicted. The left eigenfunctions can be calculated
in the way described above, with the exception that one begins with the left 
eigenvalue equation $C_S^{(0)}\left(B_N - \lambda \openone \right)=0$
for the row vector $C_S^{(0)}$. In the momentum representation,
the left function $\Psi^L_{\alpha,l}(k)$ coincides with
the right one, mirrored relative to $k=0$. (Note that in coordinate space,
the absolute values of both states are identical.) In other words,
it corresponds to a scattering state with zero amplitude of the
outgoing wave. Since for the right wave function a decay in the positive
$k$-direction is faster than the increase of the left eigenfunction (being
inverted, the same is valid in the negative $k$-direction),
the scalar product of the left and right eigenfunctions is
finite. In our numerical calculation we typically calculate both
functions in the momentum representation and then normalize them
according to
\begin{equation}
\label{2f2}
\int{\rm d}k \Psi^L_{\alpha,l}(k)\Psi^R_{\beta,n}(k)
=\langle\Psi_{\alpha,l}|\Psi_{\beta,n}\rangle
=\delta_{\alpha,\beta}^{l,n} \;.
\end{equation}
(Here and below we use the Dirac notation for the left and right wave
functions.) Let us also recall the relations 
\begin{equation}
\label{2f3}
\Psi_{\alpha,l}(x)=\Psi_{\alpha,0}(x-2\pi l)
\end{equation}
for the wave functions in the coordinate representation and 
\begin{equation}
\label{2f4}
\Psi_{\alpha,l}(k)=\exp({\rm i}2\pi lk)\Psi_{\alpha,0}(k)
\end{equation}
in the momentum space. Thus it is enough to normalize the function
for $l=0$. Then the normalization of the other functions for
$l\ne0$ will hold automatically. For the purpose of future reference
we also display a general (not restricted to the first energy
Brillouin zone) relation between the Wannier-Bloch and Wannier-Stark states
\begin{equation}
\label{2f5}
\Psi_{\alpha,l}=\int_{-1/2}^{1/2} {\rm d}\kappa
\exp(-{\rm i}2\pi l\kappa)\Phi_{\alpha,\kappa} 
\end{equation}
(compare with Eq.~(\ref{1b3}).

It is interesting to compare the resonance Wannier-Stark states with
those predicted by the tight-binding and single-band models.
Such a comparison is given in Fig.~\ref{fig2d}, where the ground
Wannier-Stark state for the potential $V(x)=\cos x$ 
is depicted for three different values of the static force $F$.
As expected, for small $F$, where the resonance is long-lived,
both approximations yield a good correspondence with the exact
calculation. (In the limit of very small $F$ the single-band model
typically gives a better approximation than the tight-binding model.)
In the unstable case, where the resonance state has a visible tail
due to the decay, the results differ in the negative direction.
On logarithmic scale one can see that the order of magnitude up to 
which the results coincide is given by the decay tail of the resonances. 
In the positive $x$-direction the resonance wave functions 
tend to be stronger localized. It should be noted that in Fig.~\ref{fig2d}
we considered the ground Wannier-Stark states only for moderate values
of the static force $F<0.1$. For larger $F$, because of the exponential 
divergence, the comparison of the resonance Wannier-Stark states with
the localized states of the single-band model loses its sense.
The same is also true for higher ($\alpha>0$) states. Moreover, the
value of $F$, below which the comparison is possible, rapidly decreases
with increase of band index $\alpha$.

\chapter{Interaction of Wannier-Stark ladders}
\label{sec3}

In this chapter we give a complete description of
the dependence of the width $\Gamma$ of the Wannier-Stark 
resonances on the parameters of the Wannier-Stark Hamiltonian.
In scaled units, the Hamiltonian has two independent
parameters, the scaled Planck constant $\hbar$ and the
field strength $F$. In our analysis we fix the value of $\hbar$
and investigate the width as a function of the field strength.
The calculated lifetimes $\tau=\hbar/\Gamma$ are
compared with the experimentally measured lifetimes of the Wannier-Stark
states.

\section{Resonant tunneling}
\label{sec3a}

To get a first glimpse on the subject, we calculate the resonances for the 
Hamiltonian (\ref{2a0a}) with $V(x)=\cos x$ for $\hbar=1$.
For the chosen periodic potential
the field-free Hamiltonian has two bands with energies
well below the potential barrier. For the third band, the
energy $\epsilon_2(\kappa)$ can be larger than the potential
height. Therefore, with the field switched on, one expects two 
long-lived resonance states in each potential well, which are 
related to the first two bands.
\begin{figure}[p]
\begin{center}
\includegraphics[width=11cm,height=14cm]{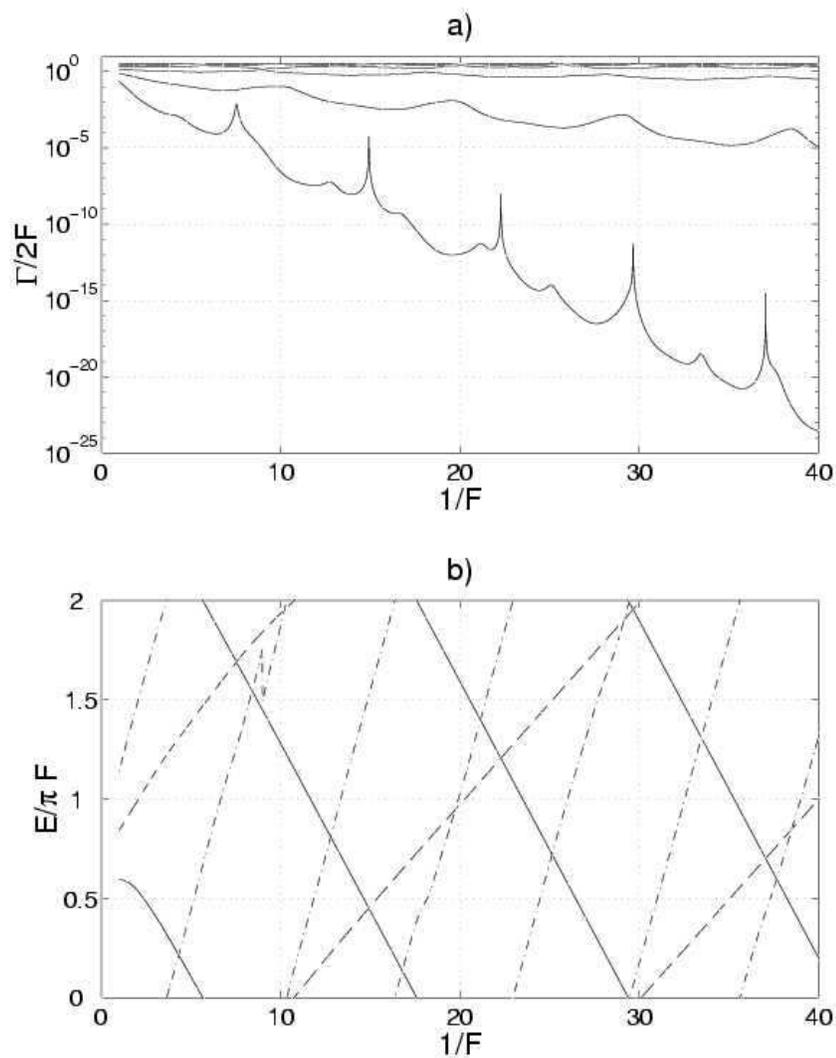}
\end{center}
\caption{\it a) Resonance width of the $6$ most stable resonances as a function 
of the inverse field strength $1/F$. 
b) Energies of the $3$ most stable resonances as a function
of $1/F$ (solid line: most stable resonance, dashed line: 
first excited resonance, dashed dotted line: 
second excited resonance). Parameters are $V(x)=\cos x$ and $\hbar = 1$.}
\label{fig3a}
\end{figure}

Figure \ref{fig3a}(a) shows the calculated widths of the six
most stable resonances as a function of the inverse field 
strength $1/F$. The two most stable resonances are clearly 
separated from the other ones. The second excited resonance can
still be distinguished from the others, the lifetime of 
which is similar. Looking at the lifetime of the most stable state, 
the most striking phenomenon is the existence of very sharp resonance-like
structures, where within a small range
of $F$ the lifetime can decrease up to six orders of magnitude.
In Fig.~\ref{fig3a}(b), we additionally depict the
energies of the three most stable resonances as a function of the
inverse field strength. As the Wannier-Stark resonances are arranged
in a ladder with spacing $\Delta E = 2\pi F$, we show only the
first energy Brillouin zone $0 < E/F < 2\pi$. 
Let us note that the mean slope of the lines in Fig.~\ref{fig3a}(b)
defines the {\em absolute} position $E^*_\alpha$ of the
Wannier-Stark resonances in the limit $F\rightarrow0$. As follows
from the single band model, these absolute positions can be approximated by
the mean energies $\bar{\epsilon}_\alpha$ of the Bloch bands. Depending on the 
value of $E^*_\alpha$, we can identify a particular Wannier-Stark resonance either 
as under- or above-barrier resonance.\footnote{This classification holds only in
the limit $F\rightarrow0$. In the opposite limit all resonances are
obviously above-barrier resonances.} 

Comparing Fig.~\ref{fig3a}(b) with Fig.~\ref{fig3a}(a), we observe 
that the decrease in lifetime coincides with crossings 
of the energies of the Wannier-Stark resonances. All three 
possible crossings manifest themselves in the lifetime: Crossings of the
two most stable resonances coincide with the sharpest peaks in the 
ground state width. The smaller peaks can be found at crossings
of the ground state and the second excited state. Finally, crossings of the
first and the second excited state fit to the peaks in the width of the
first excited state.
\begin{figure}[t]
\begin{center}
\includegraphics[width=10cm]{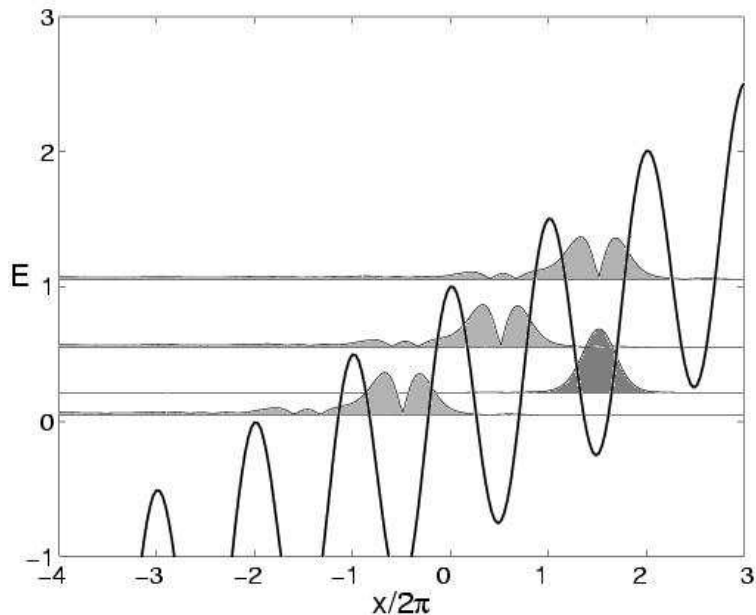}
\end{center}
\caption{\it Wannier-Stark resonances in different minima
of the potential $V(x)=\cos(x) + Fx$: The most stable resonance and some 
members of the first excited Wannier-Stark ladder are shown. 
The parameters are $\hbar = 1.0$ and $F=0.08$.}
\label{fig3b}
\end{figure}
%
The explanation of this effect is the following: Suppose we have a set of
resonances which localize in one of the $2\pi$-periodic 
minima of the potential $V(x) = \cos x + Fx$. 
Let $\Delta E_{\alpha,\beta} = E_\alpha - E_{\beta}$ be the
energy difference between two of these states.  
Now, due to the periodicity of the cosine, each resonance is 
a member of a Wannier-Stark ladder of resonances, i.e.~of a set of resonances
with the same width, but with energies separated by $\Delta E = 2 \pi F$. 
Figure \ref{fig3b} shows an example: The two most stable resonances
for one potential minimum are depicted, furthermore two other
members of the Wannier-Stark ladder of the first excited resonance.
To decay, the ground state has to tunnel three barriers.
Clearly, if there is a resonance with nearly the same energy in one 
of the adjacent minima, this will enhance the decay due to
phenomenon of resonant tunneling.
The strongest effect will be given for degenerate energies, i.e.~for 
$2 \pi F l = \Delta E_{\alpha,\beta}$, which can be achieved by properly 
adjusting $F$, because the splitting $\Delta E_{\alpha,\beta}\approx
E^*_\beta-E^*_\alpha$ is nearly independent of the field strength. For the 
case shown in Fig.~\ref{fig3b}, such a degeneracy will occur, e.g., for 
a slightly smaller value $F \approx 1/14.9$ (see Fig.~\ref{fig3a}). 
Then we have two resonances with the same energies, which are 
separated by two potential barriers. In the next section we formalize
this intuitive picture by introducing a simple two-ladder model.

\section{Two interacting Wannier-Stark ladders}
\label{sec3b}

It is well known that the interaction between two resonances can be well 
modeled by a two-state system \cite{Grec95,Wagner,Hern00,Phil00}. 
In this approach the problem reduces to the diagonalization of a $2\times2$ 
matrix, where the diagonal matrix elements correspond to the
non-interacting resonances. In our case, however, we have ladders of
resonances. This fact can be properly taken into account by introducing the 
diagonal matrix in the form \cite{PRL2,JOB2}
\begin{equation}
U_0=\exp\left(-{\rm i}\,\frac{H_0}{F}\right) \;,\quad
H_0=\left(\begin{array}{cc}
E_0-{\rm i}\Gamma_0/2 & 0 \\ 0 & E_1-{\rm i}\Gamma_1/2
\end{array}\right) \;.
\label{3b0}
\end{equation}
It is easy to see that the eigenvalues $\lambda_{0,1}(F)
=\exp[-{\rm i}(E_{0,1}-{\rm i}\Gamma_{0,1}/2]/F)$ of $U_0$ correspond to
the relative energies of the Wannier-Stark levels and, thus, the matrix 
$U_0$ models two crossing ladders of resonances.\footnote{
The resonance energies in Eq.~(\ref{3b0}) actually depend on $F$ but, 
considering a narrow interval of $F$, this dependence can be neglected.} 
Multiplying the matrix $U_0$ by the matrix
\begin{equation}
U_{\rm int}=\exp\left[{\rm i}\epsilon\left( \begin{array}{cc}
0 & 1 \\ 1 & 0 
\end{array}\right) \right] 
=\left(\begin{array}{cc}
\cos\epsilon & {\rm i}\sin\epsilon \\ {\rm i}\sin\epsilon & \cos\epsilon
\end{array}\right) \;,
\label{3b1}
\end{equation}
we introduce an interaction between the ladders. The matrix $U_0 U_{\rm int}$
can be diagonalized analytically, which yields
\begin{eqnarray}
\lambda_\pm =
\frac{\lambda_0+\lambda_1}{2}\,\cos\epsilon  
\pm\left[\left(\frac{\lambda_0+\lambda_1}{2}\right)^2\cos^2\epsilon
-\lambda_0\lambda_1\right]^{1/2} \;,\quad
\lambda_\pm=\exp\left(-{\rm i}\,\frac{E_\pm-{\rm i}\Gamma_\pm/2}{F}\right) \;. 
\label{3b1a}
\end{eqnarray}
Based on Eq.~(\ref{3b1a}) we distinguish the cases of
weak, moderate or strong ladder interaction.
\begin{figure}[t]
\begin{center}
\includegraphics[width=11cm]{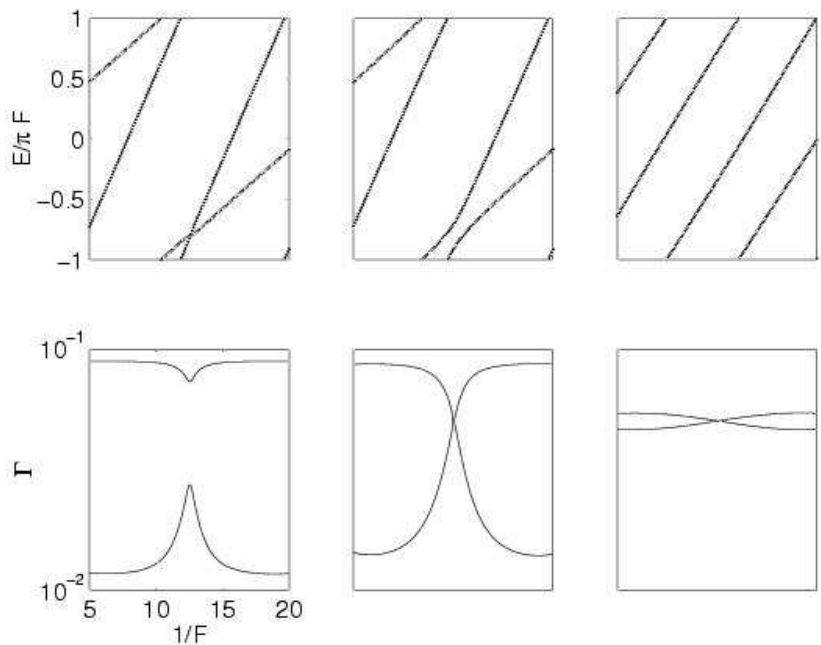}
\caption{\it Illustration to the two-ladder model. Parameters are
$\mathcal{E}_0=0.3-{\rm i}1.1\cdot10^{-2}$, 
$\mathcal{E}_1=0.8-{\rm i}0.9\cdot10^{-1}$, and $\epsilon=0.2$ (left column),
$\epsilon=0.4$ (center), and $\epsilon=\pi/2-0.1$ (right column).
Upper panels show the energies $E_\pm$, lower panels the widths $\Gamma_\pm$.}
\label{fig3c1}
\end{center}
\end{figure}

The value $\epsilon=0$ obviously corresponds to non-interacting ladders. 
By choosing $\epsilon\ne0$ but $\epsilon\ll\pi/2$ we model the case of
weakly interacting ladders. In this case the ladders show true crossing
of the real parts and ``anticrossing'' of the imaginary parts.
Thus the interaction affects only the stability of the ladders. Indeed,
for $\epsilon\ll\pi/2$ Eq.~(\ref{3b1a}) takes the form
\begin{equation}
\lambda_\pm=\lambda_{0,1}\left(1\pm\frac{\epsilon^2}{2}
\frac{\lambda_{0}+\lambda_{1}}{\lambda_{1}-\lambda_{0}}\right) \;.
\label{3b2}
\end{equation}
It follows from the last equation that at the points of crossing (where the phases
of $\lambda_0$ and $\lambda_1$ coincide) the more stable ladder (let it be
the ladder with index 0, i.e. $\Gamma_0<\Gamma_1$ or $|\lambda_0|>|\lambda_1|$)
is destabilized ($|\lambda_+|<|\lambda_0|$) and, vice versa, the less stable 
ladder becomes more stable ($|\lambda_-|>|\lambda_1|$). The case of weakly
interacting ladders is illustrated by the left column in Fig.~\ref{fig3c1}.

By increasing $\epsilon$ above $\epsilon_{\rm cr}$,
\begin{equation}
\sin^2\epsilon_{\rm cr}=\left(
\frac{|\lambda_0|-|\lambda_1|}{|\lambda_0|+|\lambda_1|}\right)^2\;,
\label{3b3}
\end{equation}
the case of moderate interaction, where the true crossing of the real parts
$E_\pm$ is substituted by an anticrossing, is met. As a consequence, 
the interacting Wannier-Stark ladders exchange their stability index at the point 
of the avoided crossing (see center column in Fig.~\ref{fig3c1}). 
The maximally possible interaction is achieved by 
choosing $\epsilon=\pi/2$. Then the eigenvalues of the matrix $U_0 U_{\rm int}$ are 
$\lambda_\pm=\pm {\rm i}(\lambda_0\lambda_1)^{1/2}$ which corresponds to the 
``locked'' ladders
\begin{equation}
E_\pm=(E_0+E_1)/2 \pm \pi F/2  \;,\quad
\Gamma_\pm=(\Gamma_0+\Gamma_1)/2  \;.
\label{3b4}
\end{equation}
In other words, the energy levels of one Wannier-Stark ladder are located
exactly in the middle between the levels of the other ladder (right column
in Fig.~\ref{fig3c1}).

\section{Wannier-Stark ladders in optical lattices}
\label{sec3c}

In the following two sections we give a comparative analysis of the ladder
interaction in optical and semiconductor superlattices.
It will be shown that the character of the interaction can be qualitatively deduced
from the Bloch spectrum of the system.
\begin{figure}[t]
\begin{center}
\includegraphics[width=10cm]{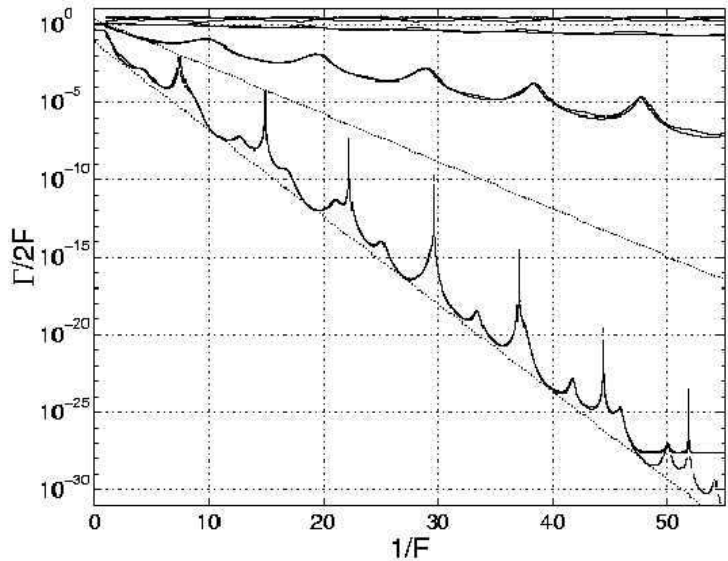}
\caption{\it Widths of the 6 most stable resonances as a function of the
inverse field $F$ for $\hbar = 1.0$ (solid lines) compared with
the fit data (dashed lines).}
\label{fig3c0}
\end{center}
\end{figure}

We begin with the optical lattice, which realizes the case of a
cosine potential (see Sec.~\ref{sec1d}). A characteristic feature
of the cosine potential is an exponential decrease of the band gaps 
as $E\rightarrow\infty$ [see Fig.~\ref{fig1a}(a), for example].
In order to get a satisfactory description of the ladder
interaction for $F\ne0$, it is sufficient to consider only the under-barrier
resonances and one or two above-barrier resonances. In particular,
for the parameters of Fig.~\ref{fig3a} it is enough to ``keep track'' of the
resonances belonging to the first three Wannier-Stark ladders. It is also
seen in Fig.~\ref{fig3a} that the case of true crossings of the resonances
is realized almost exclusively, i.e. the ladders are weakly interacting 
(which is another characteristic property of the cosine potential). 
The behavior of the resonance widths $\Gamma_\alpha(F)$ at the vicinity
of a particular crossing is captured by Eq.~(\ref{3b2}).
Moreover, extending the two-ladder model of the previous section to the 
three ladder case and assuming the coupling constants in the form
\begin{equation}
\label{3c1}
\epsilon_{\alpha}=a_{\alpha}\exp(-b_{\alpha}/F) \;,
\end{equation}
(which is suggested by the semiclassical arguments of Sec.~\ref{sec1c})
the overall behavior of the resonance width can be perfectly reproduced
(see Fig.~\ref{fig3c0}). The procedure of adjustment of the model
parameters $a_{\alpha}$ and $b_{\alpha}$ is carefully 
described in Ref.~\cite{JOB2}.
\begin{figure}[t]
\begin{center}
\includegraphics[width=10cm]{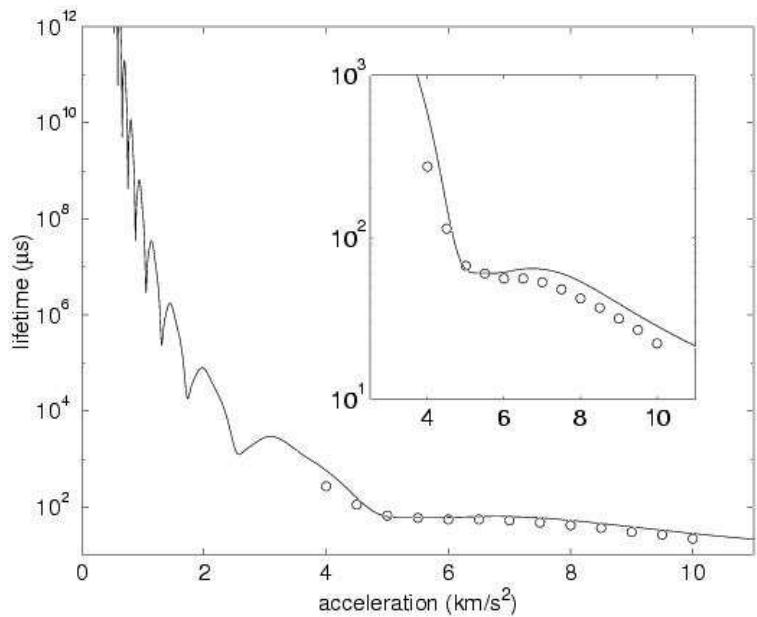}
\caption{\it Lifetime of the ground Wannier state as a function of the 
external field. The solid line is the theoretical prediction, 
the circles are the experimental data of Ref.~\cite{Bhar97}. The insert 
blows up the interval $4000m/s^2 < a < 10000m/s^2$ considered in the 
cited experiment.}
\label{fig3c}
\end{center}
\end{figure}

The lifetime of the Wannier-Stark states (given by 
$\tau=\hbar/\Gamma_\alpha$) as the function of static force was
measured in an experiment with cold sodium atoms in a laser field \cite{Bhar97}.
The setting of the experiment \cite{Bhar97} yields the accelerated cosine
potential (the inertial force takes the role of the static field) and
an effective Planck constant $\hbar=1.671$. For this value of the
Planck constant one has only one under-barrier resonance, and the
two-ladder model of Sec.~\ref{sec3b} is already a good approximation
of the real situation. Figure \ref{fig3c} compares the experimental
results for the lifetime of the ground Wannier-Stark states with the
theoretical results. The axes are adjusted to the experimental parameters. 
Namely, the field strength in our description is related to the 
acceleration in the experiment by the formula $F \approx 0.0383 a$, 
where $a$ is measured in $km/s^2$, and the unit of time in our 
description is approximately $1.34 \mu s$. The experimental data
follow closely the theoretical curve. (Explicitly, the analytical form of 
the displayed dependence is given by Eq.~(\ref{3b2}) with 
$\epsilon=a\exp(-b/F)$, $a=1.0$, $b=0.254$.) In particular we note that 
the theory predicts a local minimum of the lifetime at $a = 5000 m/s^2$, which
corresponds to the crossing of the ground and the first excited 
Wannier levels in neighboring wells. Unfortunately, the
experimental data do not extend to smaller accelerations, where
the theory predicts much stronger oscillations of the lifetime.

\section{Wannier-Stark ladders in semiconductor superlattices}
\label{sec3d}

We proceed with the semiconductor superlattices. As mentioned in Sec.~\ref{sec1d},
the semiconductor superlattices are often modeled by the square-box potential 
(\ref{1d0a}), where $a$ and $b=d-a$ are the thickness of the alternating semiconductor
layers. For the square-box potential (\ref{1d0a}) the width of the band gaps
decreases only {\em inversely proportional} to the gap's number.
Because of this, one is forced to deal with infinite number of interacting
Wannier-Stark ladders. However, as was argued in Ref.~\cite{PRB}, this
is actually an over-complication of the real situation. Indeed,
the potential (\ref{1d0a}) is only a first approximation for the superlattice
potential, which should be a smooth function of $x$. This fact can 
be taken into account by smoothing the rectangular step in (\ref{1d0a}) as 
\begin{equation}
\label{3c2}
V(x)= \tanh \sigma(x+a\pi/2d)
-\tanh \sigma(x-a\pi/2d) -1 
\end{equation}
for example. (Here we use scaled variables, where the potential is 
$2\pi$-periodic and $|V(x)|\le1$.) The parameter $\sigma^{-1}$ defines the 
size of the transition
region between the semiconductor layers and, in natural units, it cannot
be smaller than the atomic distance. The smoothing
introduces a cut-off in the energy, above which the gaps
between the Bloch bands decrease exponentially.
Thus, instead of an infinite number of ladders associated with the
above-barrier resonances, we may consider a finite number of them.
The interaction of a large number of ladders originating from
the high-energy Bloch bands was studied in some details in Ref.~\cite{PRB}. It was 
found that they typically form pairs of locked [in the sense of 
Eq.~(\ref{3b4})] ladders which show anticrossings with each other.

\begin{figure}[t]
\begin{center}
\includegraphics[width=11cm,clip,angle=0]{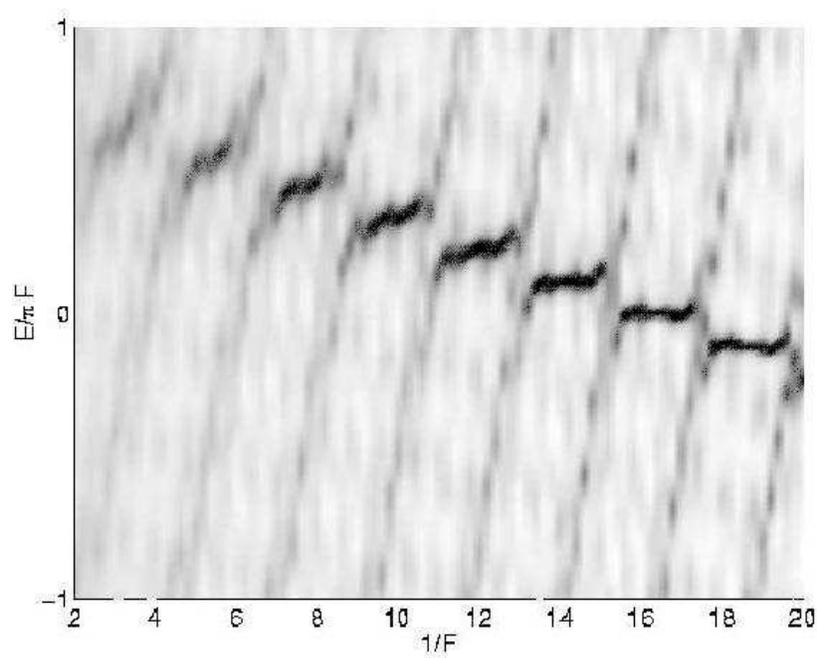}
\end{center}
\caption{\it Grey-scaled map of the Wigner delay time (\ref{3c3}) for the smoothed
square-box potential (\ref{3c2}). The parameters are $\hbar=3.28$,
$a/b=39/76$ and $\delta=0.25$.}  
\label{fig3d}
\end{figure}

Since the lifetime of the above-barrier resonances is much shorter
than the lifetime of the under-barrier resonances one might imagine
that the former are of minor physical importance. Although this is partially 
true, the above-barrier resonances cannot be ignored because they strongly
affect the lifetime of the long-lived under-barrier resonances. This is
illustrated in Fig.~\ref{fig3d}, where the resonance structure of the
Wannier-Stark Hamiltonian with a periodic potential given by Eq.~(\ref{3c2})
and $\hbar=3.28$ is depicted as a gray-scaled map of the Wigner delay
time (\ref{2c10}). In terms of Fig.~\ref{fig3a}, this way of presentation of the 
numerical results means that each line in the lower panel has a 
``finite width'' defined by the value $\Gamma$ in the upper panel.
In fact, asssuming a Wigner relation \cite{Fyod97} we get
\begin{equation}
\label{3c3}
\tau(E)=\tau_0+\sum_\alpha\left(\sum_{l=-\infty}^\infty{\rm Im}\left[
\frac{\hbar }{\mathcal{E}_\alpha+2\pi Fl-E}\right]\right) \;,  
\end{equation}
where each term in the sum over $\alpha$ is just a periodic sequence of
Lorentzians with width $\Gamma_\alpha$. (We recall that, by definition,
$\tau(E)$ is a periodic function of the energy.\footnote{The quantity 
(\ref{3c3}) can be also interpreted as the 
fluctuating part of the (normalized) density of states of the system.}) 
In the case of a large number of interacting ladders
(i.e.~in the case currently considered here, where more than 10 above-barrier 
resonances contribute to the sum over $\alpha$) we find this presentation more 
convenient because it reveals only narrow resonances, while the wide resonances 
contribute to the background compensated by the constant $\tau_0$.
For the chosen value of the scaled Planck constant, $\hbar=3.28$,
the periodic potential (\ref{3c2}) supports only
one under-barrier resonance, seen in the figure as a broken line
going from the upper-left to the lower-right corners. Wide above-barrier
resonances originating from the second and third Bloch bands and showing 
anticrossings with the ground resonance can be still identified,
but the other resonances are indistinguishable because of their large widths.
Nevertheless, the existence of these resonances is confirmed indirectly
by the complicated structure of the ``visible lines''.

In conclusion, in comparison with the optical lattices, the structure of the
Wannier-Stark resonances in semiconductor superlattices is complicated by the 
presence of large number of above-barrier resonances. Besides this, in the 
semiconductor superlattices a strong interaction  between the ladders is 
the rule, while the case of weakly interacting ladders is typical for 
optical lattices.

\chapter{Spectroscopy of Wannier-Stark ladders}
\label{sec4}

In this chapter we discuss the spectroscopy of Wannier-Stark ladders
in optical and semiconductor superlattices. We show how the different 
spectroscopic quantities (measured in a laboratory experiment) can be directly 
calculated by using the formalism of the resonance Wannier-Stark states.

\section{Decay spectrum and Fermi's golden rule}
\label{sec4a}

\begin{figure}[t]
\begin{center}
\includegraphics[width=11cm]{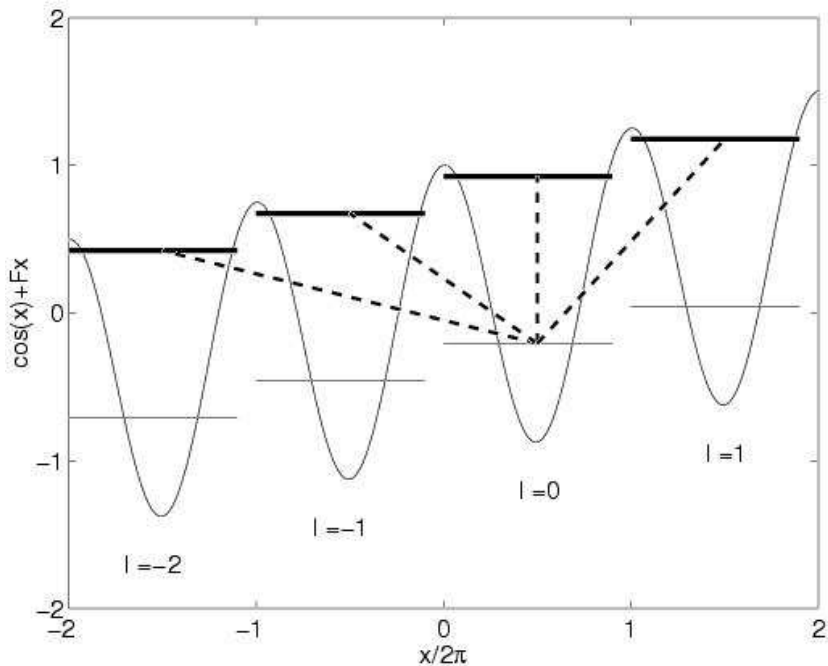}
\end{center}
\caption{\it Schematic illustration of the transitions induced
by a periodic driving. The positions of the ground and the
first excited Wannier-Stark ladder are shown for $F = 0.04$ and
$\hbar = 1.5$. The width of the states is symbolized by the
different strength of the lines.}
\label{fig4a}
\end{figure}

The spectroscopy approach assumes that one probes a quantum system by
a weak ac field $F_\omega x\cos(\omega t)$ with tunable frequency $\omega$.
In our case, the system consists of different Wannier-Stark ladders of 
resonances, the two most stable of which are schematically depicted in 
Fig.~\ref{fig4a}. The driving induces transitions between the ground and 
the excited states\footnote{Actually, transitions within the same
ladder are also induced, but their effect is important only for
$\omega\sim\omega_B=2\pi F/\hbar$. Here we shall mainly consider the case
$\omega\gg\omega_B$, where the transitions within the same ladder can be
ignored.}. Scanning the frequency $\omega$ sequentially activates the different
transition paths and the different Wannier states of the
excited ladder are populated. Because the excited states are
typically short-lived, they decay before the
driving can transfer the population back to the ground state,
i.e.~before a Rabi oscillation is performed. Then the decay rate
of the ground state is determined by the transition rate $D(\omega$) to the 
excited Wannier-Stark ladder.  The width is written as
\begin{equation}
\label{4a0a}
\Gamma_0(\omega)=\Gamma_0+D(\omega) \;,
\end{equation}
where $\Gamma_0$ takes into account the decay in the absence of driving.
In what follows we shall refer to the quantity $\Gamma_0(\omega)$ as the
induced decay rate or the decay spectrum. In Sec.~\ref{sec5} we calculate the
induced decay rate rigorously by using the formalism of quasienergy 
Wannier-Stark states. It will be shown that the decay spectrum is given by 
\begin{equation}
\Gamma_0(\omega)=\Gamma_0+\frac{F^2_\omega}{2} \sum_{\beta>0}\sum_{L}
{\rm Im}\left[ \frac{V_{0,\beta}^2(L)}
{(E_{\beta,l}+2\pi FL-E_{0,l}-\hbar\omega)-{\rm i}\Gamma_\beta/2}\right]  \;,  
\label{4a2}
\end{equation}
where $F_\omega$ and $\omega$ are the amplitude and frequency of 
the probing field and
\begin{equation}
V_{0,\beta }^2(L)=
\langle \Psi _{0,l}|x|\Psi _{\beta,l+L}\rangle 
\langle \Psi _{\beta,l+L}|x|\Psi _{0,l}\rangle   
\label{4a3}
\end{equation}
is the square of the dipole matrix element between an 
arbitrary ground Wannier-Stark state $\Psi_{0,l}(x)$ and the upper  
Wannier-Stark state $\Psi_{\beta,l+L}(x)$ shifted by $L$ lattice periods.
We would like to stress that, because for the resonance wave
functions $\langle \Psi _{\alpha,l}|x|\Psi _{\beta,l'}\rangle\ne
\langle \Psi _{\beta,l'}|x|\Psi _{\alpha,l}\rangle^*$, the square of   
the dipole matrix element $V^2_{0,\beta}(L)$ is generally a complex
number.

To understand the physical meaning of Eq.~(\ref{4a2}),
it is useful to discuss its relation to Fermi's golden rule,
which reads
\begin{equation}
D(\omega )\approx \pi F^2_\omega \int{\rm d}E
\,\Big|\int{\rm d}x\Psi^*_E(x)x\Psi_{E_{0,l}}(x)\Big|^2 
\rho(E)\delta(E-E_{0,l}-\hbar\omega) \;.
\label{4a0}
\end{equation}
in the notations used. In Eq.~(\ref{4a0}), the $\Psi_E(x)$ are the hermitian eigenfunctions of
the Hamiltonian (\ref{2a0}) (i.e., $E$ is real and continuous) and $\rho(E)$ 
is the density of states. For the sake of simplicity we also
approximate the ground Wannier-Stark resonance  by the discrete level 
$E_{0,l}$. Then Eq.~(\ref{4a0}) describes the decay of a discrete level 
into the continuum. Assuming, for a moment, that the continuum is dominated by 
the first excited Wannier-Stark ladder, the density of states $\rho(E)$ 
is given by a periodic sequence of Lorentzians with width $\Gamma_1$, 
i.e.
\begin{equation}
\rho(E)\approx \frac{1}{2\pi}\sum_L
\frac{\Gamma_1}{(E-E_{1,l+L})^2+\Gamma_1^2/4}\,.  
\end{equation}
Substituting the last equation into Eq.~(\ref{4a0}) and integrating over $E$
we have
\begin{equation}
D(\omega )\approx \frac{F^2_\omega}{2}\,
\Big|\int{\rm d}x\Psi^*_{E_{0,l}+\hbar\omega}(x)x\Psi_{E_{0,l}}(x)\Big|^2 
\sum_L \frac{\Gamma_1}{(E_{1,l+L}-E_{0,l}-\hbar\omega)^2+\Gamma_1^2/4} \;.
\label{4a1}
\end{equation}
In the case $\Gamma_1\ll 2\pi F$ the Lorentzians in the right-hand side
of Eq.~(\ref{4a1}) are $\delta$-like functions of the argument
$\hbar\omega=E_{1,l}+2\pi FL-E_{0,l}$. Thus the transition matrix 
element can be moved under the summation sign, which gives
\begin{equation}
D(\omega )\approx \frac{F^2_\omega}{2} \sum_{\beta>0}\sum_{L} 
|\widetilde{V}_{0,\beta}|^2(L)
\frac{\Gamma_\beta}{(E_{\beta,l}+2\pi FL-E_{0,l}-\hbar\omega)^2+\Gamma_\beta^2/4} \;,
\label{4a2a}
\end{equation}
where
\begin{equation}
|\widetilde{V}_{0,\beta }|^2(L)=
\Big|\int{\rm d}x\Psi^*_{E_{\beta,l+L}}(x)x\Psi_{E_{0,l}}(x)\Big|^2 
\label{4a3a}
\end{equation}
(here we again included the possibility of transitions to the higher Wannier 
ladders, which is indicated by the sum over $\beta$). It is seen that that
the obtained result coincides with Eq.~(\ref{4a2}) if
the coefficients $|\widetilde{V}_{0,\beta }|^2(L)$ are
identified with 
the squared dipole matrix elements (\ref{4a3}). Obviously, this 
holds in the limit $F\rightarrow 0$, when the resonance wave functions
can be approximated by the localized states. For a strong field, however,
Eq.~(\ref{4a2a}) is a rather poor approximation of the decay spectrum.
In particular, it is unable to predict the non-Lorentzian shape of the 
lines, which is observed in the laboratory and numerical 
experiments and which is correctly captured in Eq.~(\ref{4a2}) by the complex 
phase of the squared dipole matrix elements $V_{0,\beta }^2(L)$.

To proceed further, we have to calculate the squared matrix elements
(\ref{4a3}). A rough estimate for $V_{0,\beta }^2(L)$ can be obtained on
the basis of Eq.~(\ref{1b6}), which approximates the resonance Wannier-Stark
state by the sum of the localized Wannier states:
$\Psi_{\alpha,l}=\sum_m J_{m-l}(\Delta_\alpha/4\pi F)\psi_{\alpha,m}$.
The typical experimental settings (see Sec.~\ref{sec4c}) correspond to
$\Delta_0/4\pi F\ll 1$ and $\Delta_\beta/4\pi F>1$.
Then the values of the matrix elements are approximately
\begin{equation}
V_{0,\beta}^2(L)\approx |\widetilde{V}_{0,\beta }|^2(L)\approx
|\langle\psi _{0,l}|x|\psi _{\beta,l}\rangle|^2 
J^2_L\left(\frac{\Delta_\beta}{4\pi F}\right) \;,
\label{4a4}
\end{equation}
which contribute mainly in the region $L<\Delta_\beta/4\pi F$,
the localization length of the excited Wannier-Stark states.
The degree of validity of this result is discussed in the next subsection.

\section{Dipole matrix elements}
\label{sec4b}

In this subsection we calculate the dipole matrix elements
\begin{equation}
V_{\alpha,\beta }(l-l')=
\langle \Psi _{\alpha,l}|x|\Psi _{\beta,l'}\rangle 
\label{4b0}
\end{equation}
beyond the tight-binding approximation. We shall use Eq.~(\ref{2f5})
\begin{equation}
\Psi _{\alpha,l}(x)=\int {\rm d} \kappa \,{\rm e}^{-{\rm i}2\pi l\kappa} 
\Phi_{\alpha,\kappa}(x) \;,\quad
\Phi_{\alpha,\kappa}(x)={\rm e}^{{\rm i}\kappa x}\chi_{\alpha,\kappa}(x) \;,\quad
\chi_{\alpha,\kappa}(x)=\chi_{\alpha,\kappa}(x+2\pi) \;,
\label{4b1}
\end{equation}
which relates the Wannier-Stark states $\Psi_{\alpha,l}(x)$ to 
the Wannier-Bloch states $\Phi_{\alpha,\kappa}(x)$. As follows from the
results of Sec.~\ref{sec2}, the function $\chi_{\alpha,\kappa}(x)$ can be
generated from $\chi_{\alpha,0}(x)$ by propagating it in time
\begin{equation}
\label{4b2}
|\chi_{\alpha,\kappa}\rangle
=\exp\left({\rm i}\,\frac{\mathcal{E}_\alpha t}{\hbar}\right)
\widetilde{U}(t)|\chi_{\alpha,0}\rangle \;,
\end{equation}
where $\widetilde{U}(t)$ is the continuous version of the operator
$\widetilde{U}$ defined in Eq.~(\ref{2b1a}) and the quasimomentum $\kappa$
is related to time $t$ by $\kappa=-Ft/\hbar$. Substituting Eq.~(\ref{4b1}) 
and Eq.~(\ref{4b2}) into Eq.~(\ref{4b0}) we obtain the dipole
matrix elements as the Fourier image
\begin{equation}
\label{4b3}
V_{\alpha,\beta}(l-l')=
2\pi l\,\delta_{\alpha,\beta}^{l,l'}+\int {\rm d}\kappa \,
{\rm e}^{{\rm i}2\pi(l-l')\kappa}X_{\alpha,\beta}(\kappa) 
\end{equation}
of the periodic function
\begin{equation}
\label{4b4}
X_{\alpha,\beta}(\kappa)={\rm i}\,\langle\chi_{\alpha,\kappa}
|\,\frac{\partial}{\partial\kappa}\chi_{\beta,\kappa}\rangle 
=\frac{1}{F}\,\langle\chi_{\alpha,\kappa}|\,\frac{(p+\hbar \kappa)^2}{2}+V(x)\,|\chi_{\beta,\kappa }\rangle 
-\frac{\mathcal{E}_\alpha}{F}\,\delta_{\alpha,\beta}   \;.
\end{equation}
The last two equations provide the basis for numerical calculation
of the transition matrix elements. We also recall that one actually
needs the square of the matrix elements (\ref{4a3}) but not the
matrix elements themselves (which are defined up to an arbitrary phase).
Thus we first calculate $V_{\alpha,\beta}(L)$ and
$V_{\beta,\alpha}(L)$ for $L=0,\pm1,\ldots$ and then multiply them
term by term.
\begin{figure}[t]
\begin{center}
\includegraphics[width=12cm]{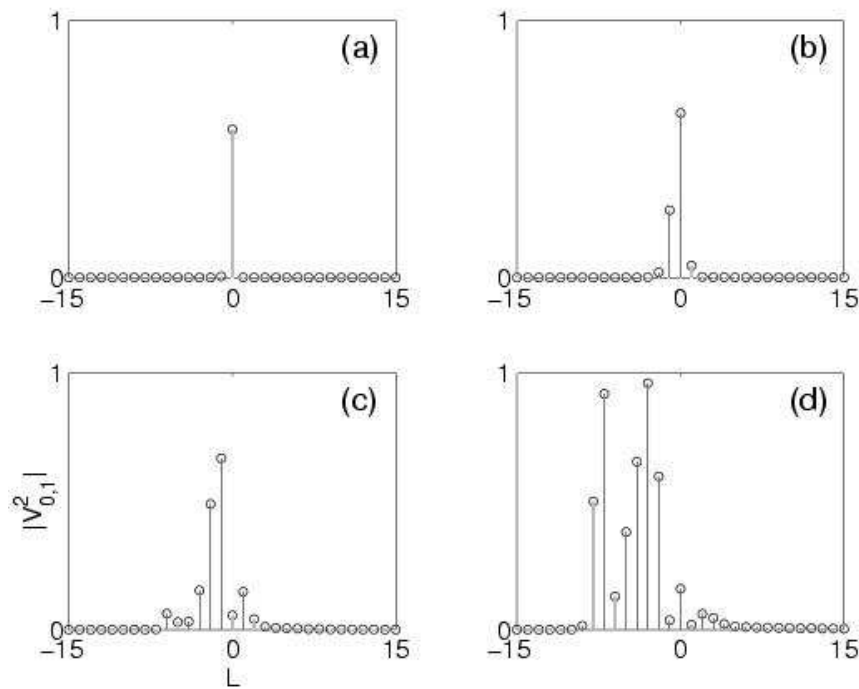}
\caption{\it The absolute values of the squared dipole matrix elements
(\ref{4a3}) for $V(x)=\cos x$, $F=0.04$ and $\hbar=1$ (a), $1.5$ (b),
$2$ (c), and $2.5$ (d).}
\label{fig4b}
\end{center}
\end{figure}

In Fig.~\ref{fig4b} we depict the squared dipole matrix elements
between the ground and first excited Wannier-Stark states
for $V(x)=\cos x$, a moderate values of the static force $F=0.04$ and 
values of the scaled Planck constant in the interval 
$1\le\hbar\le2.5$. For $\hbar=1.0$ the Bloch
bands width $\Delta_1\approx 0.05$ is much smaller than $4\pi F\approx 0.5$
and the upper Wannier-Stark state is essentially localized within single
potential well.\footnote{The ground Wannier-Stark state is localized within one well
for all considered values of the scaled Planck constant.} 
Then only ``vertical" transitions, $L=0$,  are possible between the ground 
and first excited Wannier ladders. 
By increasing $\hbar$ the localization length of the upper
state grows (proportional to the band width) and more than one matrix 
element may differ from zero. Simultaneously,
the Wannier levels move towards the top of the potential barrier
(for $\hbar>1.6$ the upper Wannier level is already above the 
potential barrier) and the Wannier state looses its stability
($\Gamma_1=1.90\cdot10^{-15}$, $1.35\cdot10^{-2}$, $5.24\cdot10^{-2}$, 
and $1.14\cdot10^{-1}$, for $\hbar=1$, 1.5, 2, and 2.5). Because for 
short-lived resonances the tight-binding result (\ref{1b6}) is a rather poor 
approximation of the resonance wave functions,
we observe an essential deviation from Eq.~(\ref{4a4}). In particular
we note a strong asymmetry of the matrix elements with respect to $L$. 
It appears that the transitions ``down the ladder" are enhanced
in comparison with the transitions ``up the ladder". At the same time,
for weak far transitions ($L\gg1$) the situation is reversed 
[see Fig.~\ref{fig4b}(d) and Fig.~\ref{fig4d}(b) below].
\begin{figure}[t]
\begin{center}
\includegraphics[width=5.2cm,height=6.5cm]{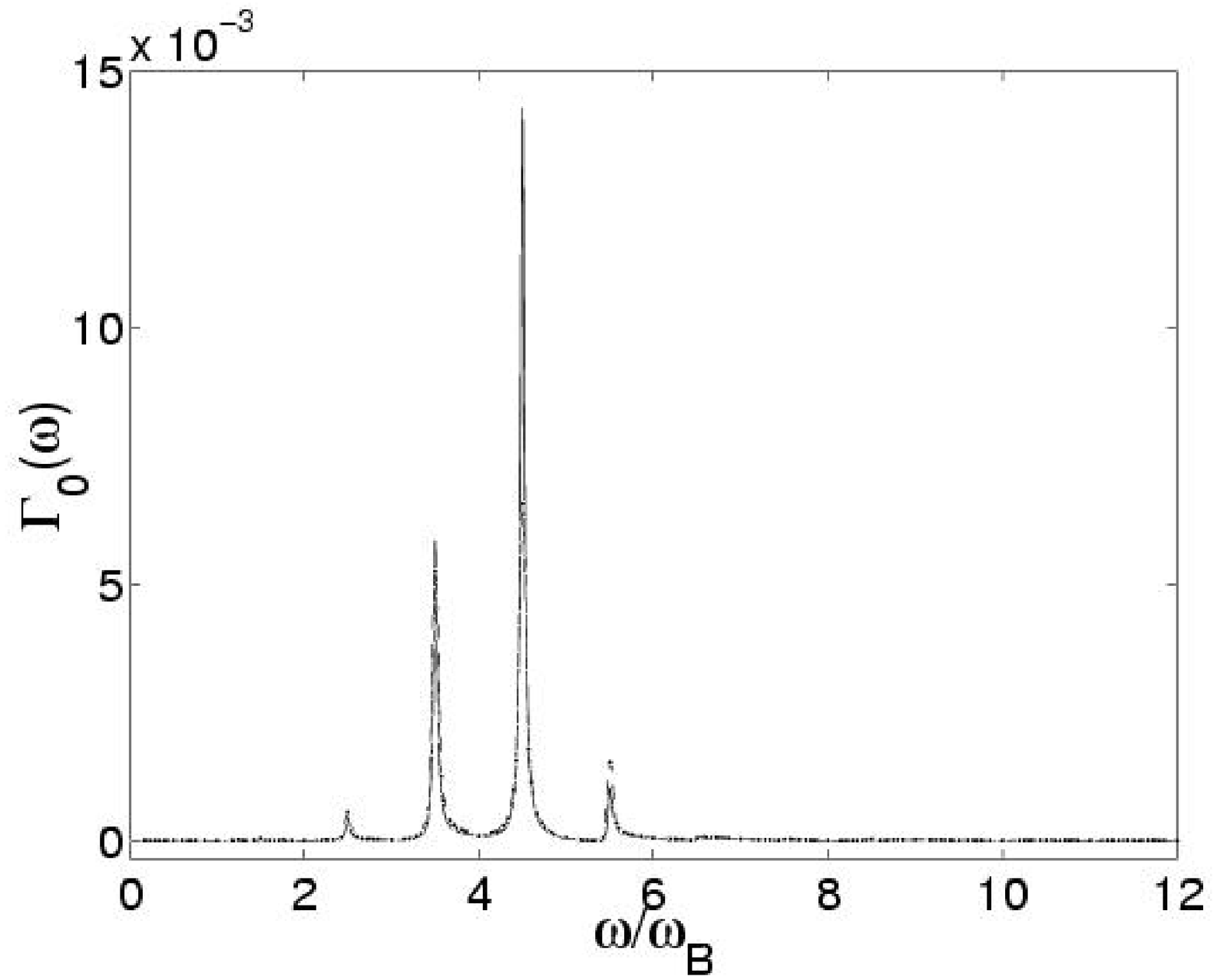}
\includegraphics[width=5.2cm,height=6.5cm]{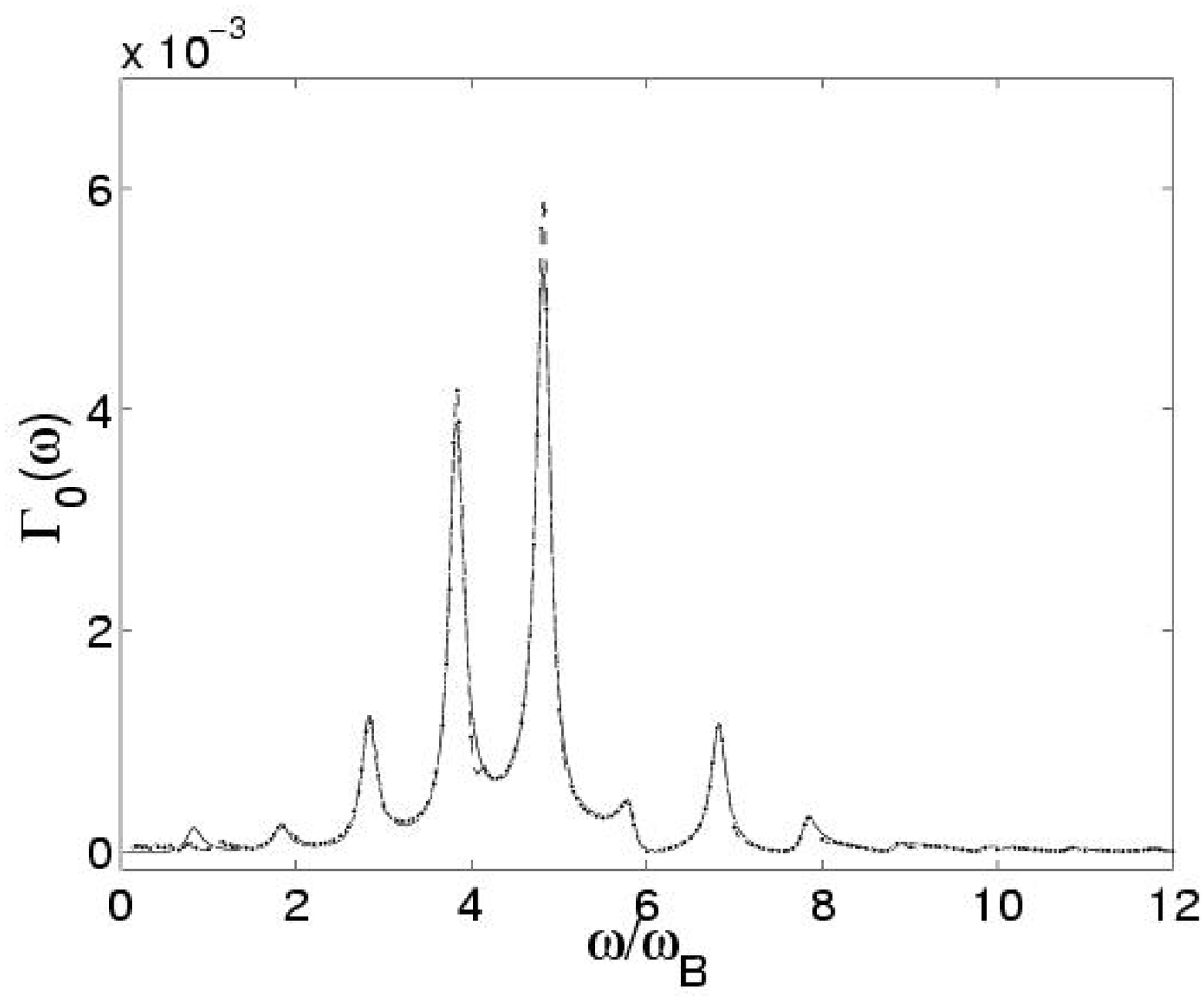}
\includegraphics[width=5.2cm,height=6.5cm]{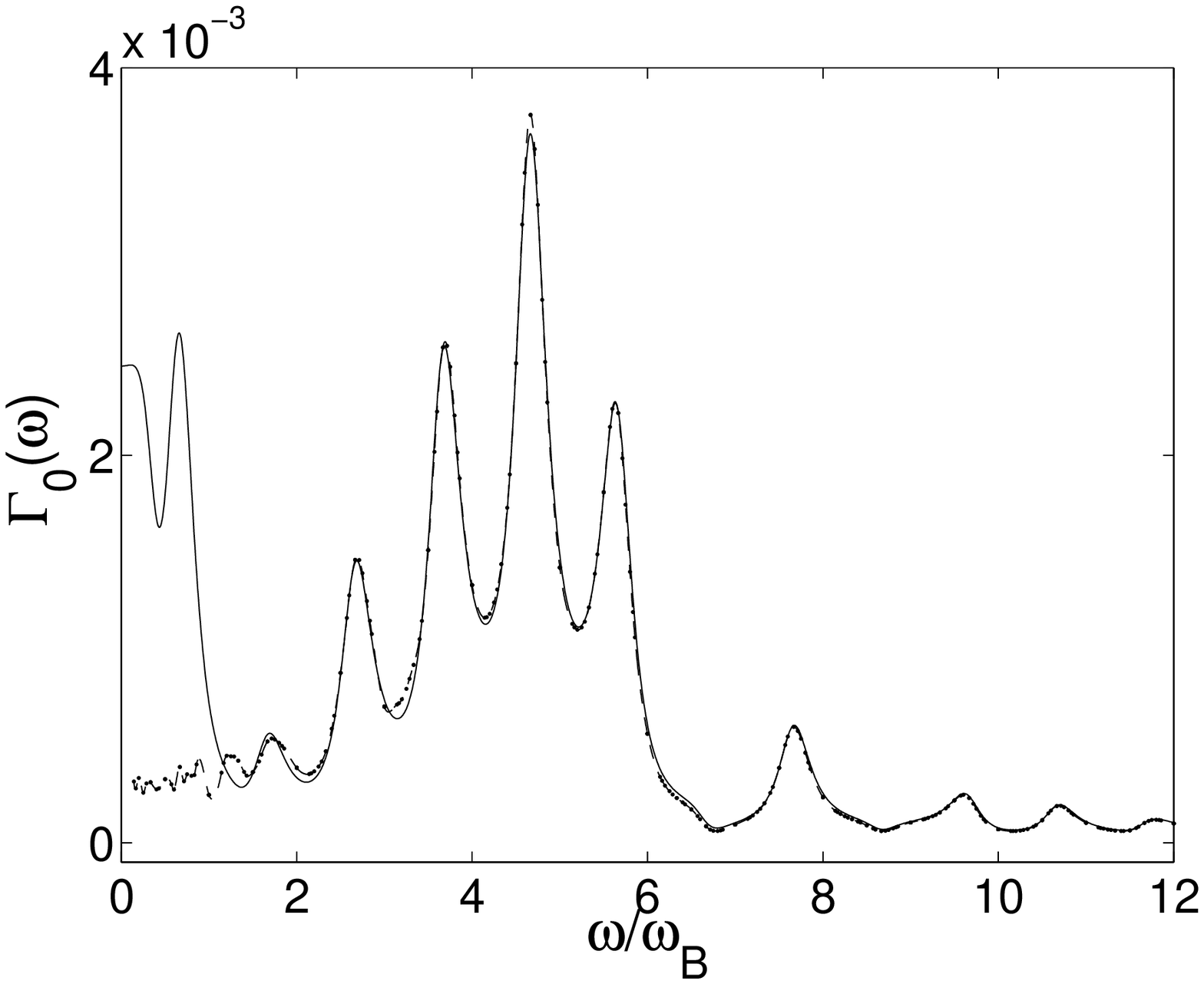}
\end{center}
\caption{\it Comparison of the $\omega$-dependence in (\ref{4a2}) (solid 
line) with the exact numerical calculation of the induced decay rate 
(dashed line). Parameters are $F=0.04$, $F_\omega=0.02$ and $\hbar=1.5$
(left), $\hbar=2.0$ (middle) and $\hbar=2.5$ (right panel).}
\label{fig4c}
\end{figure}

Substituting the calculated matrix elements into Eq.~(\ref{4a2}), we
find the decay spectra of the system. The solid line in Fig.~\ref{fig4c}
shows the decay spectra for $\hbar=1.5$, 2.0, 2.5. As expected, 
$\Gamma_0(\omega)$ has number of peaks with the same width $\Gamma_1$ 
separated by the Bloch frequency $\omega_B$.
The relative heights of the peaks are obviously given by the absolute values
of the squared dipole matrix elements shown in Fig.~\ref{fig4b},
while the shape of the lines is defined by the phase of $V_{0,\beta}^2(L)$. 
As mentioned above, the phases of the squared dipole matrix elements are 
generally not zero and, therefore, the shape of the lines is generally 
non-Lorentzian. In other words, we meet the case of 
Fano-like resonances \cite{Fano61}. For the sake of comparison the dashed lines
in Fig.~\ref{fig4c} show the results of an {\em exact} numerical calculation 
of the decay rate. A good correspondence is noticed.
The discrepancy in the region of small driving frequency is due to the
rotating wave approximation (which is implicitly assumed in the Fermi
golden rule) and the effect of the diagonal matrix elements $V_{\alpha,\alpha}^2(L)$
(which are also ignored in the Fermi golden rule approach). In principle, 
the region of small driving frequency requires a separate analysis.
\begin{figure}[t]
\begin{center}
\includegraphics[width=12cm]{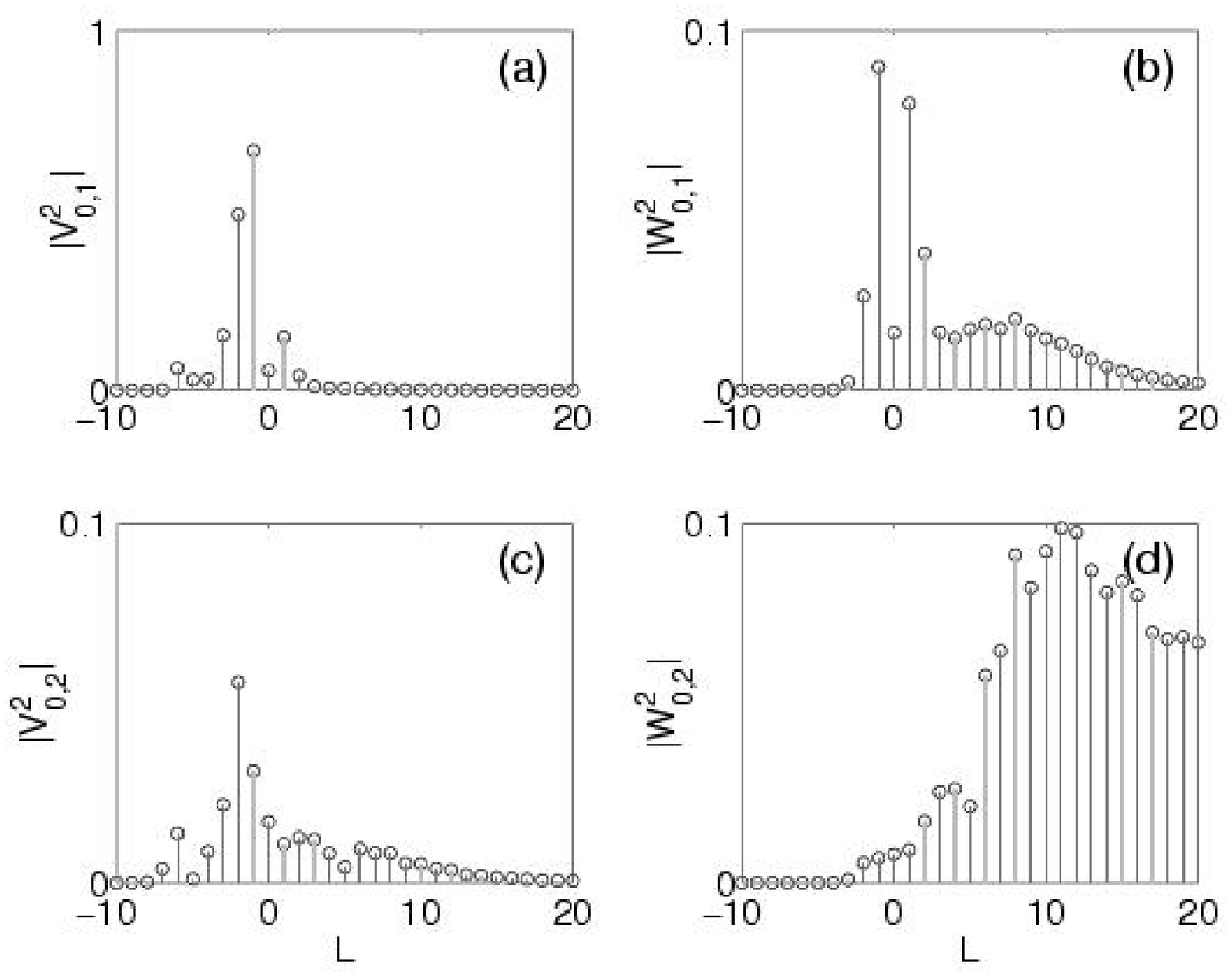}
\end{center}
\caption{\it The absolute values of the transition coefficients $V_{0,1}^2(L)$ 
(a), $W_{0,1}^2(L)$ (b), $V_{0,2}^2(L)$ (c), and $W_{0,2}^2(L)$ (d).}
\label{fig4d}
\end{figure}

In conclusion, we discuss the effect of direct transitions to 
the second excited Wannier ladder. For the case $\hbar=2$ the 
squared dipole matrix elements $V_{0,1}^2(L)$ and $V_{0,2}^2(L)$ are compared 
in the left column of Fig.~\ref{fig4d}. It is seen that the main lines in 
Fig.~\ref{fig4d}(c) are ten times smaller than those in Fig.~\ref{fig4d}(a).
Thus the effect of higher transitions can be neglected. We note, 
however, that this is not always the case. In the next section
we consider a situation when the direct transitions to the second
excited Wannier ladder can not be ignored.

\section{Decay spectra for atoms in optical lattices}
\label{sec4c}

\begin{figure}[t]
\begin{center}
\includegraphics[width=5.2cm,height=6.5cm]{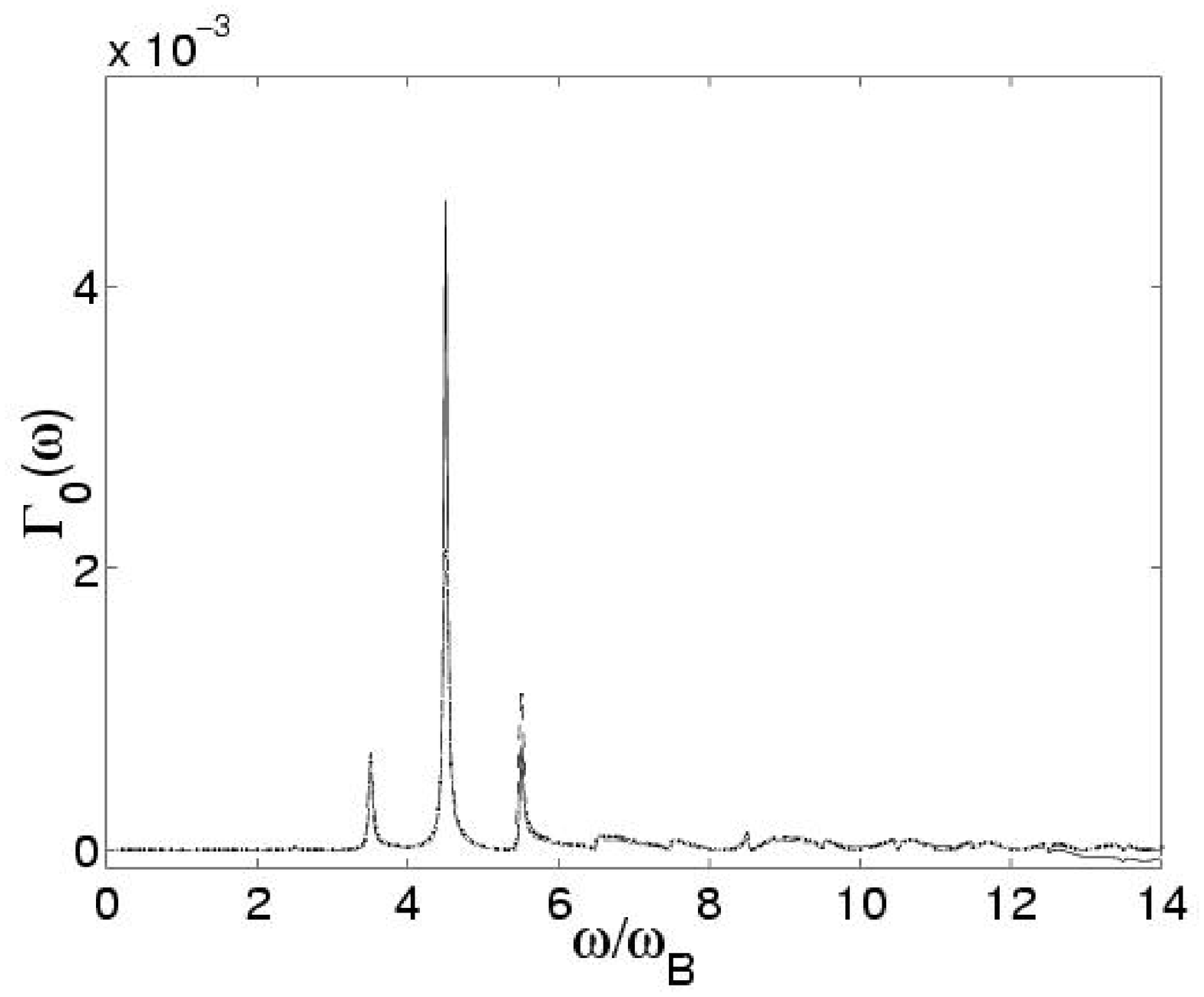}
\includegraphics[width=5.2cm,height=6.5cm]{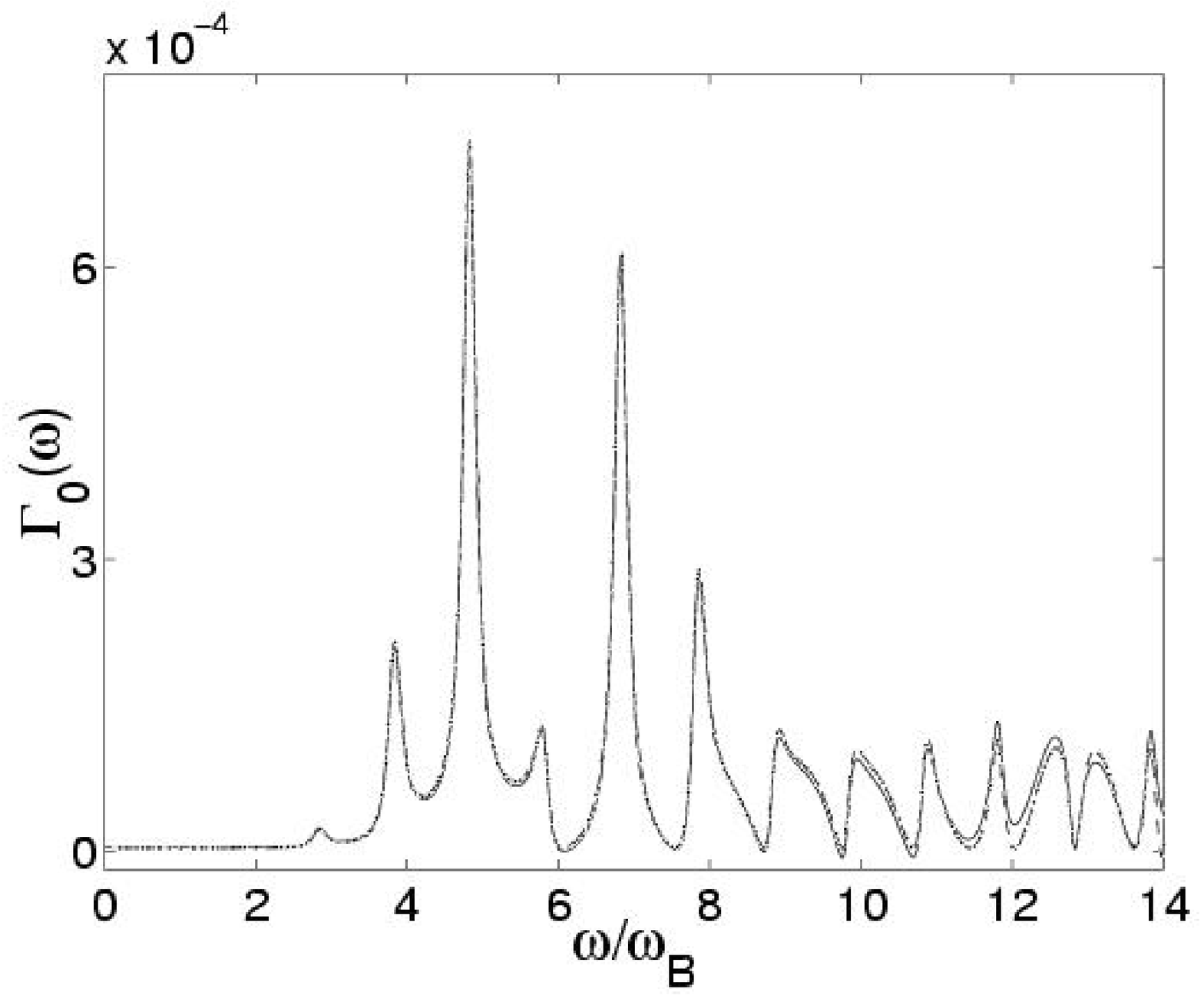}
\includegraphics[width=5.2cm,height=6.5cm]{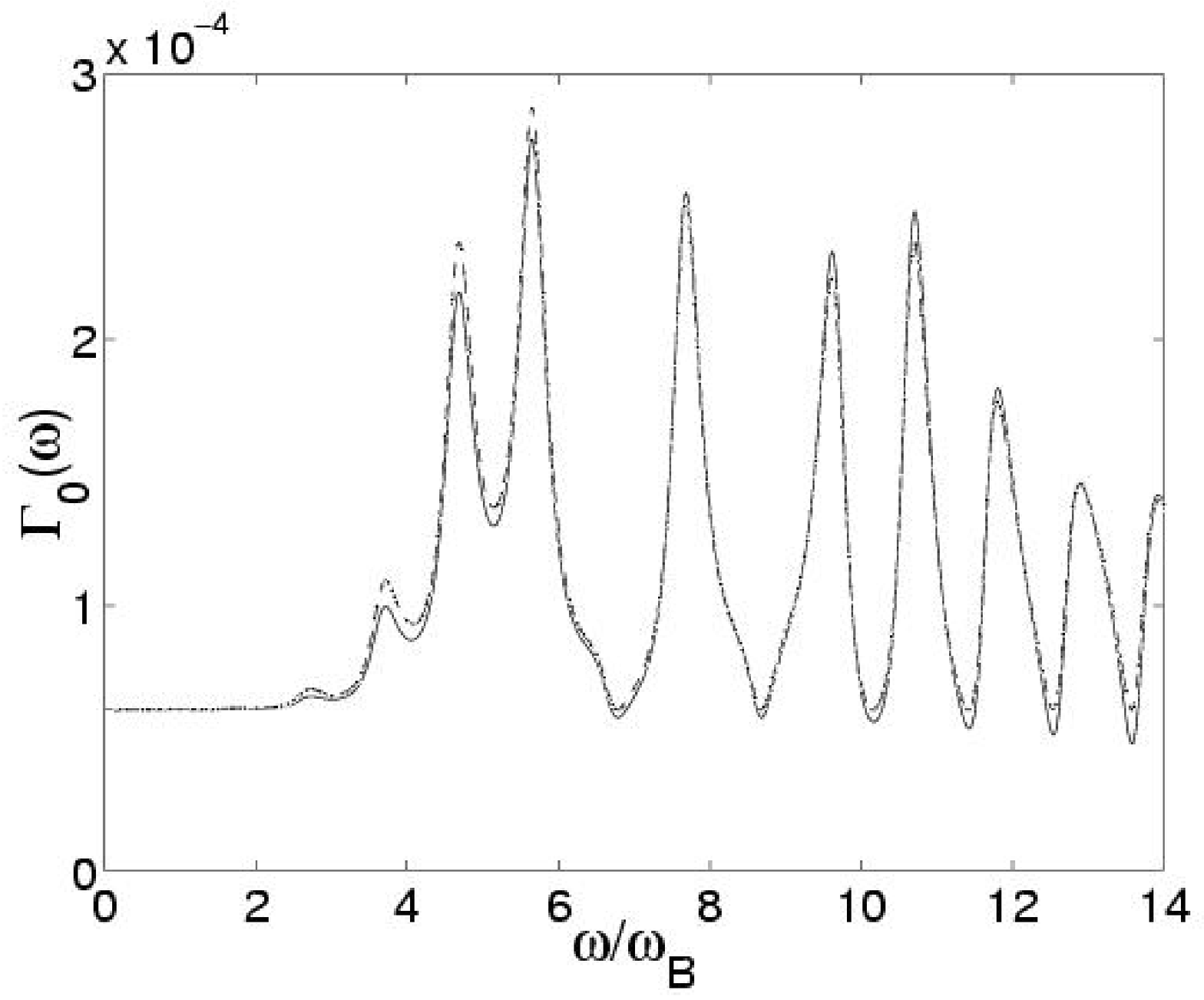}
\caption{\it Decay spectra as a function of the driving frequency 
$\omega$. Parameters are $F = 0.04$, $\varepsilon = 0.02$ and $\hbar = 1.5$ 
(left), $\hbar = 2.0$ (middle) $\hbar = 2.5$ (right panel).
The exact numerical calculation (dashed lines) are compared to the model 
prediction (solid lines). Note a complicated structure of the decay spectra
in the high-frequency region caused by the interference of the transitions to
the first and second excited Wannier-Stark ladders.}
\label{fig4f}
\end{center}
\end{figure}

The induced decay rate $\Gamma_0(\omega)$ was measured for the system of cold 
atoms in the accelerated standing laser wave \cite{Wilk96,Madi99}. Because the atoms
are neutral, the periodic driving of the system was realized by means of a
phase modulation of the periodic potential: 
\begin{eqnarray}
\label{4c0}
H = \frac{p^2}{2} + \cos[x + \varepsilon \cos(\omega t)] + F x \;,
\end{eqnarray}
Using the Kramers-Henneberger transformation 
\cite{Kram50,Kram56,Henn68,Choi74}
\footnote{The Kramers-Henneberger transformation is a canonical
transformation to the oscillating frame. In the
classical case it is defined by the generating function 
${\cal F}(p',x,t)=[p'+\epsilon\omega\sin(\omega t)][x+\epsilon\cos(\omega t)]$. 
In the quantum case one uses a substitution
$\psi(x,t)=\exp[-iF_\omega\sin(\omega t)x/\hbar\omega)\tilde{\psi}(x,t)$
together with the transformation  $x'=x+\epsilon\cos(\omega t)$.}
the Hamiltonian (\ref{4c0}) can be presented in the form
\begin{eqnarray}
\label{4c1}
H = \frac{p^2}{2} + \cos(x) + Fx+F_\omega x\cos(\omega t) \;,\quad
F_\omega=\varepsilon\omega^2 \;.
\end{eqnarray} 
Thus, the phase modulation is equivalent to the effect of an ac field.
Considering the limit of small $\varepsilon$, where 
$\cos[x+\varepsilon\cos(\omega t)]\approx\cos x+\varepsilon\sin x\cos(\omega t)$,
we can adopt Eq.~(\ref{4a2}) of the previous section to cover the
the case of phase modulation. Namely, the amplitude $F_\omega$ in 
Eq.~(\ref{4a2}) should be substituted by $\varepsilon$ and the squared dipole 
matrix elements (\ref{4a3}) by the squared matrix elements
\begin{equation}
W_{0,\beta }^2(L)=
\langle \Psi _{0,l}|\sin x|\Psi _{\beta,l+L}\rangle 
\langle \Psi _{\beta,l+L}|\sin x|\Psi _{0,l}\rangle   \;.
\label{4c3}
\end{equation}
Moreover, according to  the commutator relation for the Hamiltonian 
of the non-driven system
\begin{equation}
\hbar^{-2}[H_W,[H_W,x]]=-\sin x +F \;,
\label{4c4}
\end{equation}
the squared matrix elements $W_{0,\beta }^2(L)$  are related to 
the squared dipole matrix elements $V_{0,\beta }^2(L)$  by
\begin{eqnarray}
\label{4c5}
W^2_{0,\beta}(L)=\left| 
\frac{\mathcal{E}_{\beta,l+L} - \mathcal{E}_{0,l}}{\hbar}\right|^4 
\, V^2_{0,\beta}(L) \;.
\end{eqnarray}
It follows from the last equation that the way of driving realized in the
optical lattices suppresses the transition down the ladder and
enhances the transition up the ladder. This is illustrated in
Fig.~\ref{fig4d}, where we compare the squared matrix elements 
$W^2_{0,\beta}(L)$ and $V^2_{0,\beta}(L)$ for $\beta=1,2$ calculated
on the basis of Eq.~(\ref{4c3}) and Eq.~(\ref{4a3}), respectively.
It is seen that the practically invisible tail of far transitions in Fig.~\ref{fig4d}(a) 
shows up in Fig.~\ref{fig4d}(b). Besides this, for $L\gg1$ the squared
matrix elements between the ground and second excited Wannier-Stark states are larger
than those between the ground and first excited one. Because the width
of the second excited Wannier-Stark resonance $\Gamma_2$ is lager than $\Gamma_1$
(and actually larger than the Bloch energy), the transition
to the first and second excited Wannier ladders may interfere.
Indeed, this is the case usually observed in the high-frequency regime of
driving (see Fig.~\ref{fig4f}, which should be compared with Fig.~\ref{fig4c}).

\begin{figure}[t]
\begin{center}
\includegraphics[width=7cm,height=8cm]{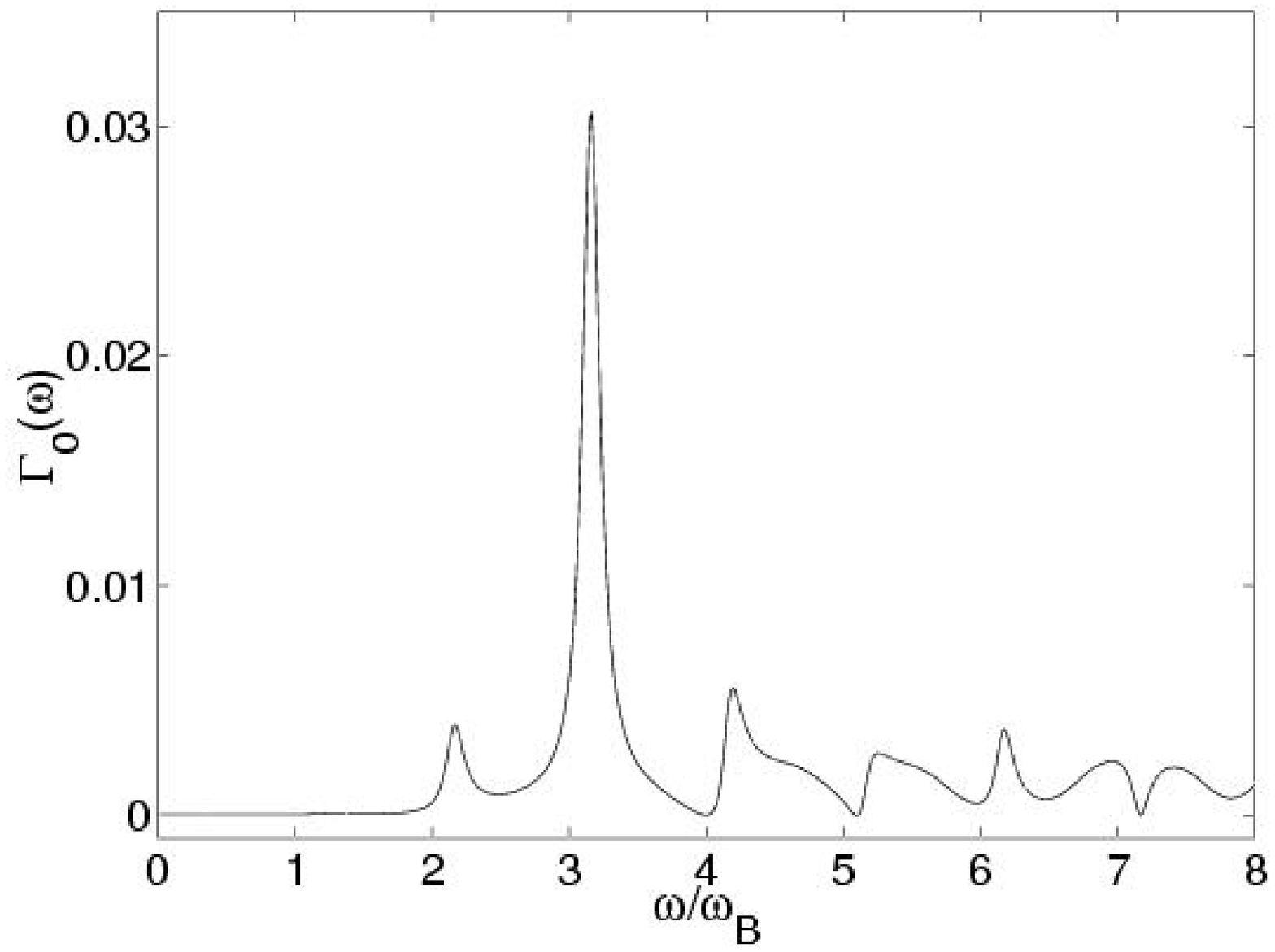}
\includegraphics[width=7cm,height=8cm]{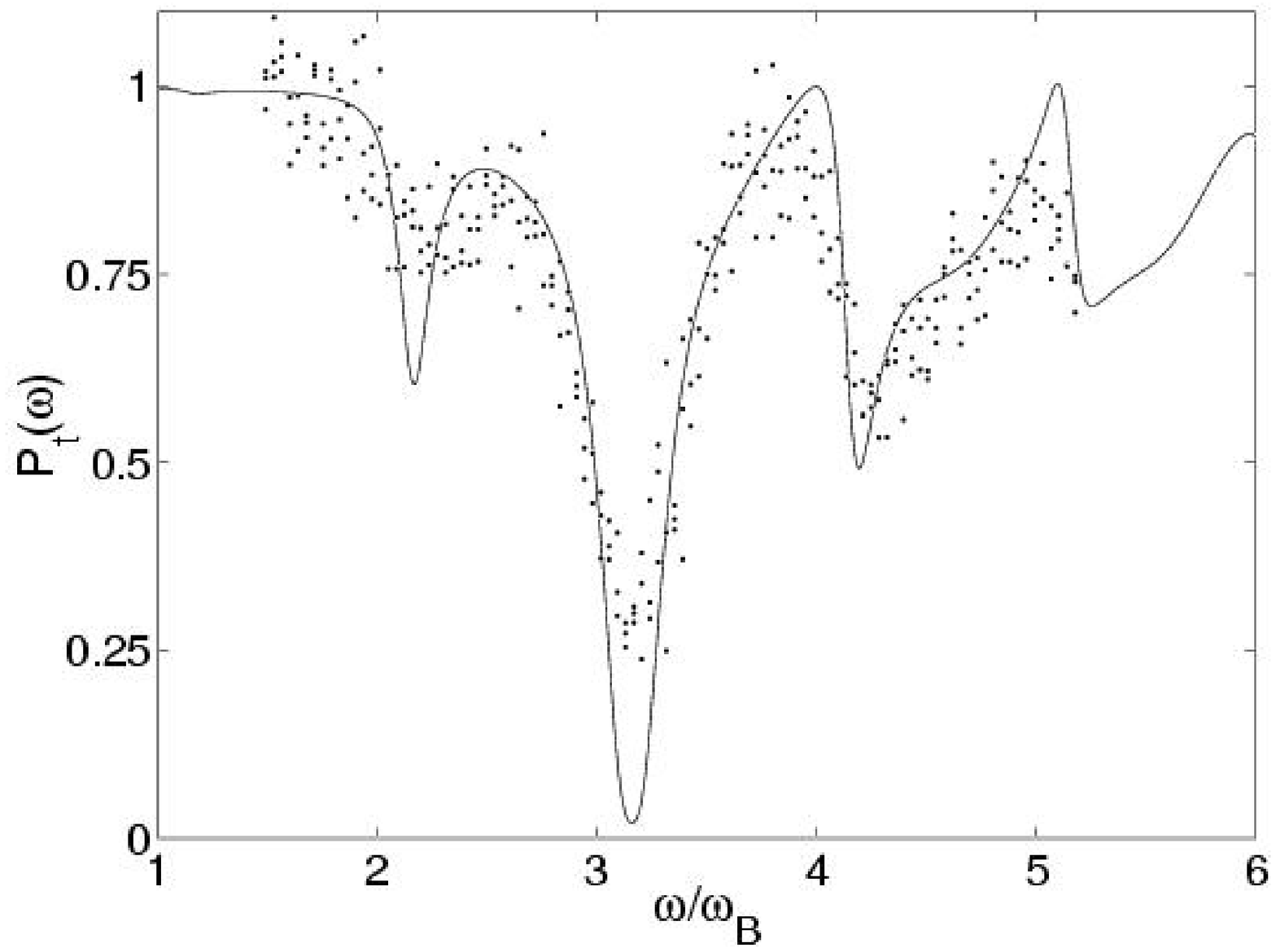}
\caption{\it The left panel shows the induced decay rate of the ground
Wannier-Stark state as function of the driving frequency for
$\hbar = 1.709$, $F = 0.0628$ and $\varepsilon = 0.096$. The right
panel compares the experimental data from \cite{Wilk96}
with the calculated survival probability $P_t(\omega)$ for 
$t = 300\mu s$.}
\label{fig4e}
\end{center}
\end{figure}

We proceed with the experimental data
for the spectroscopy of atomic Wannier-Stark ladders \cite{Wilk96} 
(note also the improved experiment \cite{Madi99}). 
The setup in the experiment \cite{Wilk96} is as follows. Sodium 
atoms were cooled and trapped in a far-detuned optical lattice. 
Then, introducing a time-dependent phase difference 
between the two laser beams forming the lattice, 
the lattice was accelerated (see Sec.~\ref{sec1d}). After some time, only atoms in 
the ground Wannier-Stark states survived, i.e.~a superposition of ground ladder 
Wannier-Stark states was prepared. Then an additional phase driving of
frequency $\omega$ was switched on and the survival probability, 
\begin{eqnarray}
\label{4c6}
P_t(\omega) = \exp\left( - \frac{\Gamma_0(\omega) t}{\hbar}\right) \;, 
\end{eqnarray} 
was measured. The experiment was repeated for different values of $\omega$. 
In scaled units the experimental settings with $V_0/h=75\pm7 kHz$ 
(we choose the value $V_0/h = 68 kHz$, which is used in all numerical
simulations in \cite{Wilk96}) and $a = 1570 m/s^2$ 
correspond to $\hbar = 1.709$ and $F = 0.0628$. (For these parameters the 
ground and first excited state have the widths $\Gamma_0 = 2.38 \cdot 10^{-5}$
and $\Gamma_1 = 6.11\cdot 10^{-2}$, respectively.) The timescale in the
experiments is $1.37 \mu s$, and the Bloch frequency is 
$\omega_B/2\pi=26.85 kHz$. The driving amplitude was $\varepsilon = 0.096$. 
The left panel of Fig.~\ref{fig4e} shows the decay spectra
as a function of the frequency in this case. The vertical transition 
dominates the figure, accompanied by the two transitions with $L = \pm 1$ 
and a tail of transitions with positive $L\gg1$. 
In the right panel, the experimental data for the survival probability 
$P_t(\omega)$ are compared to our numerical data.
The time $t$ is taken as an adjustable parameter and chosen such
that the depth of the peaks approximately coincide. The curve shows 
the survival probability at $t = 300 \mu s$ corresponding to 
$t = 219$ in scaled units. A good correspondence between 
experiment and theory is noticed. The minima of the 
survival probability appear when the driving frequency 
fits to a transition. The relative depth of the minima reflecting the
size of the transition matrix elements agrees reasonably. 
Furthermore, the asymmetric shape of the minimum between
$4 \omega_B$ and $5 \omega_B$ is reproduced.
Note that the experimental data also allow to extract the width of the first 
excited state from the width of the central minimum: 
$\Gamma_1 \approx 0.3 \omega_B \approx 6.9 \cdot 10^{-2}$, which
is in reasonable agreement with the numerical result 
$\Gamma_1 = 6.11\cdot10^{-2}$. 

\section{Absorption spectra of semiconductor superlattices}
\label{sec4d}

Equation (\ref{4a2}) of Sec.~\ref{sec4a} can be generalized to describe the
absorption spectrum $D(\omega)$ of undoped semiconductor superlattices 
\cite{PRB}. This generalization has the form
\begin{equation}
D(\omega )\sim\sum_{\alpha ,\beta }\sum_{L}
{\rm Im}\left[ \frac{I_{\alpha,\beta }^{2}(L)}
{(E_{\beta,l}^e-E_{\alpha,l}^h+edFL+E_g-\hbar\omega)
-{\rm i}(\Gamma_\beta^e+\Gamma_\alpha^h)/2}\right]  \;,  
\label{4d1}
\end{equation}
where the upper indices $e$ and $h$ refer to the electron and hole
Wannier-Stark states, respectively, $E_g$ is the energy gap between the 
conductance and valence bands in the bulk semiconductor, and
\begin{equation}
I_{\alpha,\,\beta }^{2}(L)=
\langle \Psi _{\alpha,\,l}^{h}|\Psi _{\beta,\,l+L}^{e}\rangle 
\langle \Psi _{\beta,\,l+L}^{e}|\Psi _{\alpha,\,l}^{h}\rangle   
\label{4d2}
\end{equation}
is the square of the overlap integral between the hole and electron wave 
functions. Repeating the arguments of Sec.~\ref{sec4a} it is easy to show
that in the low-field limit Eq.~(\ref{4d1}) is essentially the same as
the Fermi golden rule equation
\begin{equation}
D(\omega )\sim\int\int{\rm d}E^e{\rm d}E^h\Big|\int{\rm d}x\Psi^e(x;E^e)\Psi^h(x;E^h)
\Big|^2 \rho^e(E^e)\rho^h(E^h) \delta(E^e-E^h+E_g-\hbar\omega) \;,
\label{4d0}
\end{equation}
where $\rho^e(E^e)$ and $\rho^h(E^h)$ are the one-dimensional electron and hole
densities of states. According to Ref.~\cite{Glut99,Leo00} the quantity $D(\omega)$, 
which can be interpreted as the probability of creating the electron-hole pair 
by a photon of energy $\hbar\omega$ (the electron-hole Coulomb interaction is 
neglected), is directly related to the absorption spectrum of the semiconductor 
superlattices measured in the laboratory experiments. 

It follows from Eq.~(\ref{4d1}) that the structure of the absorption spectrum 
depends on the values of the squared overlap integral Eq.~(\ref{4d2}) which, in turn,
depend on the value of the static field. In the low-field regime the Wannier-Stark
states are delocalized over several superlattice periods and many 
transition coefficients $I_{\alpha,\,\beta }^{2}(L)$ differ from zero. 
In the high-field regime the Wannier-Stark states tend to be 
localized within a single well and the vertical transitions $L=0$ become
dominant. We would like to stress, however, that the process of localization of 
the Wannier-Stark states is always accompanied by a loss of their stability.
As mentioned above, the latter process restricts the validity of
the tight-binding results concerning a complete localization of 
the Wannier-Stark states in the limit of strong static field.
\begin{figure}[t]
\begin{center}
\includegraphics[width=12cm,clip,angle=0]{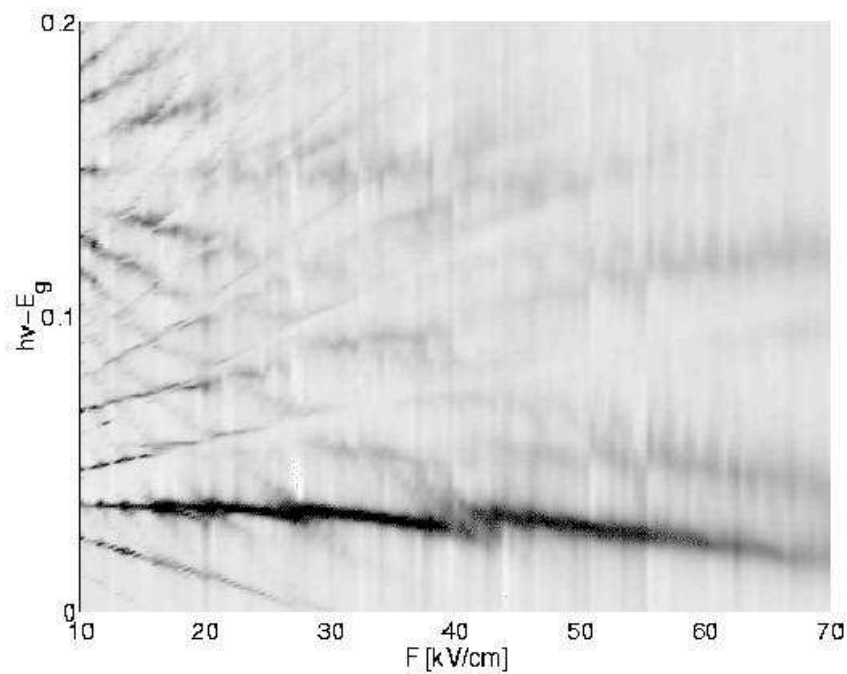}
\end{center}
\caption{\it Grey-scaled map of the one-dimensional absorption spectra (\ref{4d1})
as a function of the static field $F$ and photon energy $h\nu$.} 
\label{fig4g}
\end{figure}

As an illustration to Eq.~(\ref{4d2}), Fig.~\ref{fig4g} shows the absorption 
spectrum of the semiconductor superlattice studied in the experiment 
\cite{Leo00}.\footnote{The superlattice parameters are $V_0=0.0632$ eV  
($V_0=-0.0368$ eV) for the electron (hole) potential barrier, 
and $m^*=0.067m_e$ ($m^*=0.45m_e$) for effective electron 
(hole) mass. These parameters correspond to the value of the scaled ``electron''
and ``hole'' Planck constants $\hbar=3.28$ and $\hbar=1.64$, respectively.}
(This should be compared with the absorption spectrum calculated in Ref.~\cite{Glut99} 
by using a kind of finite-box quantization method.)
The depicted result is a typical example of a Wannier-Stark fan diagram. 
By close inspection of the figure one
can identify at least four different fans associated with the transitions
between $\alpha=0,1$ hole and $\beta=0,1$ electron states. However, in the
region of strong static fields considered here, the majority of these transitions are 
weak and the whole spectrum is dominated by the vertical $L=0$ transition
between the ground hole and electron states. Note a complicated structure
of the main line resembling a broken feather. Recalling the results of 
Sec.~\ref{sec3d} (see Fig.~\ref{fig3d}), this structure originates from
avoided crossings between the (ground) under-barrier and (first) above-barrier
electron resonances. Such a ``broken feather'' structure was well observed 
in the cited experiment \cite{Leo00}.

\chapter{Quasienergy Wannier-Stark states}
\label{sec5}

In the following chapters we investigate Wannier-Stark ladders
in combined ac and dc fields. Then the Hamiltonian of the system is
\begin{eqnarray}
\label{5a0a}
H = \frac{p^2}{2} + V(x) + F x +F_\omega x \cos(\omega t) \, , 
\end{eqnarray}
or, as described in Sec.~\ref{sec4c}, equivalently given by
\begin{eqnarray}
\label{5a0}
H = \frac{p^2}{2} + V[ x + \varepsilon \cos(\omega t)] + F x \, ,\quad
\varepsilon = F_\omega/\omega^2 \;.
\end{eqnarray}
Depending on the particular analytical approach
we shall use either of these two forms. Let us also note that the 
Hamiltonian (\ref{5a0}) can be generalized to include the case of
arbitrary space- and time-periodic potential $V(x,t)=V(x+2\pi,t)=V(x,t+T_\omega)$.

\section{Single-band quasienergy spectrum}
\label{sec5a}

For time-dependent potentials the period of the potential sets an
additional time scale. In order to define a Floquet-Bloch operator
with properties similar to the time-independent case, we 
have the restriction that the period $T_\omega$ of the potential 
and the Bloch time $T_B$ are commensurate, i.e.
\begin{equation}
\label{5a1}
p T_\omega = q T_B\equiv T\;.
\end{equation}
In this case the Floquet operator $U(T)$ 
over the common period $T$ can be presented as
\begin{eqnarray}
\label{5a2}
U(T) = {\rm e}^{-{\rm i}q x} \, \widetilde{U}(T) \;,\quad
\widetilde{U}(T) = \widehat{\exp}\left(- \frac{{\rm i}}{\hbar}
\int_0^{T} {\rm d}t \, \left[\frac{(p-Ft)^2}{2}+V(x,t)\right]\right) \;, 
\end{eqnarray}
(compare with Eqs.~(\ref{2b1})--(\ref{2b1a})).
Consequently the eigenstates of $U(T)$, 
\begin{eqnarray}
U(T) \, \Phi(x;\lambda,\kappa)=\lambda\Phi(x;\lambda,\kappa) \;,\quad
\lambda=\exp(-{\rm i}ET/\hbar) \;, 
\label{5a3}
\end{eqnarray}
can be chosen to be the Bloch-like states \cite{Zak93,Zak96}, i.e.
$\Phi(x+2\pi;\lambda,\kappa)
=e^{{\rm i}2\pi\kappa} \Phi(x;\lambda,\kappa)$. 
Due to the time-periodicity of the potential, 
$V(x,t+ T_\omega)=V(x,t)$, we have the relation
\begin{eqnarray}
\label{5a5}
U(T) = U(T_\omega)^p = \left[ \exp(-{\rm i} \, x\,  q/p)\,  
\widetilde{U}(T_\omega) \right]^p  \;.
\end{eqnarray}
As a direct consequence of this relation, the states
$\Phi(x;\lambda,\kappa)$ with the quasimomentum $\kappa - r/p$
($r = 0, 1, \dots, p-1$) are Floquet states with the same quasienergy.
In terms of the operator 
$U^{(\kappa)}(T)=\exp(-{\rm i}\kappa x) U(T)\exp({\rm i}\kappa x)$ 
this means that the operators $U^{(\kappa)}(T)$ are unitarily 
equivalent for these values of the quasimomentum.\footnote{We recall
that in the case of pure dc field the operators $U^{(\kappa)}(T_B)$ are 
unitarily equivalent for arbitrary $\kappa$.} Therefore,
the Brillouin zone of the Floquet operator $U(T)$
is $p$-fold degenerate. In the next section we introduce the
resonance Wannier-Bloch functions $\Phi_{\alpha,\kappa}(x)$
which satisfy the eigenvalue equation (\ref{5a3}) with the Siegert
(i.e. purely outgoing wave) boundary condition and correspond to the complex 
energy $\mathcal{E}_\alpha(\kappa)$. Then the $p$-fold degeneracy of the
Brillouin zone just means that the dispersion relation
$\mathcal{E}_\alpha(\kappa)$ is a periodic function of the
quasimomentum with period given by $p$.

It should be noted that the Wannier-Bloch functions
$\Phi(x;\lambda,\kappa)$ (hermitian boundary condition) or
$\Phi_{\alpha,\kappa}(x)$ (Siegert boundary condition) are {\em not}
the quasienergy functions of the system because the latter, by
definition, are the eigenfunctions of the evolution operator 
$U(T_\omega)$ over the period of the driving force. However, the
quasienergy functions can be expressed in terms of the Wannier-Bloch
functions as
\begin{eqnarray}
\label{5a6}
\Psi_{\alpha,\kappa}^{(n)}(x)=\frac{1}{p} \sum_{r = 0}^{p-1} 
\exp\left[-{\rm i}\,\frac{2\pi n}{p}r\right] \Phi_{\alpha,\kappa+r/p}(x) \,.
\end{eqnarray} 
Equation (\ref{5a6}) is the discrete analogue of the relation (\ref{2f5})
between the Wannier-Bloch and Wannier-Stark states in the case
of pure dc field. Since the evolution operator $U(T_\omega)$
commutes with the translational operator over $p$ lattice periods,
the quasienergy states $\Psi_{\alpha,\kappa}^{(n)}(x)$ are the
eigenfunctions of this shift operator. In particular, as easily 
deduced from Eq.~(\ref{5a6}), in the limit $\varepsilon\rightarrow0$
the function $\Psi_{\alpha,\kappa}^{(n)}(x)$ is a linear combination
of every $p$-th state of the Wannier-Stark ladder (and altogether
there are $p$ different subladders). Thus, as well as the 
Wannier-Bloch states $\Phi_{\alpha,\kappa}(x)$, the eigenstates
of $U(T_\omega)$ are extended states. Note that the Brillouin zone
is reduced now by a factor $p$, i.e the quasimomentum is restricted
to  $-1/2p\le\kappa\le 1/2p$. On the other hand, as $T_\omega=T/p$,
the energy Brillouin zone is enlarged by this factor, i.e. the
quasienergies take values in the interval 
$0\le{\rm Re}\,\mathcal{E}\le\hbar\omega$. Thus, if 
$\mathcal{E}_\alpha(\kappa)$ is the complex band of the Floquet
operator (\ref{5a2}), the complex quasienergies corresponding
to the quasienergy states (\ref{5a6}) are 
\begin{equation}
\label{5a7}
\mathcal{E}^{(n)}_\alpha(\kappa)=\mathcal{E}_\alpha(\kappa)
+\hbar\omega\,\frac{n}{p} \;,\quad \hbar\omega=2\pi F\,\frac{p}{q} \;.
\end{equation}
In the remainder of this section we discuss the dispersion relation
$\mathcal{E}_\alpha(\kappa)$ for the quasienergy bands on the basis of the 
single-band model. It is understood, however, that the single-band 
approach can describe at its best only the real part 
$E={\rm Re}\mathcal{E}$ of the spectrum.

In the single band analysis \cite{Zhao95}, it is convenient to 
work in the representation (\ref{5a0a}). Assuming that the two timescales 
are commensurate, the Houston functions (\ref{1c0a}) can be generalized to the
Wannier-Bloch functions, which yields the following result for the
quasienergy spectrum 
\begin{eqnarray}
\label{5b0}
E_\alpha(\kappa) = \frac{1}{T}\int_0^T 
\epsilon_\alpha(\kappa(t)\,) \, {\rm d}t \;,\quad
\kappa(t) = \kappa - \frac{F t}{\hbar} 
- \frac{F_\omega}{\hbar \omega}\,\sin(\omega t) \,,
\end{eqnarray}
In this equation, as before, $\epsilon_\alpha(\kappa)$ is the
Bloch spectrum of the field-free Hamiltonian $H_0 = p^2/2 + V(x)$
and $\kappa(t)$ is the solution of the classical equation of motion for 
the quasimomentum with initial value $\kappa$. 
Expanding the Bloch dispersion relation into the Fourier series
\begin{eqnarray}
\label{5b2}
\epsilon_\alpha(\kappa) = \sum_{\nu=0}^\infty 
\widetilde{\epsilon}_\alpha(\nu) \cos(2\pi \nu \kappa)
\end{eqnarray}
we obtain after some transformations
\begin{eqnarray}
\label{5b3}
E_\alpha(\kappa)=\sum_{\mu=0}^\infty \, 
J_{\mu q}\Big(\frac{\mu q F_\omega}{F}\Big) \, 
\widetilde{\epsilon}_\alpha(\mu p) \cos(2 \pi p \mu \kappa)  \;. 
\end{eqnarray} 
Thus, the dispersion relation for the quasienergies is given
by the original Bloch dispersion relation with rescaled
Fourier coefficients. For the low-lying bands, the coefficients 
$\widetilde{\epsilon}_\alpha(\nu)$ rapidly decrease with $\nu$, 
and for practical purpose it is enough to keep only
two first terms in the sum over $\mu$.
\begin{figure}[t]
\begin{center}
\includegraphics[width=11cm]{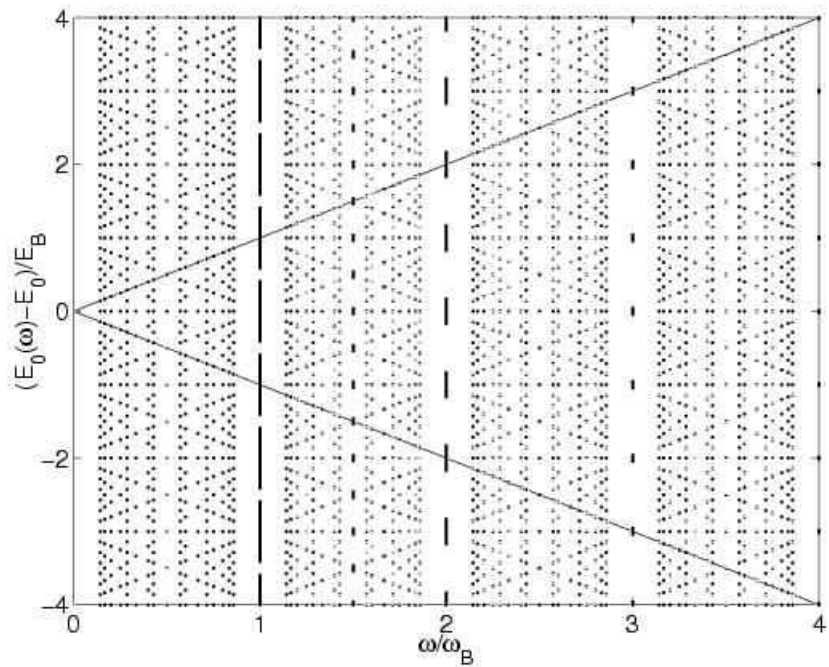}
\caption{\it The band structure of the quasienergy spectrum originating
from the ground ($\alpha=0$) Bloch band as predicted by the single-band model.
The parameters of the non-driven system are $\hbar=3$, $F=0.08$ and 
the driving amplitude is $\varepsilon=1$. Only the rational values 
$\omega/\omega_B=p/q$ with $q\le7$ are considered. The straight lines
restricts the interval $|E|\le\hbar\omega$ corresponding to two 
(quasi)energy Brillouin zones.}
\label{fig5a0}
\end{center}
\end{figure}

Because the absolute value of the Bessel 
function is smaller than unity, the width of the quasienergy
band is always smaller than the width of the parent Bloch band.
In particular, assuming $\epsilon_\alpha(\kappa)\approx
\bar{\epsilon}_\alpha+(\Delta_\alpha/2)\cos(2\pi\kappa)$ 
(as in the tight-binding approximation)
and the simplest case of the resonant driving $\omega=\omega_B$ ($p=q=1$), 
we have
\begin{eqnarray}
\label{5b4}
E_\alpha(\kappa)\approx\bar{\epsilon}_\alpha
+J_1\Big(\frac{F_\omega}{F}\Big)\frac{\Delta_\alpha}{2}\,\cos(2\pi\kappa) \;.
\end{eqnarray} 
As follows from this equation, the width of the quasienergy band 
approaches zero at zeros of the Bessel function $J_1(z)$. 
This phenomenon is often referred to in the literature as a dynamical
band suppression in combined ac-dc fields
\cite{Dunl86,Zhao91,Zhao92a,Hhong92,Zhao94,Zhao95,Dres97,Hang98}
\footnote{Actually this phenomenon (although under a different name)
was known earlier \cite{Bass86}.}.
A similar behavior in the case of a pure ac field was predicted
in \cite{Dunl86,Dres97} and experimentally observed in \cite{Madi98}.

Let us finally discuss the case of an irrational ratio of the
Bloch and the driving frequency, $\gamma = \omega / \omega_B$. 
We can successively approximate the irrational $\gamma$ by rational 
numbers $p_j / q_j$, which are the $j$-th approximants of a
continued fraction expansion of $\gamma$. Then, as for a typical $\gamma$
both $p_j, q_j \rightarrow \infty$, the bandwidth of this approximation
exponentially decreases to zero and the quasienergy spectrum turns into 
a discrete point spectrum \cite{Zhao94}. This is illustrated by
Fig.~\ref{fig5a0}, where the band structure of the quasienergy spectrum
(\ref{5a7}), calculated on the basis of Eq.~(\ref{5b3}), is
presented for $\alpha=0$ and constant value of driving amplitude
$\varepsilon=F_\omega\omega^2$. (The parameters of the non-driven system with 
$V(x)=\cos x$ are $\hbar=3$ and $F=0.08$.) Note that the quasienergy bands
have a noticeable width only for integer values of $p$.

It is an appropriate place here to note the similarity between 
the quasienergy spectrum of a driven Wannier-Stark system and the energy
spectrum of a Bloch electron in a constant magnetic field. The latter is
known to depend on the so-called magnetic matching ratio
\begin{equation}
\label{5b5}
\beta=\frac{eBd^2}{2\pi\hbar c} \;,
\end{equation}
where $d$ is the lattice period. The spectrum of the ground state energies as 
a function of $\beta$ forms the famous Hofstadter butterfly \cite{Hofs76}.
In particular, for rational control parameter $\beta=p/q$ the number of
distinct energy bands in the spectrum is given by the denominator $q$.
Note that the magnetic matching ratio can be interpreted as ratio of two
timescales, one of which is the time $d^2m/2\pi\hbar$ a particle with momentum 
$2\pi\hbar/d$ needs to cross the fundamental period $d$, and the other is
the period $eB/mc$ of the cyclotron motion.\footnote{This remark is ascribed to
F. Bloch.} Similar, the driven Wannier-Stark system has two intrinsic
timescales and the structure of the quasienergy spectrum depends on the
control parameter $\gamma=T_B/T_\omega=\hbar\omega/edF$, which is often referred 
to as the electric matching ratio.

\section{S-matrix for time-dependent potentials}
\label{sec5c}

Provided the condition (\ref{5a1}) is satisfied, the definition of a
scattering matrix closely follows that of Sec.~\ref{sec2c}. Thus
we begin with the matrix form of the eigenvalue equation (\ref{5a3}),
which reads
\begin{eqnarray}
\label{5c0}
\sum_{n} \widetilde{U}^{(\kappa)}_{m+q,n} \, G_{S}(n)=\lambda \, G_{S}(m) \,.
\end{eqnarray}
(To simplify the formulas we shall omit the quasimomentum index in
what follows.) Comparing this equation with Eq.~(\ref{2c0}), we note that 
index of the matrix $\widetilde{U}$ is now shifted by $q$. Because of
this, we have $q$ different asymptotic solutions, which should be
matched to each other. Using the terminology of the common scattering theory
we shall call these solution the {\em channels}.

It is worth to stress the difference in the notion of decay channels
introduced above and the notion of decay channels in the problem 
of above threshold ionization (a quantum particle in a single potential
well subject to a time-periodic perturbation) \cite{Bensch91}. In the latter case
there is a well defined zero energy in the problem (e.g., a ground state of the
system). Then the periodic driving originates a ladder of quasienergy
resonances separated by quanta $\hbar\omega$ of the external field and, thus, 
the number of the corresponding decay channels is infinite. 
In the Wannier-Stark system, however, the ladder
induced by the periodic driving (let us first discuss the simplest case
$p=q=1$) coincides with the original Wannier-Stark ladder. In this sense the 
driving does not introduce new decay channels. These new channels appear only 
when the induced ladder does not coincide with the original ladder. Moreover, 
in the commensurate case $\omega/\omega_B=p/q$ (because of the partial
coincidence of the ladders) their number remains finite. With this remark
reserved we proceed further.

As before, we decompose the vector $G_S$ into three parts, 
i.e.~$G_S^{(+)}$ contains all coefficients with $n > N$
and $G_S^{(-)}$ all coefficients with $n < -N-q$. The third part, 
$G_S^{(0)}$, contains all remaining coefficients with $-N-q \le n \le N$. 
The coefficients of $G_S^{(+)}$ and $G_S^{(-)}$ are
defined recursively,
\begin{eqnarray}
G_S(m) = (\lambda/u_m) \, G_S(m-q)&\quad& {\rm for}\,\, m > N\, ,\\
G_S(m-q) = (u_m/\lambda) \, G_S(m)&\quad& {\rm for}\,\, m < - N\, ,
\label{5c1}
\end{eqnarray}
where $u_m=\exp({\rm i}\hbar^2[(\kappa+m-q)^3-(\kappa+m)^3]/6F)$.
Let $W$ be the matrix $\widetilde{U}$ truncated to the
size $(2N+1)\times(2N+1)$, and, furthermore, let $O_{m,n}$
be an $m\times n$ matrix of zeros. With the help of the definition
\begin{eqnarray}
B_N = \left(\begin{array}{cc}
 O_{q,2N+1} & O_{q,q} \\[2mm]
  W & O_{2N+1,q} \\
  \end{array}\right) \;, 
\label{5c2}
\end{eqnarray}
the equation for $G_S^{(0)}$ reads
\begin{eqnarray}
\label{5c3}
(B_N - \lambda \openone )\, G_S^{(0)} = - 
\left(\begin{array}{c} 
  u_{N+q} G_S(N+q)\\ \vdots \\ 
  u_{N+1} G_S(N+1)\\[2mm] O_{2N+1,1}\\
\end{array}\right) \;.
\end{eqnarray}
The right hand side of the last equation contains $q$ subsequent 
terms $G_S(m)$ and therefore contributions
from the $q$ different incoming asymptotes. However, we can treat the
different incoming channels separately, because the sum of solutions
for different inhomogeneities yields a solution of the equation with
the summed inhomogeneity. Thus, let us rewrite (\ref{5c3}) in 
a way that separates the incoming channels. We 
define the matrices $e^q$ and $e_q$ as
\begin{eqnarray}
e^q = \left(\begin{array}{c} 
  \openone_{q,q}\\[2mm]  O_{2N+1,q}\\
\end{array}\right) \,, \quad 
e_q = \left(\, O_{q,2N+1} , \openone_{q,q}\,\right) \;,
\end{eqnarray}
where $\openone_q$ denotes a unit matrix of size $q \times q$.
Furthermore, we define the matrix $u^q$ 
as a diagonal $q\times q$ matrix $u^q$ with the diagonal 
\begin{eqnarray}
{\rm diag}(u^q) = (u_{N+q}, \dots, u_{N+1})\, 
\end{eqnarray}
and finally the column vectors $G^q$ and $G_q$ with 
the entries  $G(N+q), \dots, G(N+1)$ and 
$G(-N-1), \dots, G(-N-q)$, respectively. With the help of these
definitions the right hand side of equation (\ref{5c3}) reads
$e^qu^qG^q$, which directly leads to the following
relation between the coefficients of the incoming and the 
outgoing channels
\begin{eqnarray}
\label{inhomog4}
G_q = e_q\left[ B_N - \lambda \openone\right]^{-1}e^q 
u^q\, G^q\, .
\end{eqnarray}
In the S-matrix formula we additionally need to include the influence
of the free states, which are again discrete versions
of Airy functions. Thus, with the help of two additional diagonal matrices,
$a^q(E,N)$ and $a_q(E,N)$, which contain the contributions
of the free solutions, 
\begin{eqnarray}
{\rm diag}(a^q) =( G_{0}(N+q),\,  \dots\, , G_{0}(N+1)) \;,\quad
{\rm diag}(a_q) =( G_{0}(-N-1), \dots, G_{0}(-N-q)) 
\end{eqnarray}
with 
$G_{0}(m) = \exp\left({\rm i}\hbar^2[\kappa + m]^3/6F 
- {\rm i}E[\kappa + m]/F\right)$, we define the $q\times q$ \ S-matrix
\begin{eqnarray}
\label{5c5}
S(E) = \lim_{N\rightarrow\infty} a_q^{-1}\, 
e_q\left[ B_N - \lambda \openone\right]^{-1}e^q \, 
u^q \, a^q\;.
\end{eqnarray}
It can be proved that the matrix (\ref{5c5}) is unitary by
construction, i.e. $S^\dagger(E)S(E) = \openone$. 

Based on Eq.~(\ref{5c5}), the equation for
the resonance wave functions has the form
\begin{eqnarray}
\label{5c6}
(B_N - \lambda\openone) \, G_S^{(0)} = 0  \, .
\end{eqnarray}
In fact, as follows from the explicit form of the matrix $B_N$,
the first $q$ elements of the eigenvector are zero and, according
to Eq.~(\ref{5c1}), $G_S^{(+)}=0$. Thus, the solution of Eq.~(\ref{5c6})
satisfies the resonance-like boundary condition of empty
incoming channels. The corresponding energies are given by
$\mathcal{E}=i\hbar\ln\lambda/T$ and actually depend on $\kappa$,
which enters all equations displayed above as a parameter.

To conclude this section, we generalize the equation for the Wigner delay 
time. The generalization of (\ref{2c10}) for systems with $q$ decay 
channels reads 
\begin{eqnarray}
\label{5c7}
\tau = - \frac{\rm i\,\hbar}{q} \, 
\frac{\partial \ln[ \det S(E) ]}{\partial E}\, ,
\end{eqnarray}
or,  equivalently 
\begin{eqnarray}
\label{5c8}
\tau = \frac{1}{q} \,{\rm Tr}\,(\widehat{\tau}) \;,\quad
\widehat{\tau} = -{\rm i}\,\hbar\, S^\dagger(E) \, 
\frac{\partial S(E)}{\partial E}  \,.
\end{eqnarray}
where $\widehat{\tau}$ is the so-called Smith matrix \cite{Smit60}.
Along with the Wigner delay time, in the random matrix theory
of chaotic scattering (see chapter 7) the notion of partial
delay times, which are the eigenvalues 
of the Smith matrix, and one-channel delay times, which are the 
the diagonal elements of the Smith matrix, appear.
The sum of the partial or one-channel delay times obviously yields
the Wigner delay time.

\section{Complex quasienergy spectrum}
\label{sec5d}

Using the scattering matrix approach of the preceding section
we can calculate the complex quasienergy spectrum of the Wannier-Stark system
for arbitrary values of the parameters. In this chapter, however, we confine
ourselves to the perturbation regime of small $\varepsilon$ and relatively
large values of the scaled Plank constant $\hbar$. The opposite
case of large $\varepsilon$ and small $\hbar$ will be considered in
chapter 7.
\begin{figure}[t]
\begin{center}
\includegraphics[width=7.5cm,height=8cm]{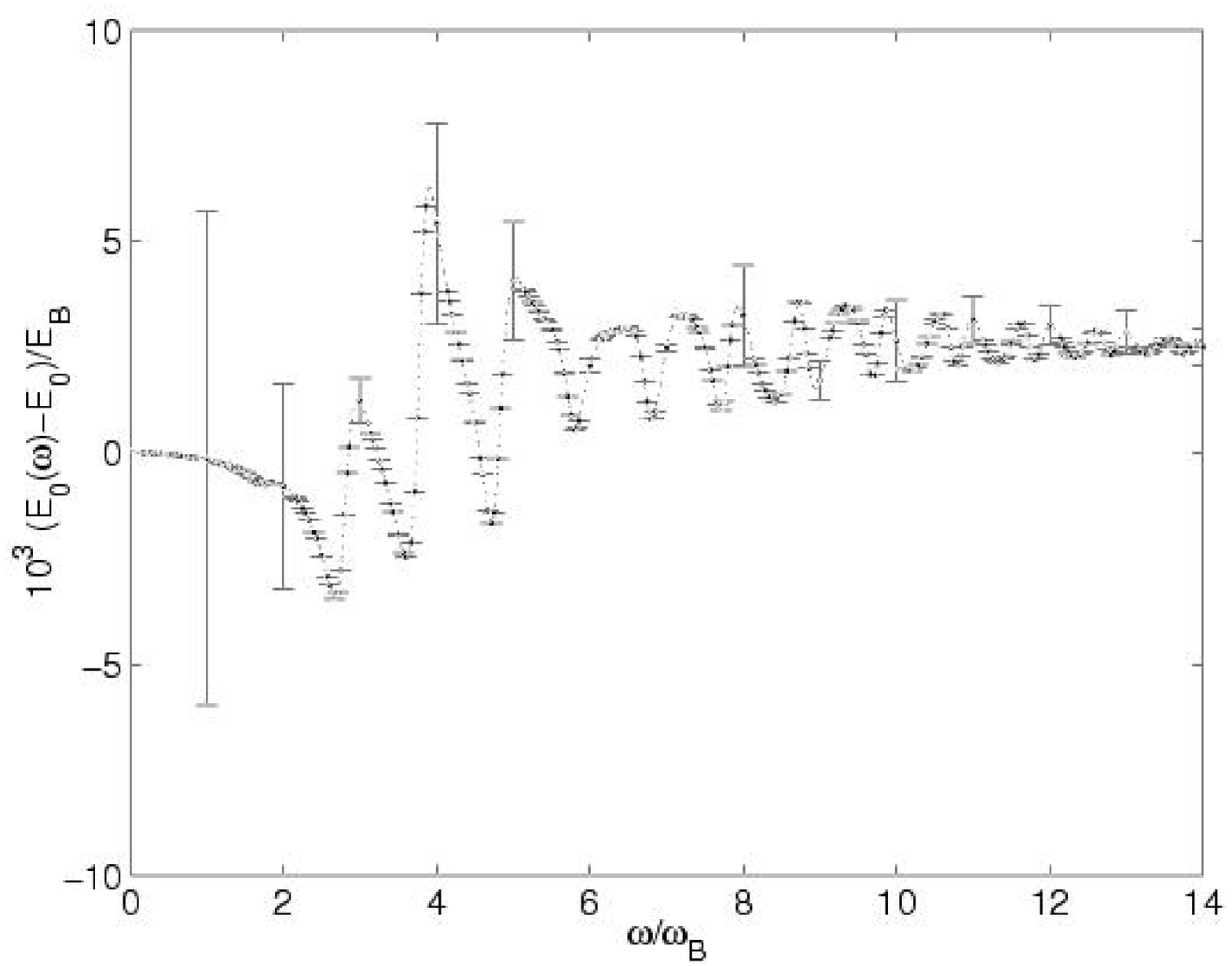}
\includegraphics[width=7.5cm,height=8cm]{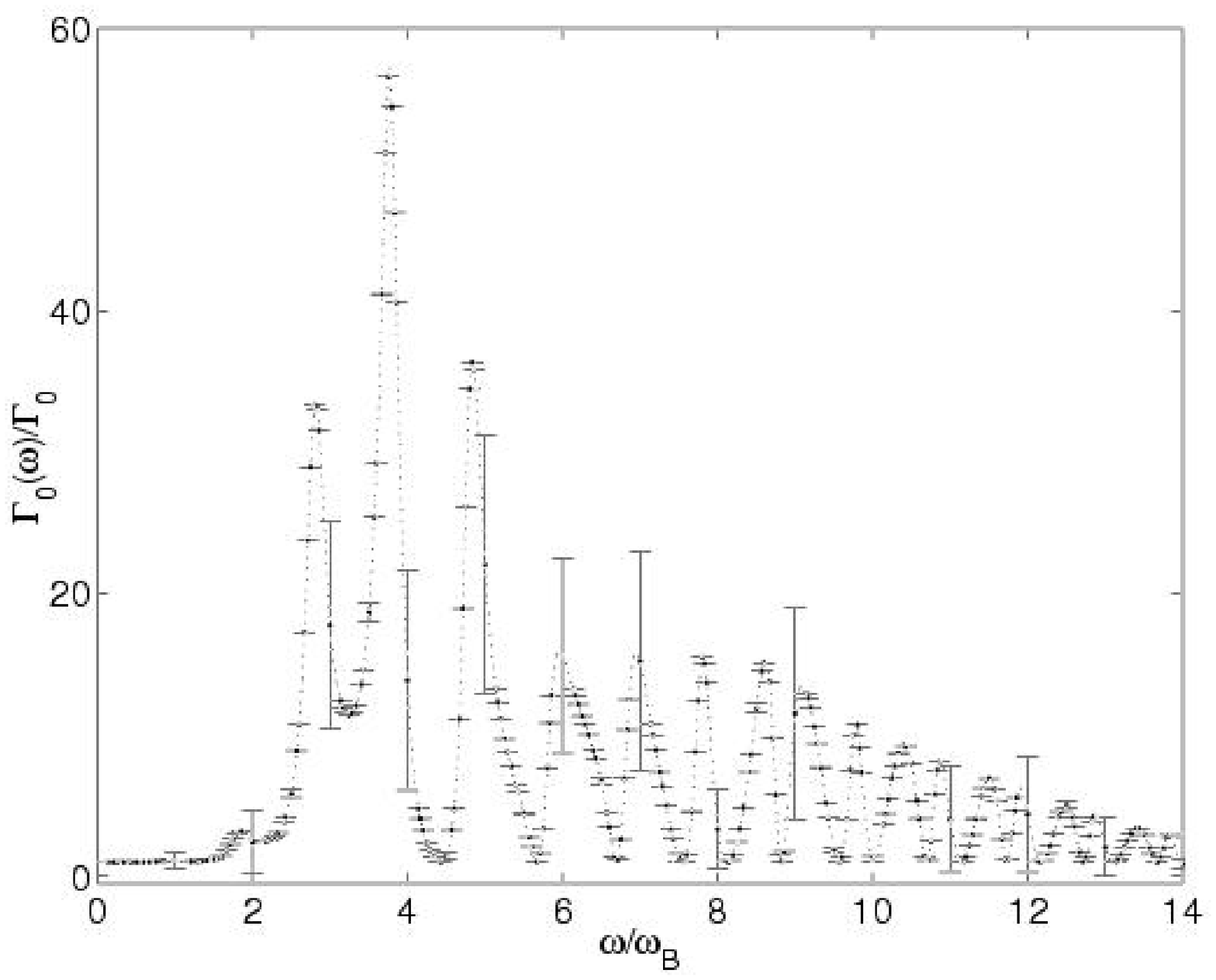}
\caption{\it The real (left panel) and the imaginary
(right panel) parts of the ground quasienergy resonances as function of the 
driving frequency $\omega$. The dashed line interpolates the average
values $\bar{E}_0(\omega)$ and $\bar{\Gamma}_0(\omega)$ 
obtained for the rational 
values $\omega/\omega_B = p/q$ with $q \le 7$ and $p \le 98$.
The ``error bars'' mark the bandwidths $\Delta^{\rm Re}_0(\omega)$ and 
$\Delta^{\rm Im}_0(\omega)$. The system parameters are $\hbar = 2$, 
$F = 0.061$ and $\varepsilon = 0.08$.}
\label{fig5a}
\end{center}
\end{figure}

We begin with the analysis of the real part of the spectrum,
$E={\rm Re}\,\mathcal{E}$. Recalling the results of Sec.~\ref{sec5a} the
real part of the quasienergy spectrum is expected to obey
\begin{equation}
\label{5d0}
E_\alpha^{(n)}(\kappa)=\bar{E}_\alpha +\frac{2\pi Fn}{q}
+\frac{\Delta_\alpha^{\rm Re}}{2}\,\cos(2\pi p\kappa) \;,\quad
n=0,\ldots,p-1 \;.
\end{equation}
The left panel in Fig.~\ref{fig5a} shows the mean position of the ground 
quasienergy bands (dots) and the band widths (marked as error bars) calculated
for some rational values  of the driving frequency $\omega$
(only the bands with $n=0$ are shown).
The parameters of the non-driven system with $V(x)=\cos x$ are
$\hbar=2$ and $F=0.061$. For these parameters the widths of two
first resonances are $\Gamma_0 = 1.24 \cdot 10^{-4}$ and 
$\Gamma_1 = 1.30 \cdot 10^{-1}$. The distance between the real parts
of the resonances is $E_1 - E_0 =3.784 \hbar\omega_B$. 
It is seen in the figure that the band widths 
$\Delta_0^{\rm Re}=\Delta_0^{\rm Re}(\omega)$ are large only for $\omega=p\omega_B$,
in qualitative agreement with the estimate (\ref{5b3}).
We would like to stress, however, that estimate (\ref{5b3}) is obtained 
within the single-band approximation and, because of this, the actual 
bandwidths deviate from this dependence.
(We shall discuss the conditions of validity of Eq.~(\ref{5b3}) later on 
in Sec.~\ref{sec5f}.) The second deviation from the predictions
of single-band model is the dependence of the mean quasienergy 
band position $\bar{E}_0$ on $\omega$. As shown below, this
dependence reflects the presence of the other quasienergy states,
originating from the higher ($\alpha>0$) Bloch bands. Let us also
note that the mean position $\bar{E}_\alpha=\bar{E}_\alpha(\omega)$
is, unlike the band width $\Delta_\alpha^{\rm Re}=\Delta_\alpha^{\rm Re}(\omega)$, 
a continuous function of the frequency.

The right panel in Fig.~\ref{fig5a} shows the imaginary part
$\Gamma=-2{\rm Im}\,\mathcal{E}$ of the quasienergy spectrum.
In the perturbation regime $\varepsilon\rightarrow0$ a  behavior similar
to (\ref{5d0}),
\begin{equation}
\label{5d2}
\Gamma_\alpha(\kappa)\approx \bar{\Gamma}_\alpha 
+\frac{\Delta_\alpha^{\rm Im}}{2}\,\cos(2\pi\kappa) \;,
\end{equation}
is observed. It should be noted that the smooth function 
$\bar{\Gamma}_0=\bar{\Gamma}_0(\omega)$ approximating the mean
values of the bands is nothing else as the induced decay rate 
discussed in Sec.~\ref{sec4a}. In fact, an arbitrary initial state of the
system (which was assumed to be the ground Wannier-Stark state
$\Psi_{0,l}(x)$ in Sec.~\ref{sec4a}) can be expanded in the
basis of the quasienergy states $\Psi_\alpha^{(n)}(x)$ as
\begin{equation}
\label{5d3}
\Psi(t=0)=\sum_{\alpha,n}c_{\alpha,n}\Psi_\alpha^{(n)} \;,\quad
c_\alpha(n)=\langle\Psi_\alpha^{(n)}|\Psi(t=0)\rangle \;.
\end{equation}
(Here we assume that $\omega/\omega_B$ is an irrational number and,
therefore, the quasienergy functions are localized function
with discrete spectrum.) During the time evolution the coefficients
$c_{\alpha,n}(t)$ decay as $\exp(-\Gamma_\alpha t/2\hbar)$. 
Since $\Gamma_\alpha>\Gamma_0$ ($\alpha>0$), the projection of
the wave function back to the initial state decays (after a short
transient) exponentially with an increment given by 
$\bar{\Gamma}_0(\omega)/2$. This is the underlying argument of our
numerical method of calculating the decay spectrum of the system.
Namely, to obtain the decay spectrum discussed in chapter 4 we
calculated the mean imaginary values of the quasienergy bands for
a number of rational $\omega/\omega_B$ and then interpolate them
for an arbitrary $\omega$.

\begin{figure}[t]
\begin{center}
\includegraphics[width=7cm,height=8cm]{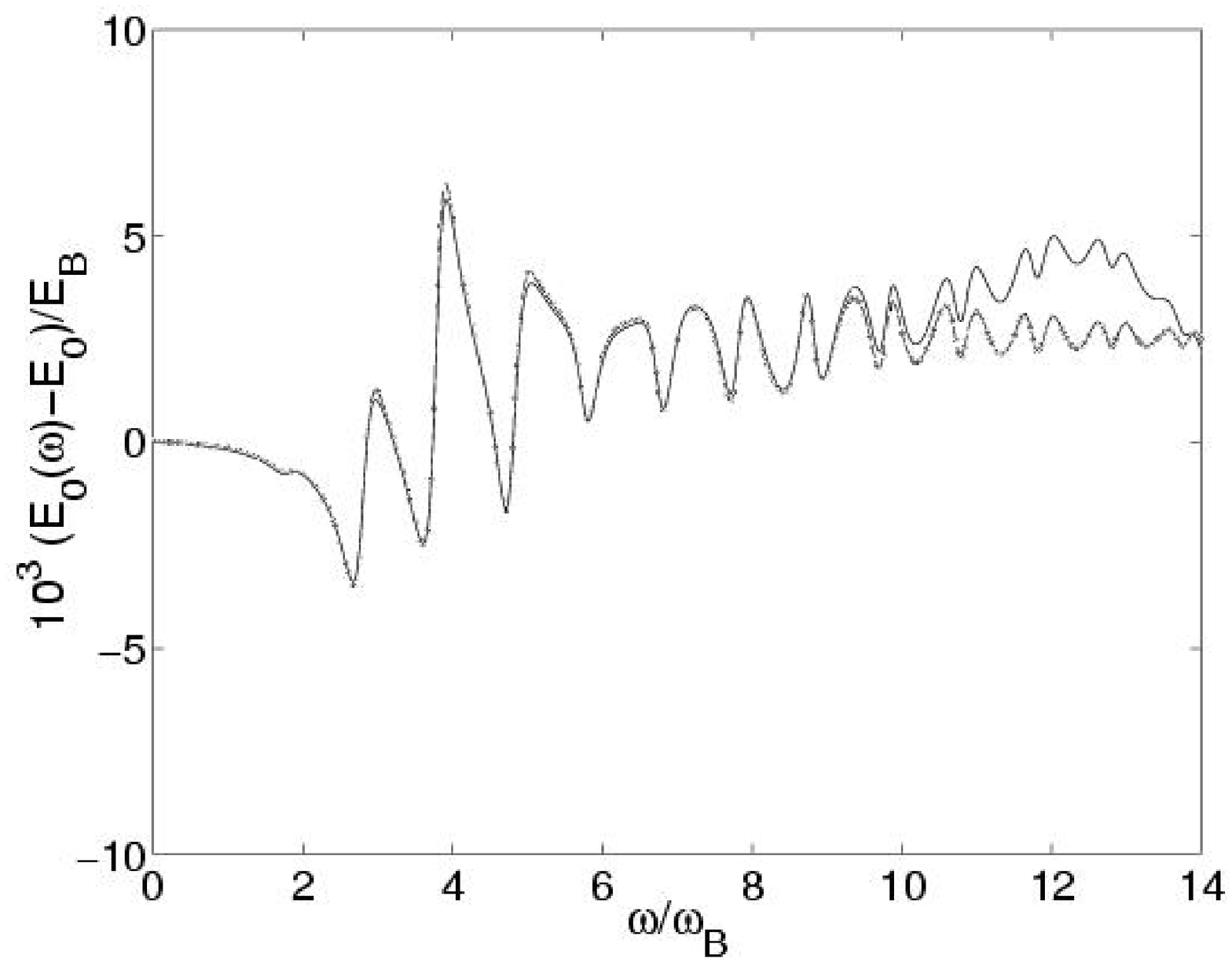}
\includegraphics[width=7cm,height=8cm]{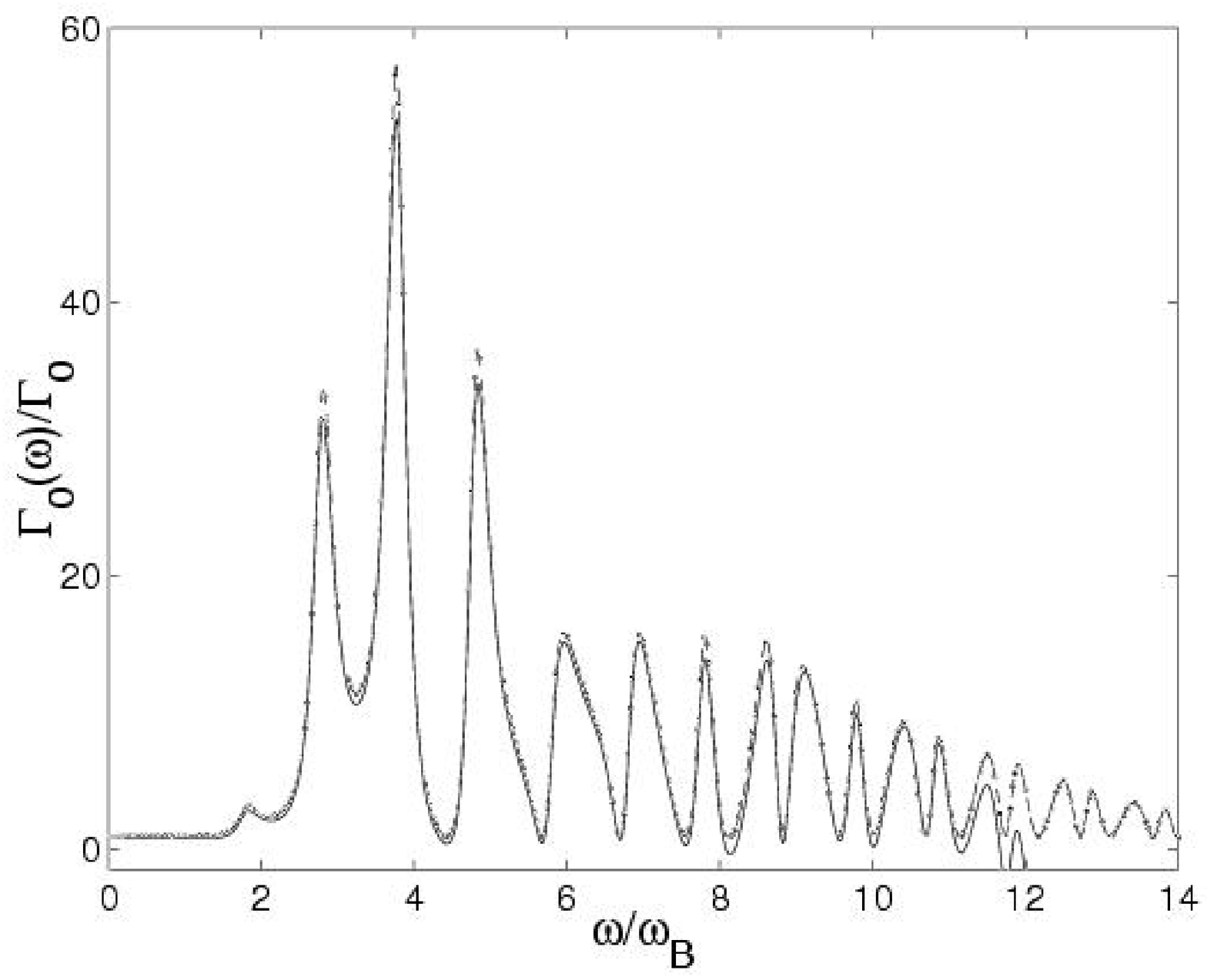}
\caption{\it Corrections to the ground state energy from 
Fig.~\ref{fig5a} (dashed line) compared to approximations based on
equation (\ref{5d5}) (solid line). The left panel shows the real part of the
ground state energy, the right panel the imaginary part.}
\label{fig5a1}
\end{center}
\end{figure}

Let us now discuss the $\omega$-dependence of the smooth
functions $\bar{E}_0(\omega)$, $\bar{\Gamma}_0(\omega)$.
Because we analyze the case of weak driving, these functions can
be obtained by using perturbation theory. In fact, assuming
again an irrational value of $\omega/\omega_B$, the zero order
approximation of the most stable quasienergy function is the ground
Wannier-Stark state $\Psi_{0,n}(x)$. According to the common
perturbation theory, the first order correction is
\begin{equation}
\label{5d4}
\Psi_0^{(n)}=\Psi_{0,n}+F_\omega\sum_{\alpha,l}\sum_\pm
\frac{\langle\Psi_{0,n}|x|\Psi_{\alpha,l}\rangle} 
{\mathcal{E}_{\alpha,l}-\mathcal{E}_{0,n}\pm\hbar\omega}\,\Psi_{\alpha,l} \;.
\end{equation}
Correspondingly, the second oder correction to the energy is
\begin{equation}
\label{5d5}
\mathcal{E}_0^{(n)}=\mathcal{E}_{0,n}+\frac{F^2_\omega}{2}\sum_{\alpha,l}
\sum_\pm \frac{V^2_{0,\alpha}(l-n)} 
{\mathcal{E}_{\alpha}-\mathcal{E}_{0}+(l-n)\hbar\omega_B\pm\hbar\omega} \;.
\end{equation}
In Eq.~(\ref{5d5}) we used the notation (\ref{4a3}) for the squared dipole
matrix elements and took into account that the energies of the Wannier-Stark
states form the ladder $\mathcal{E}_{\alpha,l}=\mathcal{E}_\alpha+l\hbar\omega_B$.
Equation (\ref{5d5}) is illustrated in Fig.~\ref{fig5a1}, where the real (left
panel) and imaginary (right panel) parts of the quasienergy calculated on
the basis of this equation (solid line) are compared to the numerical data
of Fig.~\ref{fig5a} (dots, interpolated by a dashed line). For small (relative
to $\omega_B$) frequencies both curves coincide almost perfectly, but deviate
for large $\omega$. This deviation can be attributed to the slow convergence of the
perturbation series over $\alpha$ in the high-frequency region. (For the presented
results, the upper limit for the sum over the Wannier-Stark ladders 
is taken as $\alpha=3$.) 

The concluding remark of this section concerns the relation between
Eq.~(\ref{4a2}) [i.e. the imaginary part of Eq.~(\ref{5d5})] and 
the ``$\varepsilon$-version'' of Eq.~(\ref{4a2}) used to analyze the decay 
spectrum of atoms in optical lattices in Sec.~\ref{sec4c}. The 
difference is the use of the squared matrix elements (\ref{4c3}) instead of 
the squared dipole matrix elements (\ref{4a3}). 
However, recalling the relation $\varepsilon=F_\omega /\omega^2$
and relation (\ref{4c5}), this difference can be shown to be
within the accuracy of the second order perturbation theory.
The advantage of the $\varepsilon$-version over the $F_\omega$-version is a better 
convergence in the high-frequency region.

\section{Perturbation theory for rational frequencies}
\label{sec5f}

Discussing the perturbation approach in the previous section we excluded
the case of rational ratio of the driving and Bloch frequencies. Let us now turn
to it. To be concrete, we restrict ourselves by the simplest but
important case $\omega=\omega_B$. In this case the periodic driving couples
the Wannier-Stark states belonging to the same Wannier-Stark ladder and, 
therefore, the extended Wannier-Bloch function $\Phi_{\alpha,\kappa}$ is an
appropriate zero order approximation to the quasienergy function.

As described in the beginning of this chapter, the complex quasienergies 
of the system are found by solving the eigenvalue equation
\begin{eqnarray}
\label{5f0}
U(T_\omega) \Phi_{\alpha,\kappa} = \exp[ -{\rm i}\mathcal{E}_\alpha(\kappa) 
T_\omega/\hbar]\,  \Phi_{\alpha,\kappa}  \;.
\end{eqnarray}
Let us approximate the Hamiltonian (\ref{5a0}) by the first order
of the Taylor expansion in $\varepsilon$,
$H \approx  H_W - \varepsilon \sin(x) \cos(\omega t)$ (here $H_W$ is
the Wannier-Stark Hamiltonian (\ref{2a0}) and $V(x)=\cos x$ is assumed 
for simplicity). Then we can calculate the effect of the 
periodic driving in the interaction representation of the Schr\"odinger equation. 
Explicitly, we get
\begin{eqnarray}
\label{5f1}
U(T_\omega) \approx U_\varepsilon(T_\omega) \,U_W(T_\omega) 
\end{eqnarray}
where the operator $U_\varepsilon(T_\omega)$ reads
\begin{eqnarray}
\label{5f2}
U_\varepsilon(T_\omega) = \widehat{\exp}\left( \frac{{\rm i}\varepsilon}{\hbar}
\int_0^{T_\omega} {\rm d}t \, \cos(\omega t) \, 
U^\dagger_W(t) \,  \sin(x) \, U_W(t) \right) \;,
\end{eqnarray}
and the operator $U_W(t)$ is the evolution operator for the unperturbed system.
According to common perturbation theory,
the first order correction is given by the diagonal elements 
of the operator $U_\varepsilon(T_\omega)$,
\begin{eqnarray}
\label{5f3}
\exp[-{\rm i}\Delta \mathcal{E}_\alpha(\kappa) T_\omega /\hbar] 
= \langle \Phi_{\alpha,\kappa}|\,U_\varepsilon(T_\omega)\,|
\Phi_{\alpha,\kappa}\rangle \, .
\end{eqnarray}
Let us approximate this formula further. Expanding the operator
exponent in a series in $\varepsilon$ and keeping only the first term, 
the correction to the quasienergy reads
\begin{eqnarray}
\label{5f4}
\Delta E_\alpha(\kappa) =-\frac{\varepsilon}{T_\omega} 
\int_0^{T_\omega} {\rm d}t \, \cos(\omega t) \, 
\langle\Phi_{\alpha,\kappa}|\,U^\dagger_W(t) \,\sin(x) \, U_W(t)\,|
\Phi_{\alpha,\kappa}\rangle \;.
\end{eqnarray}
Using the solution $U_W(t) \Phi_{\alpha,\kappa} = 
\exp( -{\rm i}\mathcal{E}_\alpha t/\hbar)\, \Phi_{\alpha,\kappa-Ft/\hbar}$
and substituting 
${\rm d}t/T_\omega = - {\rm d}\kappa$, \\ Eq.~(\ref{5f4}) takes the form
\begin{eqnarray}
\label{5f5}
\Delta \mathcal{E}_\alpha(\kappa) = -\varepsilon\int_0^{1} {\rm d}\kappa' 
\cos(2\pi\kappa') \, \langle \Phi_{\alpha,\kappa+\kappa'} |
\sin(x) | \Phi_{\alpha,\kappa+\kappa'} \rangle \;.
\end{eqnarray}
Finally, using the symmetry property of the Wannier-Bloch function,
the integral (\ref{5f5}) can be presented in the form
\begin{eqnarray}
\label{5f6}
\Delta \mathcal{E}_\alpha(\kappa) = 
\left(\frac{\Delta^{\rm Re}_\alpha+{\rm i}\Delta^{\rm Im}_\alpha}{2}\right) 
\cos(2\pi \kappa) \;,
\end{eqnarray}
where
\begin{eqnarray}
\label{5f6a} 
\Delta^{\rm Re}_\alpha+{\rm i}\Delta^{\rm Im}_\alpha =
-2\varepsilon \int_0^{1} {\rm d}\kappa \,\cos(2\pi \kappa) \, 
\langle \Phi_{\alpha,\kappa} | \sin(x) | \Phi_{\alpha,\kappa} \rangle \, .
\end{eqnarray}
(The special notation for the band width stresses that the integral on
the right hand side of Eq.~(\ref{5f6a}) is a complex number.)
Thus, a weak periodic driving removes the degeneracy of the
Wannier-Bloch bands which then gain a finite width. Moreover, 
there are corrections both to the real and imaginary part of the quasienergy.

In conclusion, let us briefly discuss the relation between the formulas
(\ref{5f6}), (\ref{5f6a}) and the tight-binding result (\ref{5b4}). As was stated
many times, the single-band model neglects the interband tunneling,
which is justified in the limit $F\rightarrow0$. In this limit the
quasienergy band width can be estimated as
\begin{eqnarray}
\label{5f7}
\Delta^{\rm Re}_\alpha=\varepsilon\frac{4\pi^2F\Delta_\alpha}{\hbar^2}
=\Delta_\alpha\frac{F_\omega}{F} \;,
\end{eqnarray}
where $\Delta_\alpha$ is the width of the Bloch band.
Indeed, using Eq.~(\ref{4c4}), the band width in Eq.~(\ref{5f6a}) can be
expressed in terms of the dipole matrix elements as
\begin{eqnarray}
\label{5f8}
\Delta^{\rm Re}_\alpha+{\rm i}\Delta^{\rm Im}_\alpha  = 
\varepsilon \frac{(2\pi F)^2}{\hbar^2} \, \Big( \, 
\langle \Psi_{\alpha,1}| x |\Psi_{\alpha,0}\rangle 
+ \langle \Psi_{\alpha,0}| x |\Psi_{\alpha,1}\rangle \, \Big) \;.
\end{eqnarray}
Then, using the tight-binding approximation (\ref{1b6}) for
the resonance Wannier-Stark states $\Psi_{\alpha,1}(x)$, we obtain
the estimate  (\ref{5f7}). (Alternatively, we can approximate 
$\chi_{\alpha,\kappa}(x)$ in Eq.~(\ref{4b4}) by the periodic part of the Bloch 
function.) It is seen, that the estimate (\ref{5f7}) coincides with 
Eq.~(\ref{5b4}) in the limit $F_\omega/F\rightarrow0$. We would like
to stress, however, that the actual perturbation parameter of the problem
is $\varepsilon\sim F_\omega/F^2$ and not $F_\omega/F$, as it could be
naively expected on the basis of the tight-binding model.

\section{Selective decay}
\label{sec5e}

This section serves as an illustration to the perturbation theory
of Sec.~\ref{sec5f} and discusses some important limitations of
the perturbation approach. In order to 
reduce the number of relevant resonance states, we choose 
the parameters of the unperturbed system as $\hbar = 2$, $F = 0.08$. 
In this case we have to take into account mainly two resonances with
energies $\mathcal{E}_0 = 9.42 \cdot 10^{-2} - {\rm i} 5.60 \cdot 10^{-4}$ 
and $\mathcal{E}_1 = 4.18 \cdot 10^{-2} - {\rm i}8.81 \cdot 10^{-2}$. All
other resonances are very unstable and approximately do not 
influence the results. The frequency of the time-periodic
perturbation is given by $\omega = 2\pi F/\hbar \approx 0.251$. 
\begin{figure}[p]
\begin{center}
\includegraphics[width=12cm,height=8cm]{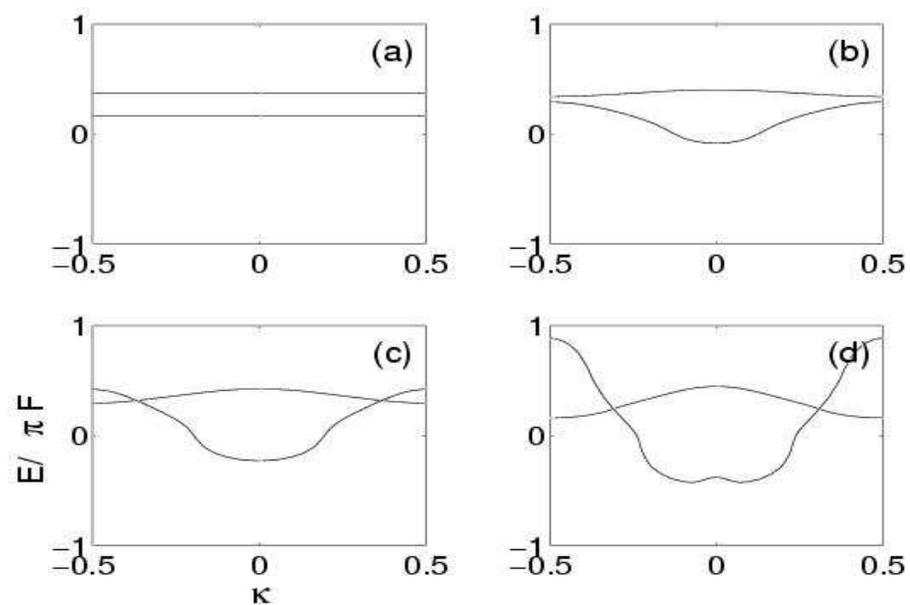}
\caption{\it Real part of the two most stable (quasi-)energy Wannier-Bloch
bands for $\varepsilon = 0$ (a), $\varepsilon = 0.2$ (b), $\varepsilon = 0.4$ (c)
and $\varepsilon = 1$ (d). The other system parameters are $\hbar = 2$, 
$F = 0.08$ and $\omega = 2\pi F/\hbar \approx 0.251$.}
\label{fig5b}
\end{center}
\end{figure}
\begin{figure}[p]
\begin{center}
\includegraphics[width=10cm,height=6.5cm]{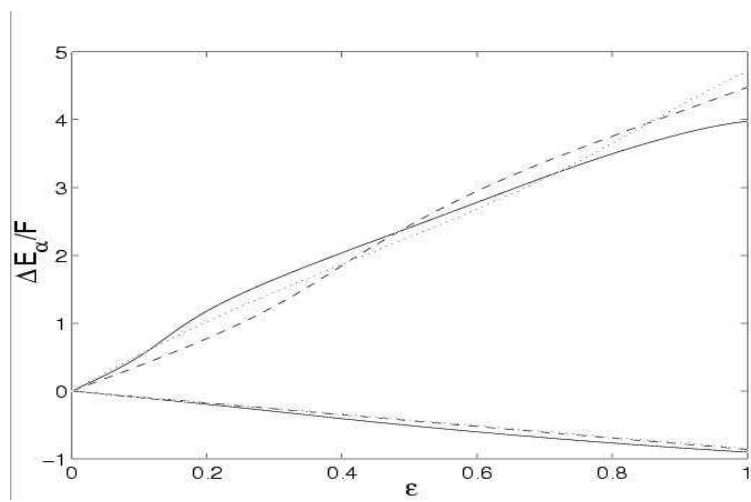}
\caption{\it Band width of the two most stable Wannier-Bloch bands
as a function of the perturbation parameter $\varepsilon$ for $F = 0.08$ 
(solid line), $F = 0.04$ (dashed line) and $F = 0.02$ (dotted line).}
\label{fig5c}
\end{center}
\end{figure}
\begin{figure}[p]
\begin{center}
\includegraphics[width=12cm,height=8cm]{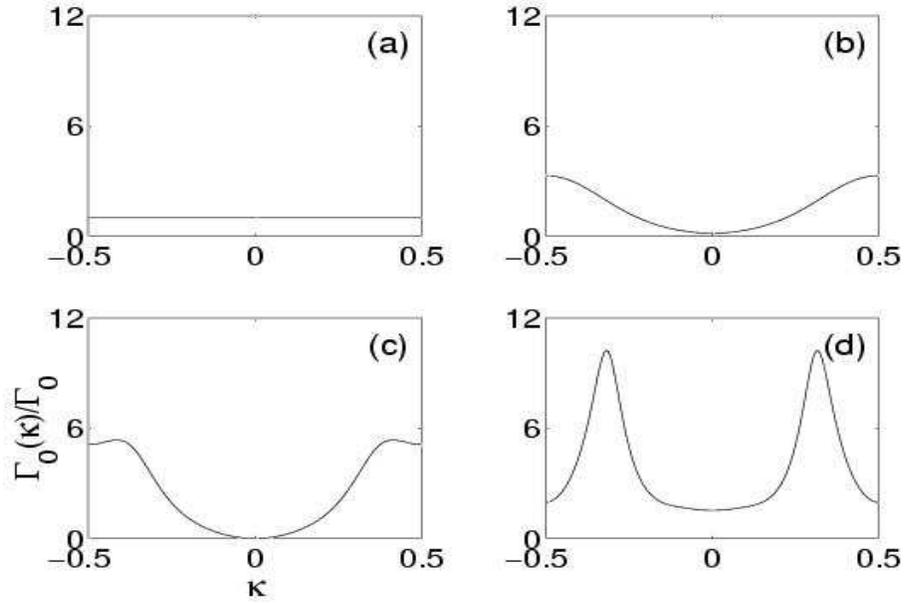}
\caption{\it Width (decay rate) of the ground state as a function of the Bloch index $\kappa$
for the cases studied in Fig.~\ref{fig5b}. The width is normalized
with respect to the width at $\varepsilon = 0$.}
\label{fig5d}
\end{center}
\end{figure}
\begin{figure}[p]
\begin{center}
\includegraphics[width=10cm,height=6.5cm]{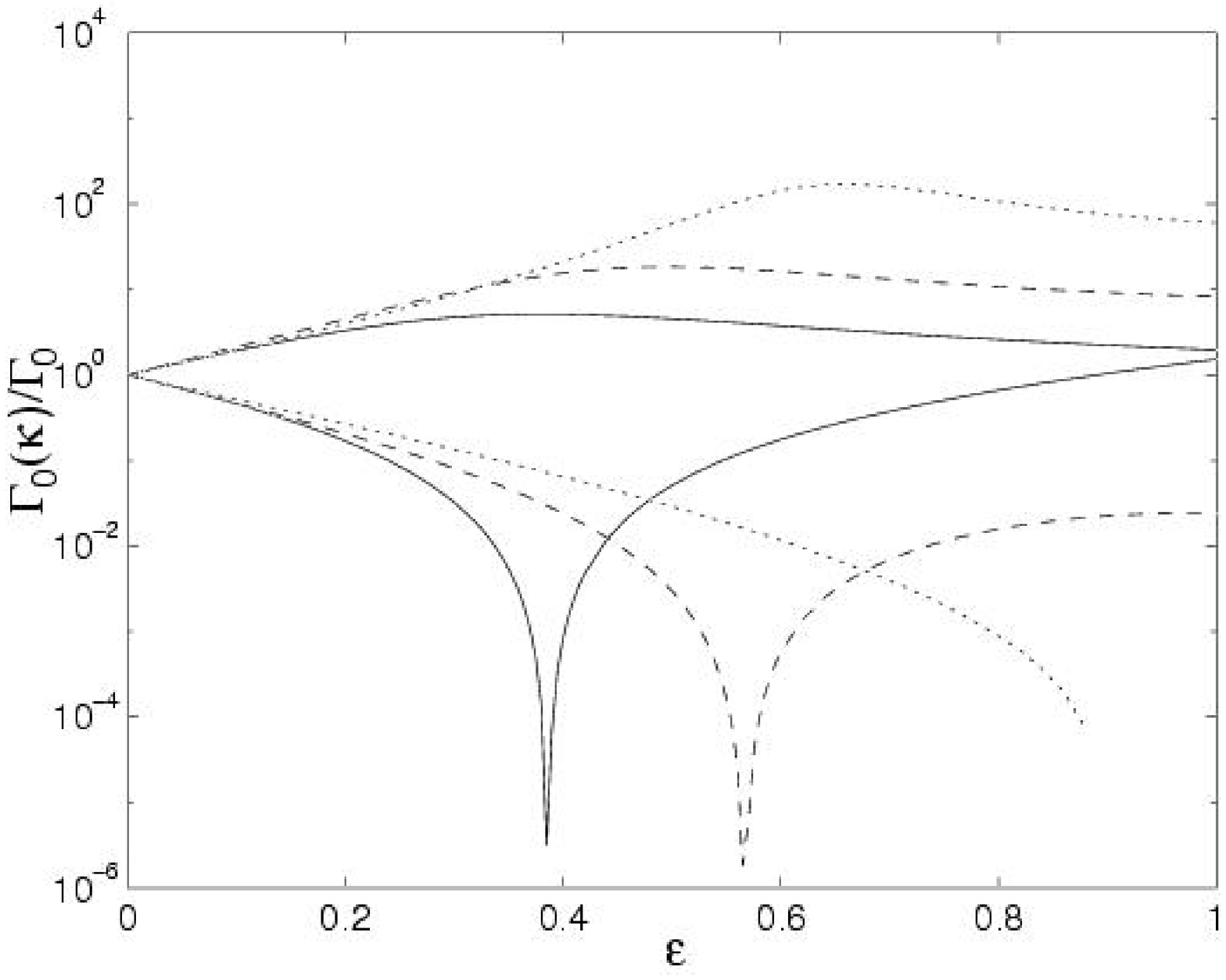}
\caption{\it Decay rate of the most stable Wannier-Bloch state
at $\kappa = 0$ (lower family of curves) and $\kappa = \pm 1/2$ (upper
family of curves) for the same parameters as in Fig.~\ref{fig5c}, 
i.e.~$F = 0.08$ (solid line), $F = 0.04$ (dashed line) and $F = 0.02$
(dotted line). The rate is normalized against the decay rate 
at $\varepsilon = 0$.}
\label{fig5f}
\end{center}
\end{figure}

Figure \ref{fig5b} shows the real parts of the quasienergies of the
two most stable Wannier-Bloch resonances for different amplitudes 
$\varepsilon$. In panel (a) we have the unperturbed case with flat bands. 
When the ac driving is added, the dispersion relation of the
ground band is well described by the theoretical cosine dependence. 
The first excited band follows this relation only up to $\varepsilon = 0.2$. 
If the amplitude is increased further, deviations from the cosine
appear, and for $\varepsilon = 1$ other effects strongly influence the 
band (note that in this case $F_\omega/F \approx 0.79$, 
thus we are still far away from the parameter range where the tight-binding 
model predicts dynamical band suppression). Furthermore, for $\varepsilon > 0.2$ 
the bands cross, and then we cannot neglect their interaction. 

Next we investigate the bandwidth, i.e.~the difference between the
extrema of the real parts of the quasienergies,
$\Delta E\!=\!E(\kappa \!=\!1/2)-E(\kappa\!=\!0)$.
Figure \ref{fig5c} shows the width of the two most stable
bands as a function of the amplitude $\varepsilon$ for three different
field strengths $F = 0.02$, $F = 0.04$ and $F = 0.08$. 
It is seen that in all cases the bandwidth grows approximately linearly. 
Again, the agreement is much better for the ground band; for the
first excited band one observes an oscillation around the linear growth.
Note that the slope is proportional to $F$ as expected on the basis
of the perturbation theory [see Eq.~(\ref{5f7})].

We proceed with the analysis of the imaginary part of the quasienergy spectrum.
Figure \ref{fig5d} shows the width of the ground state as a 
function of the Bloch index for the parameters of Fig.~\ref{fig5b}.
For $\varepsilon = 0$ the band is flat as predicted from the theory. 
For $\varepsilon = 0.2$ the width can be approximated by a cosine, 
however, the mean is shifted to approximately twice the unperturbed 
width. If we further increase $\varepsilon$, additional structures appear. 
In comparison with Fig.~\ref{fig5b}, we see that the bandwidth is
increased where the (real part of the quasienergy of the) ground band crosses 
the first excited band. Therefore, we can clearly assign the
increase of the width to the band crossings. Recalling the
results of chapter 3, we again observe effects of resonant tunneling, 
now as a function of the quasimomentum. As shown in Ref.~\cite{PLA2,Thesis}, 
the two-state model of Sec.~\ref{sec3c} can be adopted to the present case 
and yields good correspondence to the numerical data.
 
In Fig.~\ref{fig5d} we can see that 
the perturbation can both increase and decrease the width and
thus the rate of decay of the quasienergy states. In the case considered,
for small $\varepsilon$ the decay is enhanced at the edges of the
Brillouin zone and suppressed in its center.\footnote{The regions of
enhanced and suppressed decay depend on the difference between the phase
of the driving force and the phase of the Bloch oscillation. For example,
the change of $\cos(\omega t)$ in the Hamiltonian (\ref{5a0}) to 
$\sin(\omega t)$ shifts the displayed dispersion relation by a quarter
of the Brillouin zone.} Let us therefore take these two quasimomenta to 
further investigate the dependence on the perturbation parameter $\varepsilon$. 
The results of a calculation of the widths at $\kappa = 0$ and $\kappa = \pm 1/2$
as a function of the amplitude $\varepsilon$ are shown in figure \ref{fig5f}. 
For small $\varepsilon$ the dependence is nearly linear, 
but for larger values it is highly nontrivial. In particular, we 
would like to draw the attention to the behavior of the solid
line at $\varepsilon \approx 0.4$ and of the dashed line at 
$\varepsilon \approx 0.54$. Here the decay rate is suppressed by
more than a factor $10^5$! This tremendous decrease of the
decay rate has enormous consequences on the global dynamics. 
For example, let us initially take the most stable Wannier-Stark state
and then add the ac driving. Then the survival probability is given by
\begin{eqnarray}
\label{5e1}
P(t) = \int_{-1/2}^{1/2} {\rm d}\kappa \exp\left( 
- \frac{\Gamma_0(\kappa) \, t}{\hbar}\right) \;.
\end{eqnarray}
If we approach the critical value $\varepsilon_{\rm cr}$, the decay is suppressed
and asymptotically
\begin{eqnarray}
\label{5e2}
P(t) \sim t^{-1/2}\exp( -\Gamma_{\rm min} \, t/\hbar) \;,
\end{eqnarray}
where $\Gamma_{\rm min}$ is the minimal decay rate. Let us also note another
property. Since the decay rate of the quasienergy states depend on the
quasimomentum, after some time only the contributions with quasimomentum 
around the value with the smallest decay rate will survive.  In what follows 
we shall refer to this phenomenon as the {\em selective decay} of the
quasienergy states. Some physical consequence of this phenomenon are 
discussed in the next chapter.

\chapter{Wave packet dynamics}
\label{sec6}

In this section we address the question of the time evolution
of an {\em initially localized} wave packet. Usually this problem
is analyzed by simulating the wave packet dynamics on the basis of the
time-dependent
Schr\"odinger equation. However, this numerical approach is very time
consuming and has an upper limit for the times considered. In what
follows we describe the evolution of the wave packet in terms of
the resonance states. Besides tremendous decrease of the computational
efforts, the latter approach  also gives additional insight into
the decay process of the Wannier-Stark states.

\section{Expansion over resonance states}
\label{sec6a}

A direct expansion of a localized state in terms of 
resonances yields inappropriate results because in the 
negative $x$-direction the resonance states extend to infinity.
Therefore the description needs to be modified to take into 
account the finite extension of the initial state.
Recently this problem was analyzed for decaying quantum systems
with a finite range potential 
\cite{Dijk99,Dijk99a}.\footnote{However, this problem was already 
addressed in textbooks as, e.g., \cite{Tayl72}.}

Let us adopt the approach of \cite{Dijk99a} to describe the evolution of 
the wave packet in momentum space. In this approach, the wave function
$\psi(k,t)$ is expressed in terms of the stationary scattering states
$\Psi_S(k;E)$:
\begin{eqnarray}
\label{6a0}
\psi(k,t) = \int_{-\infty}^{\infty} {\rm d}E  \, f(E) \, \Psi_S(k;E)
\, \exp\left(-{\rm i}\,\frac{E t}{\hbar}\right)  \;,
\end{eqnarray}
where $f(E) =  \langle \Psi_S(k;E) | \psi(k,0) \rangle$. [We recall that
the states $\Psi_S(k;E)$ are normalized to $\delta$-function:
$\langle \Psi_S(k;E') | \Psi_S(k;E) \rangle=\delta(E-E')$.]
We are mainly interested in the properties of the decay tail at
$k\rightarrow-\infty$. In this region the scattering states can be
approximated by their asymptotic form [see Eq.~(\ref{2a5})]
\begin{eqnarray}
\label{6a1}
\lim_{k \rightarrow \pm \infty} \Psi_\pm(k;E) = g_\pm(E)
\,\exp\left({\rm i}\,\frac{\hbar^2 k^3}{6F} - {\rm i}\,\frac{E k}{F} \right) \;,
\quad g_\pm(E)=e^{\pm{\rm i}\varphi(E)} \;.
\end{eqnarray}
Substituting this asymptotic form into Eq.~(\ref{6a0})  we have
\begin{eqnarray}
\label{6a2}
\psi(k,t) = \exp\left( {\rm i}\,\frac{\hbar^2 k^3}{6F}\right)\,  
G_{-}\left( k + \frac{F t}{\hbar} \right) \,,\quad k\ll0 \;,
\end{eqnarray}
where 
\begin{eqnarray}
\label{6a3}
G_{-}(k) = \int_{-\infty}^{\infty} {\rm d}E \, \frac{f(E)}{g_{+}(E)} \,
\exp\left( - {\rm i}\,\frac{E k}{F} \right)  \;. 
\end{eqnarray}
If the initial wave function $\psi(k,0)$ has a finite support, 
$f(E)$ is an entire function in the complex plane. Then the function $f(E)/g_+(E)$ 
has simple poles at zeros of $g_+(E)$, i.e at the poles of the
scattering matrix $S(E)=g_-(E)/g_+(E)$. This property suggests to evaluate 
the integral (\ref{6a3}) with the help of the residuum theorem.  
Without knowing the explicit form of the function $f(E)/g_+(E)$ we have to
make some assumptions on its asymptotic behavior in order to 
proceed further. In particular, if we assume that the function $f(E)/g_+(E)$ 
does not influence the behavior of the integrand at infinity, 
the integral yields a sum over the residua located within the 
appropriate contour. Explicitly, for $k > 0$ the contour should be closed 
in the lower half of the complex plane, for $k < 0$ it contains the 
upper half.  Since all poles of the scattering matrix are located in the 
lower half of the complex plane, we get
\begin{eqnarray}
\label{6a4}
G_-(k) = 2\pi {\rm i} \, \Theta\left(k\right) 
\sum_\nu b_\nu \exp\left( - {\rm i}\,\frac{\mathcal{E}_\nu k}{F} \right) \, , 
\end{eqnarray}
where $\Theta(k)$ is the Heaviside function, the $b_\nu$ are the residua of 
$f(E)/g_+(E)$ at the poles, and $\nu=\{\alpha,l\}$. Inserting this result 
in (\ref{6a2}) yields
\begin{eqnarray}
\label{6a5}
\psi(k,t) =  \Theta\left(\hbar k + F t\right) \sum_\nu  c_\nu
\exp\left[{\rm i}\left(\frac{\hbar^2 k^3}{6F} 
- \frac{\mathcal{E}_\nu k}{F} -\frac{\mathcal{E}_\nu t}{\hbar}\right)\right] \;,
\end{eqnarray}
with $c_\nu = 2\pi {\rm i} b_\nu$. The terms of the sum are actually
proportional to the asymptotic form of the resonance wave functions 
$\Psi_\nu(k,t)$. Thus, we can equivalently present the wave function as
\begin{eqnarray}
\label{6a7}
\psi(k,t) =  \Theta\left(\hbar k + F t\right) \sum_\nu c_\nu \, 
\exp\left(-{\rm i}\,\frac{\mathcal{E}_\nu t}{\hbar}\right)\Psi_\nu(k) \;.
\end{eqnarray}
Therefore, in the Stark case we can describe the evolution of an initial 
state by a superposition of resonances, where we take into account the 
space-time decay process in the prefactor 
$ \Theta\left(\hbar k + F t\right)$. This factor truncates 
the wave function at the momentum $\hbar k = - Ft$, 
i.e.~only momenta with $\hbar k > -Ft$ contribute. 
With increasing time, the wave function extends to smaller momenta, 
where the edge moves according to the classical equation of motion.

It should be noted that the location of the edge reflects the assumption on the
behavior at infinity we made in order to explicitly evaluate the integral.
For example, the function $f(E)/g_+(E)$ may contain an additional
exponential factor $\exp({\rm i} \alpha E)$ (see the example in 
\cite{Dijk99a}). Though this factor does not 
influence the poles, it nevertheless influences the argument of the 
Heaviside function. In fact, in a realistic situation the edge will be 
shifted, because the truncation edge at $t =0$ has to reflect 
the extension of the initial state in momentum space. 
We take this into account by replacing $\hbar k$ in the argument of the 
Heaviside function by $\hbar(k + k_0)$, where $k_0$ describes 
the extension of the initial state in the negative $k$-direction.
Furthermore, if the initial state does not have a compact support
but a tail in the negative momentum direction, the edge will be smoothed
and deformed. However, the qualitative behavior remains
unchanged:  the prefactor is approximately constant for positive 
arguments of the Heaviside function, and it approximately vanishes 
for negative arguments. Therefore, we take the Heaviside description 
as a reasonable approximation to the real situation. 
Let us also note that the wave function constructed in this way can be 
normalized. Indeed, in the positive
momentum direction the resonances decrease stronger than exponentially, 
and in the negative direction the wave function is truncated. 

Now we discuss the dynamics of the wave packet
in coordinate space. If we are interested in the asymptotic
behavior for $x\ll0$, the wave function $\psi(x,t)$ can be found
by a Fourier transform of the asymptotic form of Eq.~(\ref{6a7}):
\begin{eqnarray}
\label{6a8}
\psi(x,t) = \int_{-\infty}^{\infty}\!\!{\rm d}k \,
\Theta[\hbar (k+k_0) + F t] 
\sum_\nu  c_\nu \exp\left[{\rm i}\left(\frac{\hbar^2 k^3}{6F} 
- \frac{\mathcal{E}_\nu k}{F} - \frac{\mathcal{E}_\nu t}{\hbar} 
+ k x \right)\right] \;.
\end{eqnarray}
Let us evaluate the integral in the stationary phase approximation.
The equation for the stationary phases reads
\begin{eqnarray}
\label{6a9}
\frac{{\rm d}}{{\rm d}k}\left(\frac{\hbar^2 k^3}{6F}
 - \frac{k \mathcal{E}_\nu}{F} + kx\right)=0  \;. 
\end{eqnarray}
Neglecting the imaginary part of the energy $\mathcal{E}_\nu$
\footnote{More precisely, we treat the exponential of the imaginary
part as a slowly varying function.}, the stationary phase condition is 
just the energy conservation, and the stationary points are the classical 
momenta $\hbar k_{\nu} = \sqrt{2(E_\nu - Fx)}=p_\nu(x)$.
If $p_\nu(x) \ll - \hbar k_0 -F t$, the prefactor is zero and 
the integral vanishes. On the other hand, if 
$p_\nu(x) \gg - \hbar k_0 -F t$, the integral of the contribution of 
the $\nu$-th resonance yields approximately
\begin{eqnarray}
\label{6a10}
\exp\left(-{\rm i} \frac{\mathcal{E}_\nu t}{\hbar}\right)
\,  \sqrt{\frac{2 \pi F}{\hbar p_\nu(x)}\,}\, 
\exp\left(-{\rm i} \frac{p_\nu^3(x)}{3 \hbar F} 
-\frac{\Gamma_\nu p_\nu(x)}{2 \hbar F} \right) \;, 
\end{eqnarray}
which is just the asymptotic form of the Wannier-Stark state in the
coordinate representation. The critical point is 
$p_\nu(x) = - \hbar k_0 - F t$, where the approximation
breaks down because the Heaviside function is not a slowly varying
function at this point. Actually, in the vicinity of this point 
the integral interpolates between the other two possibilities.
Let us skip a more detailed analysis here and roughly describe the
transition between both regimes by a Heaviside function
of the argument $p_\nu(x) + \hbar k_0 + F t$, or, equivalently, 
of the argument $x + F (t+t_0)^2/2 - E_\nu/F$, where $t_0 = \hbar k_0 / F$.
Then, replacing the contribution (\ref{6a10}) by $\Psi_\nu(x,t)$, 
we get
\begin{eqnarray}
\psi(x,t) = \sum_\nu \, c_\nu \, \Theta\left[x + \frac{F (t+t_0)^2}{2} 
- \frac{E_\nu}{F} \right] \, 
\exp\left(-{\rm i}\,\frac{\mathcal{E}_\nu t}{\hbar}\right) \Psi_\nu(x) \, . 
\end{eqnarray}
In comparison to equation (\ref{6a7}) there are two differences. 
First, in coordinate space the truncation depends on the energy of the 
resonances. Furthermore, the edges of the different contributions move 
with a quadratic time dependence, which reflects the classical 
(accelerated) motion in a constant external field.

\section{Pulse output from Wannier-Stark systems}
\label{sec6b}

Let us consider the dynamics of a coherent superposition of the
Wannier-Stark resonances belonging to a particular Wannier-Stark ladder
\begin{eqnarray}
\label{6b0}
\psi(k,t) = \sum_l c_l \exp\left(-{\rm i}\,\frac{\mathcal{E}_{\alpha,l} t}
{\hbar}\right) \, \Psi_{\alpha,l}(k) \;,\quad
c_l=\frac{1}{\sqrt{\pi}\sigma}\exp\left(-\frac{l^2}{\sigma^2}\right) \;.
\end{eqnarray}
(To shorten the notation, we skip 
here the truncation by the Heaviside function because the truncation 
does not influence the properties which we are going to discuss). 
This problem, as will be seen later on, is directly related to the  
experiment \cite{Ande98}, where a coherent pulse output of cold atoms was 
observed. Based on this phenomenon, a possibility of constructing an atomic laser 
is currently discussed in the literature.

According to Eq.~(\ref{2f4}) the Wannier-Stark states belonging
to the same ladder are related by 
$\Psi_{\alpha,l}(k) = \exp(-{\rm i} 2 \pi l k) \,\Psi_{\alpha,0}(k)$
and $\mathcal{E}_{\alpha,l}=\mathcal{E}_\alpha+2\pi lF$.
Combining this phase relation with the different phases due to the time 
evolution, the time evolution of the superposition is given by
\begin{eqnarray}
\label{6b1}
\psi(k,t)=\Psi_{\alpha,0}(k,t)\, \widetilde{C}\left(\frac{F t}{\hbar} + k\right) 
\;,\quad \widetilde{C}(k)= \sum_{l} c_l \,\exp(-{\rm i}2\pi l k) \;.
\end{eqnarray}
where $\Psi_{\alpha,0}(k,t) = \exp(-{\rm i}\mathcal{E}_\alpha t/
\hbar)\Psi_{\alpha,0}(k)$. 
Thus, the time evolution of the superposition is given by the 
time-evolved wave function at the mean energy, $\Psi_{\alpha,0}(k,t)$, 
times the discrete Fourier transform $\widetilde{C}(k)$ of the 
amplitudes $c_l$, which is taken at the momenta $k + Ft/\hbar$. 
Since the function $\widetilde{C}(k)$ is periodic in momentum space
the factor $\widetilde{C}(k + Ft/\hbar)$ 
is also periodic in time with the period $\hbar/F = T_B$. 
In what follows we shall refer to the function $\widetilde{C}(k)$
as amplitude modulation factor. In the considered case 
$c_l\sim\exp(-l^2/\sigma^2)$ the amplitude modulation factor is 
obviously a periodic train of Gaussians with the width $\sigma^{-1}$.

We turn to the coordinate representation. Following the derivation
of the preceding section, the wave function $\psi(x,t)$ can be shown
to obey
\begin{eqnarray}
\label{6b2}
\psi(x,t) = \Psi_{\alpha,0}(x,t) \,\widetilde{C}\left( \frac{t}{T_B} 
+ \frac{p(x)}{\hbar} \right) \;,
\end{eqnarray}
where, as before, $p(x) = \sqrt{2 (E_\alpha - F x)}$ is the classical
momentum. Because the function $\widetilde{C}(k)$ has peaks at
integer values of the arguments, the function 
$\widetilde{C}(t/T_B + p(x)/\hbar)$ has peaks at the  coordinates
\begin{eqnarray}
\label{6b3}
x = x_0 - \frac{F}{2} \, (t + m T_B)^2 \, ,
\end{eqnarray}
where $x_0 = E_\alpha /F$ is the classical turning point. Thus, as a function of
time, the peaks accelerate according to the classical equation of motion 
of a free particle subject to a constant electric field. Additionally, the
peaks broaden linearly with increasing time (or with increasing $m$). 
It is straightforward to combine the result (\ref{6b2}) with the
result of the previous sections. Generally, we have to truncate the wave 
front approximately at the coordinate $x = - F (t+t_0)^2/2$. 
\begin{figure}[t]
\begin{center}
\includegraphics[width=10cm]{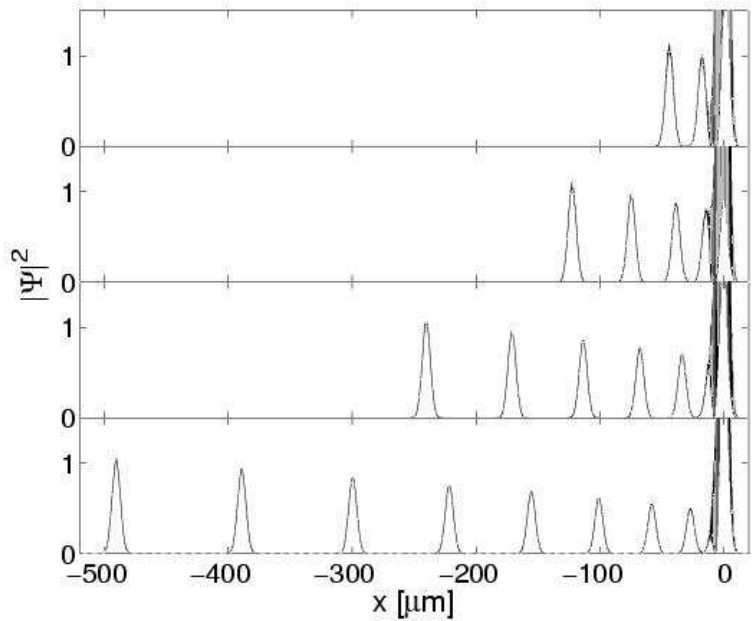}
\caption{\it Space-time decay of the wave packet for the parameters
of the experiment  \cite{Ande98}. From top to bottom, the panels correspond 
to $t = 3\,{\rm ms}$, $5\,{\rm ms}$, $7\,{\rm ms}$ and $10\,{\rm ms}$, 
respectively.\label{fig6a}}
\end{center}
\end{figure}
 
Figure \ref{fig6a} shows the evolution of the superposition of the
ground Wannier-Stark resonances for $V(x)=\cos x$. The system parameters
are $\hbar = 3.3806$, $F = 0.0661$ and $\sigma=15$, which correspond
to the setting of the experiment \cite{Ande98}.
The figure was calculated in the following way: First the ground $l=0$
Wannier-Stark state was calculated in the momentum representation. 
Then the wave function was multiplied with the 
amplitude modulation factor $\widetilde{C}(Ft/\hbar+k)$ taken at the specified 
times and truncated according to equation (\ref{6a7}). (We shifted the
truncation edge by $k_0 = 1/2$ in order to avoid a truncation 
directly at the maxima. As mentioned in Sec.~\ref{sec6a} this shift
takes into account the finite extension of the initial state.)
Finally the resulting function was Fourier transformed into coordinate 
space. The obtained result reproduces the findings of the experiment \cite{Ande98}. 
A series of pulses is formed which then accelerate according to the
free motion. At a fixed value of the coordinate, the sequence is periodic 
in time (after the first pulse passed), up to an overall exponential decay 
which reflects the fact that every pulse takes away a certain amount 
of probability. 

A few words should be added about the validity of the one-particle 
approximation. In fact, in the experiment cited, the authors used
a Bose-Einstein condensate of Rubidium atoms, uploaded in a
vertically aligned optical lattice. Thus a description of the system
with the help of the Gross-Pitaevskii equation\footnote{A detailed
introduction to the physics of Bose-Einstein condensates can be found in the review 
article \cite{Park98}.} 
\begin{eqnarray}
\label{6b4}
{\rm i}\hbar \partial_t \Psi = \left[ \frac{p^2}{2M} + V_0 \cos(2k_L z) + 
M g z + V_{\rm int} |\Psi|^2 \right] \Psi \,,
\end{eqnarray}
looks more appropriate. Equation (\ref{6b4}) was studied numerically 
in Ref.~\cite{Chio00,Berg98,Choi99,Wu00,Java99,Zoba00,Ceri00}. It was found
that for moderate densities of the condensate (realized in practice)
the pulse formation is only slightly modified by the nonlinear term in
the Gross-Pitaevskii equation. Thus the physics behind the experimentally
observed phenomenon is provided by single-particle quantum mechanics
and can be well understood in terms of
Wannier-Stark resonance states.

\section{Atom laser mode-locking}
\label{sec6c}

The crucial point for the existence of the pulse output in the
Wannier-Stark system is the fixed phase relation between the
probability amplitude $c_l$ in Eq.~(\ref{6b0}). 
In the experiment \cite{Ande98}, this fixed phase relation 
was achieved by the self-interaction of the Bose condensate.
In the following we show that one can prepare an appropriate initial state 
within single particle quantum mechanics. Explicitly, 
the statement is as follows. {\it Take an arbitrary initial state (i.e. arbitrary 
$c_l$) and drive the system for a finite time $T_{\rm int}$ with the frequency
matching the Bloch frequency $\omega_B$. If the driving amplitude is 
sufficiently large and the interaction time $T_{\rm int}$ is long enough, 
the initial state decays with a pulse output afterwards.} 

The physics behind this effect is the selective decay of the quasienergy 
Wannier-Bloch state discussed in Sec.~\ref{sec5e}. Indeed, let
the $\Phi_{\alpha,\kappa}(k)$ be the quasienergy 
states of the dc-ac Hamiltonian (\ref{5a0}).
Then we can expand the initial state $\psi(k,0)$ in this basis
\begin{eqnarray}
\label{6c0}
\psi(k,0) = 
\sum_\alpha \int_{-1/2}^{1/2} {\rm d}\kappa \, c_\alpha(\kappa) \,
\Phi_{\alpha,\kappa}(k) \,,
\end{eqnarray}
where the $c_\alpha(\kappa)$ are periodic functions of the quasimomentum.
[In particular, assuming the adiabatic switching of the field, the
initial condition $\psi(k,0)=\Psi_{0,l}(k)$ will correspond to 
$c_\alpha(\kappa)=\delta_{\alpha,0}\exp({\rm i}\kappa l)$.] Note
that the expansion (\ref{6c0}) is also valid in the case of a pure dc field 
considered in the previous section, but we preferred there the alternative 
basis of the Wannier-Stark states [see Eq.~(\ref{6b0})]. After $N$ periods
of driving  the wave function reads
\begin{eqnarray}
\label{6c1}
\psi(k,N T_B) \approx  
\int_{-1/2}^{1/2}  {\rm d}\kappa \, c_0(\kappa) \,
\exp\left(-{\rm i}\,\frac{\mathcal{E}_0(\kappa) N T_B}{\hbar}\right) \, 
\Phi_{0,\kappa}(k) \,,
\end{eqnarray}
where we assumed that all quasienergy states, excluding the ground
states $\alpha=0$, have decayed. Now the ac field is  
switched off, and we take the final state $\psi(k,N T_B)$ as the 
initial state of the pure dc dynamics. Expanding it in the basis of the 
Wannier-Stark states yields $\psi(k,N T_B) = c_0(k) \,
\exp(-{\rm i}\mathcal{E}_0(k) N T_B/\hbar)\, \Psi_{0,0}(k)$,
where the functions $c_0(\kappa)$ and $\mathcal{E}_0(\kappa)$
are treated now as the periodic function of the momentum instead of the 
quasimomentum. Then
\begin{eqnarray}
\label{6c2}
\psi(k,t>NT_B)=
\Psi_{\alpha,0}(k,t) \,\widetilde{C}\left(\frac{F t}{\hbar} + k\right)
\,,\quad \widetilde{C}(k)= c_0(k) \,
\exp\left(-{\rm i}\,\frac{\mathcal{E}_0(k) N T_B}{\hbar}\right) \;.
\end{eqnarray}
Comparing this result with Eq.~(\ref{6b1}), we notice that
the prefactor $c_0(k) \exp(-{\rm i} \mathcal{E}_0(k) N T_B/\hbar)$
takes the role of the amplitude modulation factor $\widetilde{C}(k)$
of the new initial state. Let us discuss this factor in more detail.
\begin{figure}[t]
\begin{center}
\includegraphics[width=12.5cm]{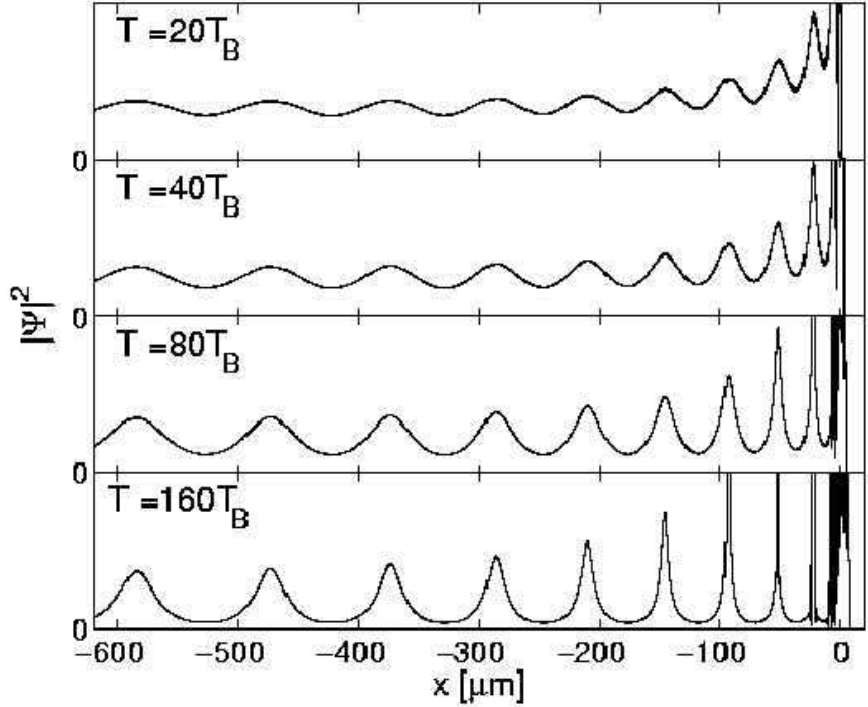}
\caption{\it Tail of the wave function after the system was driven 
for different times $T=NT_B$. The parameters are $\hbar = 3.3806$, 
$F = 0.0661$ and $\varepsilon = 0.1$.}
\label{fig6b}
\end{center}
\end{figure}

As shown in Sec.~\ref{sec5f}, for small $\varepsilon$ the 
dispersion relation of the complex quasienergy band is 
$\mathcal{E}_0(\kappa)=\mathcal{E}_0+(\Delta_0^{\rm Re}/2
+{\rm i}\Delta_0^{\rm Im}/2)\cos(2\pi\kappa)$.
Thus the absolute value of the amplitude modulation factor is given by
\begin{eqnarray}
\label{6c3}
|\widetilde{C}(k)|^2 = |c(k)|^2 \exp\left[- \frac{\Gamma_0 N T_B}{\hbar} - 
\frac{\Delta_0^{\rm Im} N T_B}{2\hbar} \cos(2 \pi k) \right] \,.
\end{eqnarray}
If the interaction time $T_{\rm int} = N T_B$ is large enough (and if $c(k)$ is
sufficiently smooth), the strong modulation of the exponential 
dominates the form of the amplitude modulation factor. 
Then the wave function is periodically peaked in momentum space. Of course, 
such a periodically peaked structure is also found for larger values of 
$\varepsilon$ where formula (\ref{5f6}) is no longer valid. In fact, due to the 
stronger modulation of $\Gamma_0(\kappa)$, it appears even for short 
interaction times.

The behavior of the wave function in coordinate space is additionally modified 
by the dispersion due to the real parts of the quasienergies. If we
approximate it by the cosine and
again apply the stationary phase approximation in the Fourier transform
of equation (\ref{6c2}), the stationary points $k_s$
are solutions of the slightly modified equation
$\hbar^2 k_s^2/2 + \pi \Delta_0{\rm Re} N \sin(2 \pi k_s)= E_0 - F x$.
 The implications are as follows. In coordinate space, the form of the 
peaks is changed compared to the dispersion-free case, in particular, 
the peaks can be broadened or narrowed. 
Note that for small $|k_s|$ there may be three instead of one 
stationary point for each branch of the square root. 
Then the wave function shows additional interferences due to 
the interaction of the three different contributions. 
However, for large $|k_s|$ (i.e.~for $x \rightarrow - \infty$),
the dispersion only slightly influences the shape of the peaks. 
Thus, for large $|x|$, the shape of the peaks of the decay tail mainly 
reflects the function $\Gamma_0(\kappa)$, which provides a method
to experimentally access of this function.
\begin{figure}[t]
\begin{center}
\includegraphics[width=12.3cm]{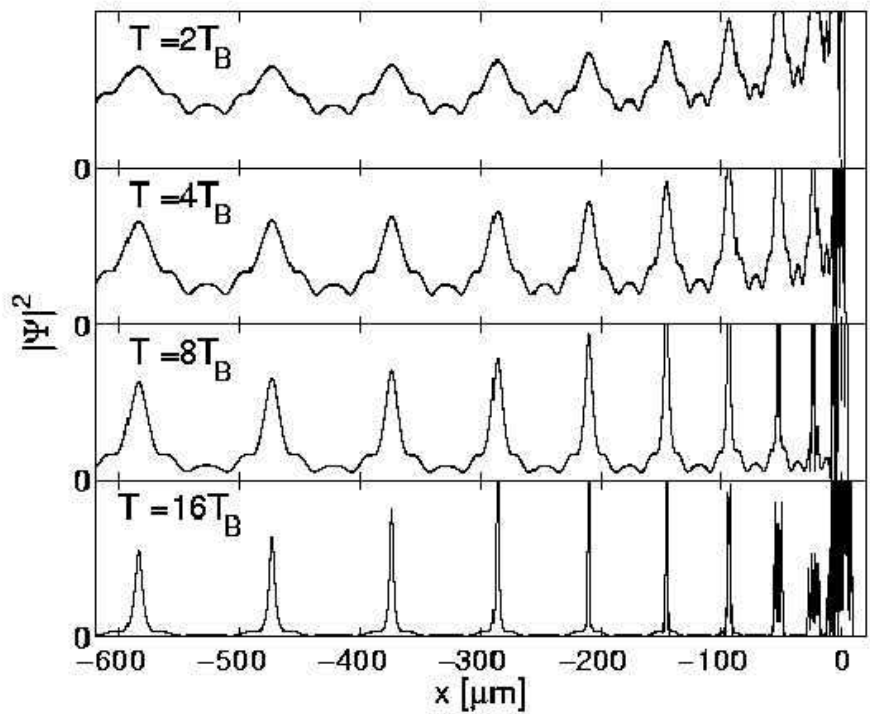}
\caption{\it The same as in Fig.~\ref{fig6b} but $\varepsilon=1.5$.}
\label{fig6c}
\end{center}
\end{figure}

To support the above analysis, Fig.~\ref{fig6b} shows the decay tails which
develop for a weak driving with $\varepsilon=0.1$. In this numerical
example we choose the ground Wannier-Stark resonance $\Psi_{0,0}(x)$ as the
initial state (the other parameters are the same as in Fig.~\ref{fig6a})
and drive the system for different interaction time. 
(Explicitly, we calculated
the resonances wave function $\Psi_{0,0}(k)$, multiplied by the
amplitude modulation factor $\exp(-{\rm i} \mathcal{E}_0(k) N T_B/\hbar)$, 
where the dispersion relation was calculated independently, 
and finally Fourier transformed to coordinate space.)
After short interaction times, 
the tail is slightly modulated. For longer interaction times, 
the modulation depth increases and pulses develop, 
which finally are clearly separated. Note that, apart from 
effects due to the dispersion, we can decrease
the width of the pulses by further increasing the interaction time, 
which provides a simple way to tune the width experimentally.

A crucial point of the weak driving regime is the long 
interaction time which is needed to generate well separated pulses. 
The relevant timescale is set by the most long-lived state 
from the ground band. For the case $\varepsilon = 0.1$, the minimum width 
is $\Gamma_{\rm min} = 7.214 \cdot 10^{-3}$, which corresponds to a 
lifetime approximately $10 \,T_B$. Thus, the
interaction time is much longer than the lifetime of the most stable
state. Consequently, a predominant part of the initial wave 
packet has already decayed before pulses are being formed. 
One can, however, surmount this problem by increasing the amplitude of the
resonant driving. Figure \ref{fig6c} shows the decay tail for 
$\varepsilon = 1.5$. Now the pulses develop after much shorter
interaction times. For $\varepsilon = 1.5$, the function $\Gamma_0(\kappa)$
has four minima, which are due to two crossings with higher excited
Wannier-Stark ladders (see Sec.~\ref{sec5f}). Note that one can directly read off
this property from the substructure of the pulses on the decay tail. 
In the lower panels of the figure one can also see the narrowing caused by the 
dispersion. In particular, the first peaks (counted from the right)
strongly oscillate, which reflects the existence of three stationary points 
in this region. However, the last peaks have approximately
the same shape, i.e.~here the narrowing effect can be neglected. 

\chapter{Chaotic scattering}
\label{sec7}

This chapter continues the analysis of Wannier-Stark system affected
by an ac field. In chapters 5-6 we have considered the case of a
relatively large scaled Planck constant $\hbar$ (see Sec.~\ref{sec1d})
and relatively small
values of driving amplitude $\varepsilon$, where the perturbation approach
can be applied to analyze the spectral and dynamical properties of the
system. Now we turn to the region of small $\hbar\rightarrow0$.
In this region even a weak driving violates the condition 
of perturbation theory which roughly reads $\varepsilon/\hbar<1$.
On the other hand, a small $\hbar$
corresponds to the semiclassical region, where the classical mechanics can 
guide the quantum-mechanical analysis. It turns out (see next section) that 
the classical dynamics of the system (\ref{5a0}) is typically chaotic. Then
the question we address sounds as ``What are the quantum manifestations 
of this chaotic dynamics?''. This question belongs to the list
of problems considered by the modern branch of quantum mechanics known
as Quantum Chaos (and actually can be considered as the definition of 
the field) \cite{Haak91}. 

A powerful tool of the theory of quantum chaos
is the random matrix theory (RMT) \cite{Haak91,Guhr98,Been97,Bohi84}.
Its application is based on the conjecture that the 
spectral properties of a classically chaotic system are similar to
those of a random matrix of the same (as the Hamiltonian) symmetry class.
Recently a considerable progress has been made in {\em nonhermitian}
random matrix theory, which aims at describing the properties of 
chaotic scattering 
systems \cite{Lehm95,Fyod97,Seba96a,Seba96b,Somm99}. In what follows we study the system 
(\ref{5a0}) from the point of view of nonhermitian random matrix theory.
In particular, we numerically calculate the distribution of the width
of the quasienergy Wannier-Stark resonances and distribution of
the Wigner delay time and compare them with the prediction 
of RMT. We would like to note that presently
there are just a few physical models which allow a detailed comparison with
analytical results of RMT.\footnote{Among the physical models, 
two-dimensional billiards with attached leads \cite{Ishi95,Wirt97,Ishi00},
simplified models of atomic and molecular systems \cite{Blue96,Mand97c,Mand97d},
the kicked rotor with absorbing boundary condition \cite{Borg91,Casa97,Casa00},
and scattering on graphs \cite{Kott97,Kott00}
could be mentioned.} In this context, the driven Wannier-Stark system
(\ref{5a0}), 
\begin{eqnarray}
\label{7}
H = \frac{p^2}{2} + \cos[ x + \varepsilon \cos(\omega t)] + F x \, ,\quad
\varepsilon = \frac{F_\omega}{\omega^2} \;,
\end{eqnarray}
(to be concrete, we choose $V(x)=\cos x$)
serves an excellent example for testing an abstract RMT.

\section{Classical dynamics}
\label{sec7a}

\begin{figure}[p]
\begin{center}
\includegraphics[width=13cm,height=8cm]{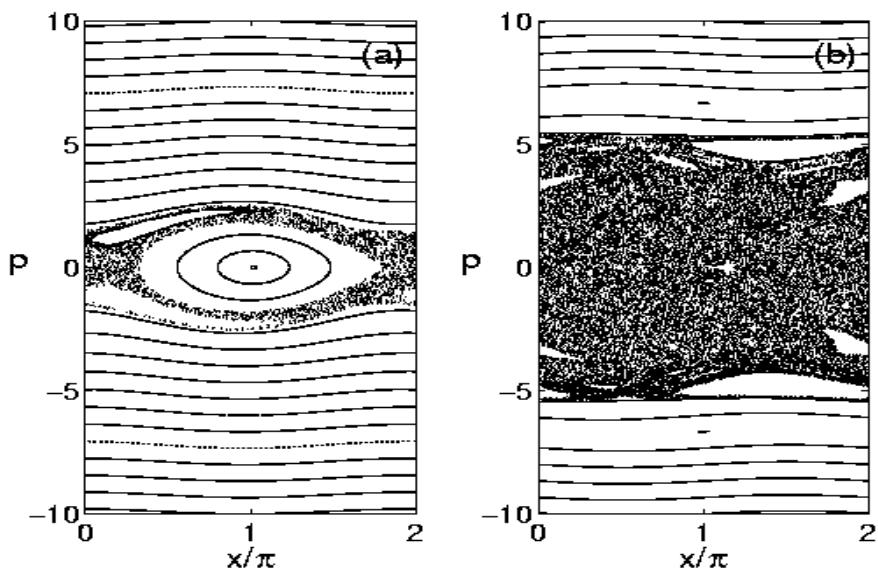}
\caption{\it Classical stroboscopic surface of section of the system
(\ref{7}) for $F=0$. The driving frequency is $\omega = 10/6$, and 
the amplitude $\varepsilon = 0.1$ (a) and
$\varepsilon = 1.5$ (b). In the first case the system is almost regular, 
in the second case a broad chaotic band appears.}
\label{fig7a1}
\end{center}
\end{figure}

We begin with the analysis of the classical dynamics of the driven
Wannier-Stark system (\ref{7}). Let us consider first the case $F=0$. 
Expanding the space- and time-periodic potential in a Fourier series yields
\begin{eqnarray}
\label{7a0}
\cos[x + \varepsilon \cos(\omega t)] &=& J_0(\varepsilon)\cos(x) 
- J_1(\varepsilon) \,[\, \sin(x + \omega t) + \sin(x - \omega t)\,] \nonumber\\
&&\quad - J_2(\varepsilon) \,[\,  \cos(x + 2\omega t) + \cos(x - 2\omega t)\,]\nonumber \\
&&\quad + J_3(\varepsilon) \,[\,  \sin(x + 3\omega t) + \sin(x - 3\omega t)\,] + \dots 
\end{eqnarray}
Then, from the perspective of the classical nonlinear dynamics
\cite{LichtLieb}, the system (\ref{5a0}) is a system of many interacting
nonlinear resonances. Depending on a particular choice of the
parameters $\omega$ and $\varepsilon$, its dynamics can be either
quasiregular or chaotic \cite{Grah91}. This is exemplified by
Fig.~\ref{fig7a1}, where the stroboscopic surface of 
section\footnote{The stroboscopic surface of section is generated by 
plotting the momentum $p(t)$ and coordinate $x(t)$, taken by modulus
$2\pi$, for $t=nT_\omega$ ($n=0,1,\ldots$).}
is shown for $\omega=10/6$ and $\varepsilon=0.1$ and $\varepsilon=1.5$.
In the quasiregular case with $\varepsilon=0.1$ only the three terms,
$\cos x$ and $\sin(x\pm\omega t)$,  in series (\ref{7a0}) are
important. The three corresponding nonlinear resonances are clearly
visible in the left panel. The main resonance of $\cos x$ appears as 
the large central island and the two other resonances correspond 
to the two smaller islands at $x\approx1$ and $p\approx\pm1.5$.
For large $\varepsilon=1.5$ many such nonlinear resonances overlap,
and a broad chaotic band appears. Assuming an initial condition in
this chaotic band, the classical motion is then confined to this
chaotic region, i.e. in the field free case $F=0$ it remains bounded in 
the momenta.
\begin{figure}[p]
\begin{center}
\includegraphics[width=10cm,height=6.5cm]{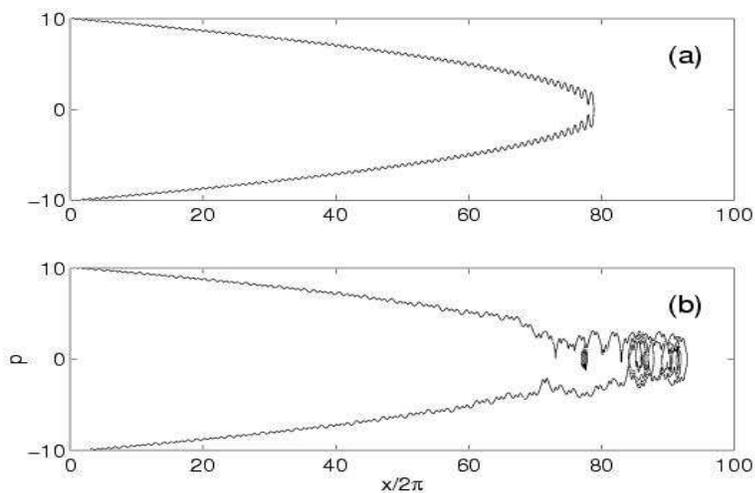}
\caption{\it Example of classical trajectories for the 
system (\ref{7}) with parameters $\omega = 10/6$, $F = 0.13$
and $\varepsilon = 0$ (a), $\varepsilon = 1.5$ (b).}
\label{fig7a2}
\end{center}
\end{figure}
\begin{figure}[p]
\begin{center}
\includegraphics[width=12cm,height=14cm]{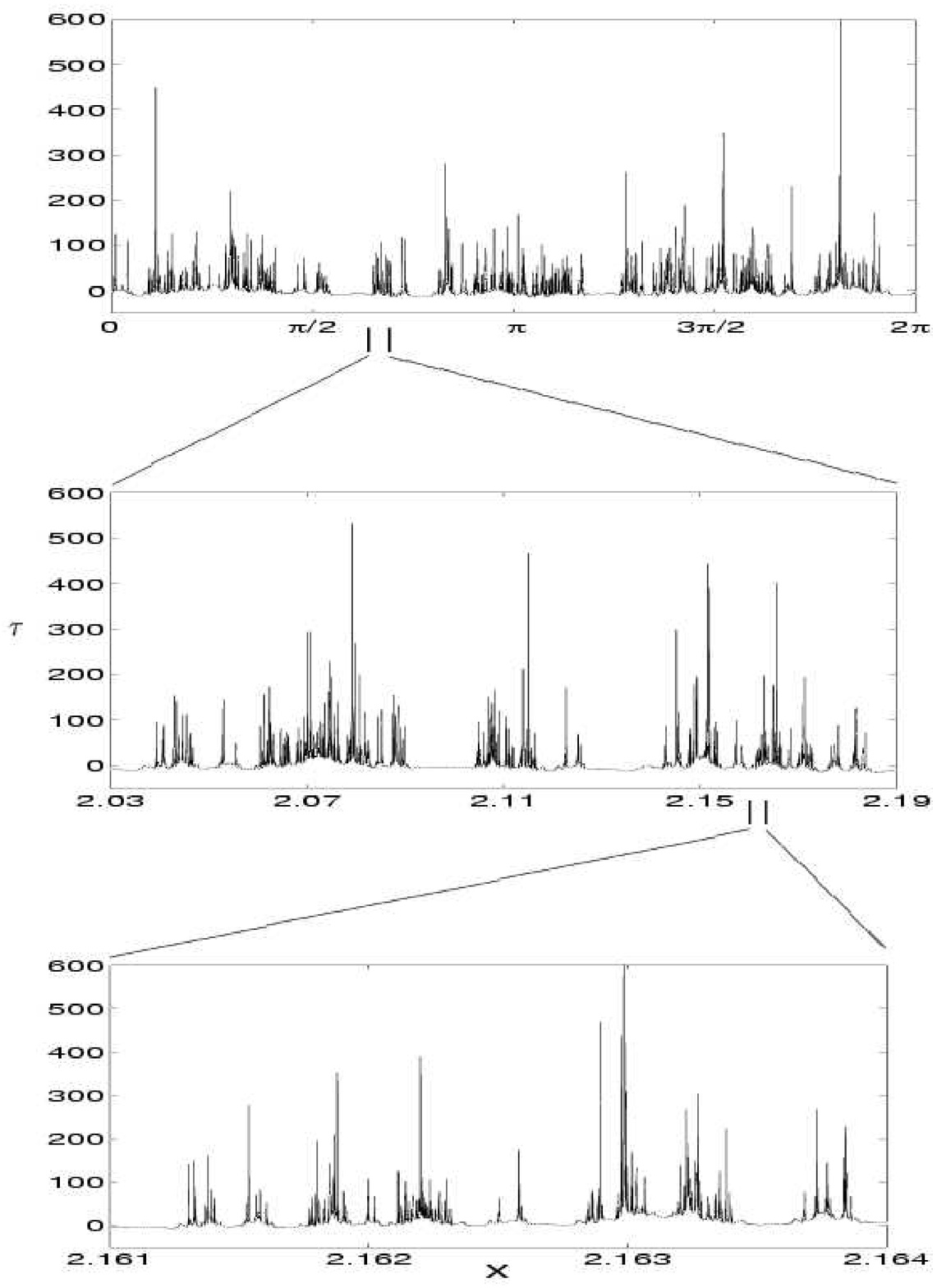}
\caption{\it Fractal structure of the classical delay time $\tau$
as a function of the initial coordinate $x$. The system parameters are
$F = 0.3$, $\omega = 10/6$ and $\varepsilon = 1.5$.}
\label{fig7a3}
\end{center}
\end{figure}

Adding a dc field changes this property, since it destroys the invariant
curves separating the chaotic component of the phase space from the outer
region of the regular motion. In fact, the static field connects the
regions of large momentum, because a particle initially localized in
the regular region of large positive momentum $p\gg p^*\approx5$ can then
move into chaotic region (small momentum $|p|<p^*$) from where it can
finally reach the region of large negative momentum. Thus the 
scattering process $p\rightarrow-p$ consists of tree stages: almost uniformly
deaccelerated motion for $p>p^*$, temporal chaotic motion $|p|<p^*$,
and accelerated motion for $p<-p^*$ (see Fig.~\ref{fig7a2}). The time
spent by the particle in the chaotic region is the delay or
dwell time $\tau$, which we define as the time gain or loss relative the
case $V(x,t)\equiv0$
\begin{eqnarray}
\label{7a1}
\tau = \lim_{p_0 \rightarrow \infty} \left[ \tau(p_0 \rightarrow -p_0)
- 2 p_0 / F \right] \, .
\end{eqnarray}
Figure \ref{fig7a3} shows the delay time (measured in periods $T_\omega$)
as function of the initial coordinate $x_0$ (the momentum $p_0$ is
kept fixed). The function is very irregular. Regions
where it is approximately constant are 
intermitted by regions of irregular peak structures. If we zoom
into such a structure, this behavior repeats on a finer scale, 
and altogether the function $\tau(x_0)$ shows a fractal behavior
which is one of the main characteristics of classical chaotic 
scattering. 
\begin{figure}[t]
\begin{center}
\includegraphics[width=11cm]{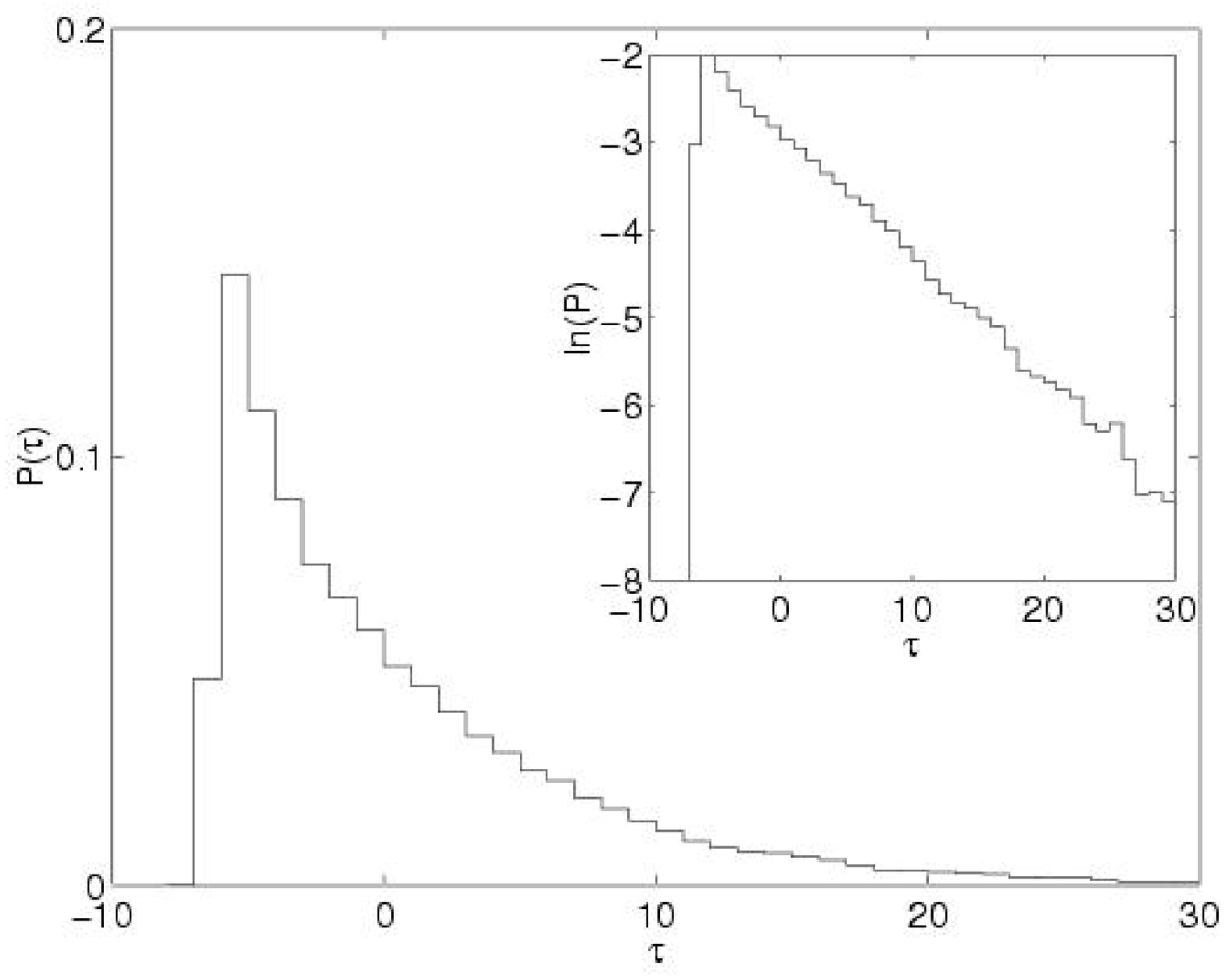}
\caption{\it Distribution of the scaled ($\tau\rightarrow F\tau$)
classical delay time for $\omega=10/6$, $\varepsilon=1.5$ and $F=0.065$.}
\label{fig7a4}
\end{center}
\end{figure}

The randomness of $\tau$ suggests its statistical analysis.
Figure \ref{fig7a4} shows the distribution $P_{cl}(\tau)$ of the classical
delay time for $\omega=10/6$, $\varepsilon=1.5$ and $F=0.065$.
It is seen that the distribution has an exponential tail
\begin{eqnarray}
\label{7a2}
P_{cl}(\tau)\sim\exp(-\nu\tau) \;,
\end{eqnarray}
which is another characteristic feature of the chaotic 
scattering\footnote{
In principle, the far asymptotic of the distribution $P_{cl}(\tau)$ may
deviate from the exponential law, which is known to be due to the effect of
the stability islands or their remnants. In our case, however, we did
not observe such a deviation.}.
The value of the decay increment $\nu$ primarily depends on $F$, and
for $F=0.13$ and $F=0.065$  (used later on in the quantum simulation)
it is $\nu\approx0.13F$ or $\nu\approx0.20F$, respectively. 
Note that the distribution of the delay times also defines the decay of
the classical survival probability $P_{cl}(t)$. Assuming an ensemble of classical
particles with initial conditions in the chaotic region, the latter
quantity is defined as the relative number of particles remaining
in the chaotic band. Obviously, the classical survival probability
(asymptotically) decreases exponentially with the same increment $\nu$,
i.e. $P_{cl}(t)\approx\exp(-\nu t)$.

\section{Irregular quasienergy spectrum}
\label{sec7b}

We proceed with the quantum mechanical analysis of the system.
Let us recall that we consider the commensurate case of a
rational ratio between the Bloch period $T_B$ and the period
$T_\omega$ of the exciting force, i.e. 
$T_B/T_\omega=\hbar \omega =p/q$ with integers $p$ and $q$
(see Sec.~\ref{sec5a}). We 
begin with the analysis of the complex quasienergy spectrum
for the simplest case $p=q=1$, where the quasienergy spectrum coincides
with the spectrum of the Floquet-Bloch operator (\ref{5a2}). 
\begin{figure}[p]
\begin{center}
\includegraphics[width=13cm,height=7cm]{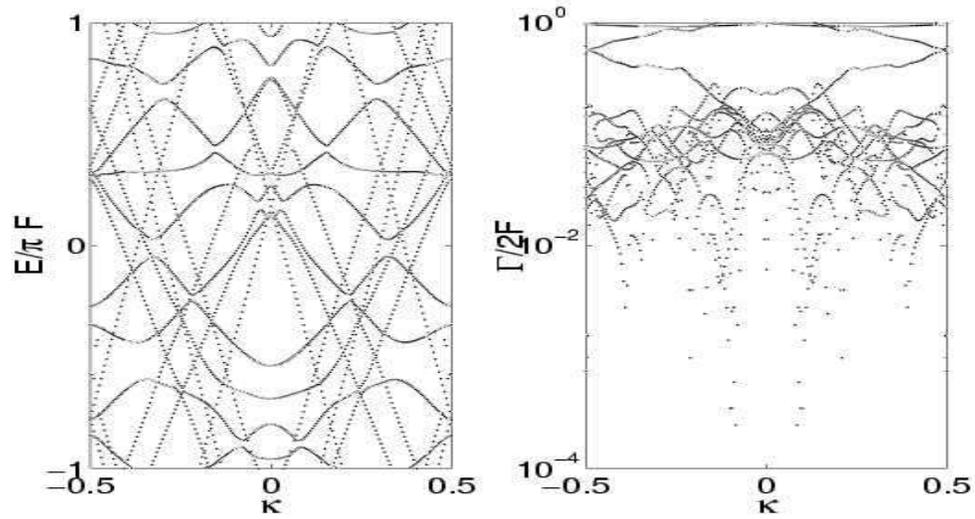}
\caption{\it Real and imaginary parts for
the quasienergy spectrum $\mathcal{E}_\alpha(\kappa)$ 
of the system (\ref{7}) with parameters $\hbar = 0.5$, 
$\omega = 10/6$, $\varepsilon = 1.5$ and $F \approx 0.13$.}
\label{fig7b1}
\end{center}
\end{figure}
\begin{figure}[p]
\begin{center}
\includegraphics[width=10cm]{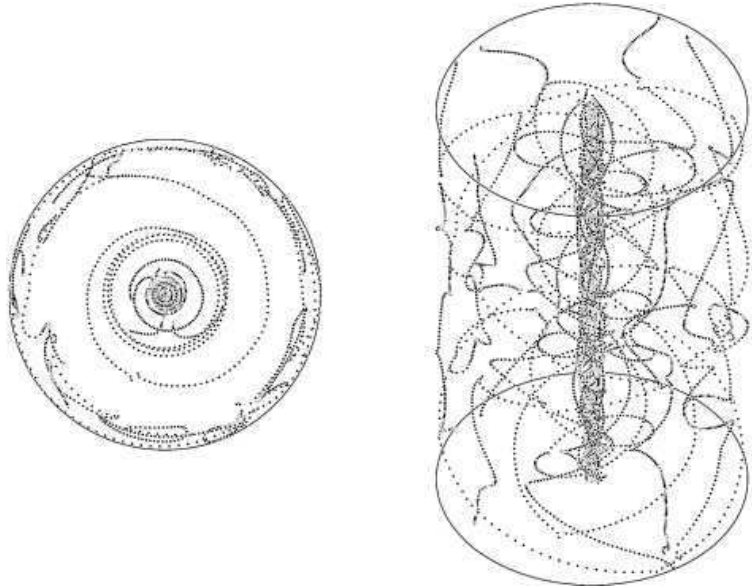}
\caption{\it Eigenvalues $\lambda_\alpha(\kappa) = 
\exp(-{\rm i}\mathcal{E}_\alpha(\kappa)/F)$ 
in a polar plot. The left panel shows the location of the eigenvalues 
inside the unit circle, the right panel additionally shows the dependence 
on the quasimomentum.}
\label{fig7b2}
\end{center}
\end{figure}

Figure \ref{fig7b1} shows the real and imaginary part of the spectrum 
$\mathcal{E}_\alpha(\kappa)$ for $\omega=10/6$, $\varepsilon=1.5$ and $\hbar=0.5$.
(The value of the static force is fixed by the resonant condition
$\omega=\omega_B=2\pi F/\hbar$, which corresponds to $F\approx0.133$.) For each value
of the quasimomentum $\kappa$, the 15 most stable resonances are plotted.
In addition to Fig.~\ref{fig7b1}, Fig.~\ref{fig7b2} shows the same spectrum
as a polar plot for the eigenvalues $\lambda_\alpha(\kappa)=
\exp(-{\rm i}\mathcal{E}_\alpha(\kappa)/F)$, where the axis of the cylinder is 
the quasimomentum axis. Now 30 resonances are depicted.
It is seen in the figures that,
apart from the symmetry $\kappa\rightarrow-\kappa$ [which reflects the
symmetry $t\rightarrow-t$, $p\rightarrow-p$ of the Hamiltonian (\ref{7})],
the spectrum looks very irregular. The formal reason for this irregularity 
is the interaction of the quasienergy bands discussed in 
Sec.~\ref{sec5f}. However, in the presently considered case of small $\hbar$,
this interaction appears to be so strong, that it makes 
an analytic description of the dispersion relation impossible.

An important result following from the numerical data is a clear
separation of the resonances according to their stability. Namely, for
every $\kappa$ there is a finite number of relatively stable
resonances which occupy the region near the unit circle in Fig.~\ref{fig7b2}.
The rest of the resonances are very unstable and they occupy the region
in the center of the unit circle. Using the phase-space representation of the resonance
wave function (for example the Husimi representation \cite{PLA3})
it can be shown that the former resonances are supported by
the chaotic region of Fig.~\ref{fig7a1} and, thus, are associated
with the chaotic component of the classical phase space. The latter
resonances are associated with the outer regular region of the classical
phase space and can be considered as a kind of ``above-barrier''
resonances. According to the Weyl rule, the total number of the relatively 
stable (chaotic) resonances can be estimated as
\begin{equation}
\label{7b0}
N=\frac{1}{2\pi\hbar}\, \oint p\,{\rm d}x\;,
\end{equation}
where the integral $\oint p{\rm d}x$ stands for the volume
of the chaotic component.\footnote{This formula
also estimates the number of under-barrier resonances for $\varepsilon=0$.
Then $\oint p{\rm d}x$ is the phase volume confined
by the separatrix.} Let us also note that these resonances have the
width of nearly the same order of magnitude. This fact and the
avoided crossings in the real part reflect the chaotic structure 
of the interaction region in classical phase space, which 
quantum mechanically results in a strong interaction of the
participating states.
\begin{figure}[t]
\begin{center}
\includegraphics[width=7cm,height=7cm]{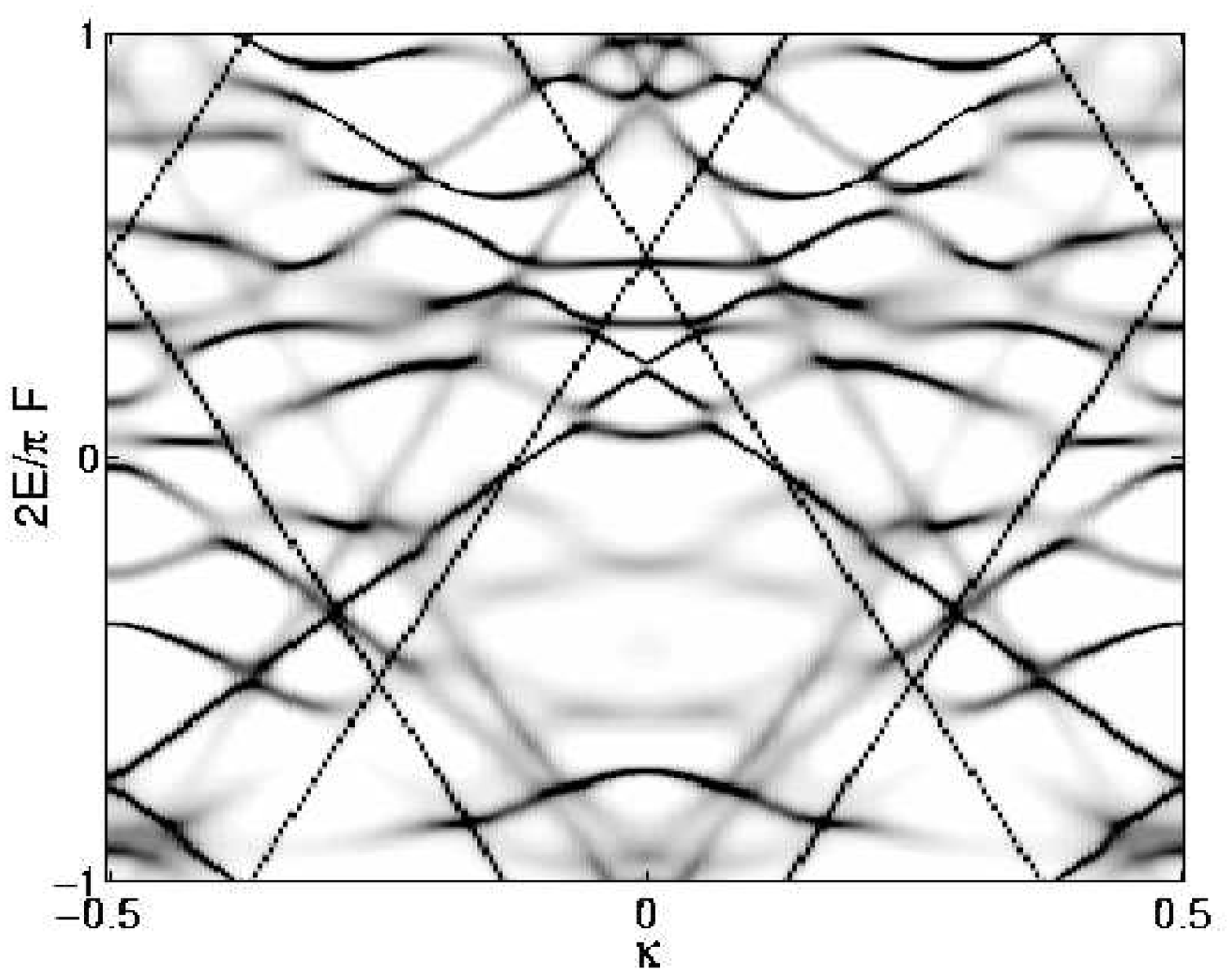}
\includegraphics[width=7cm,height=7cm]{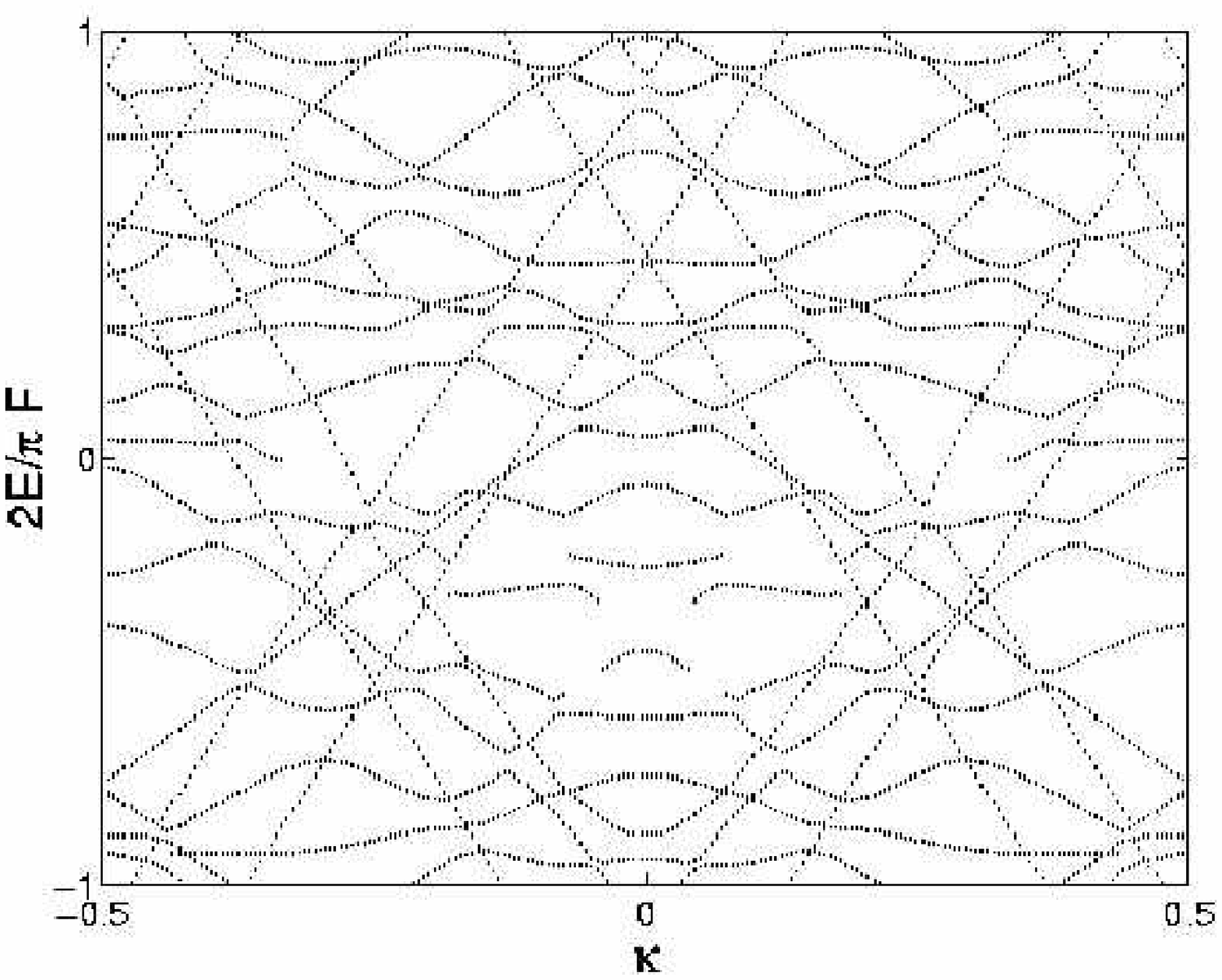}
\caption{\it The Wigner delay time (left panel) and the real part of the 
quasienergy spectrum (right panel) for the system (\ref{7}) with 
parameters $\varepsilon = 1.5$, $\omega = 10/6$ and $\hbar = 0.25$ in 
the case $p/q = 1/2$.\label{fig7b3}} 
\end{center}
\end{figure}

Additional information about the structure of the quasienergy spectrum
can be obtained by considering the Wigner delay time (\ref{5c7}). As an example,
Fig.~\ref{fig7b3} shows the Wigner delay time $\tau(E)$
for $\omega/\omega_B=1/2$. As already mentioned in Sec.~\ref{sec3d}, since
\begin{equation}
\label{7b1}
\tau(E)\sim \sum_\alpha\frac{\Gamma_\alpha}{(E-E_\alpha)^2+\Gamma_\alpha^2/4} \;,
\end{equation}
the Wigner delay times reveals only the narrow resonances. The majority
of these resonances can be identified with the chaotic resonances, which
form an irregular pattern in Fig.~\ref{fig7b3}. However, besides this
irregular pattern, a regular one in the form of a rhombus is clearly seen.
Below we show that this regular structure is due to the
stability islands of the classical phase space.

In fact, let us consider an arbitrary term in Eq.~(\ref{7a0}). This term
corresponds to classical nonlinear resonance at $p\approx m\omega$. Assuming 
that the interaction between the nonlinear resonances does not
completely destroy this particular resonance, the dynamics of the system
in the vicinity of its stable periodic point is locally governed by
the effective Hamiltonian
\begin{equation}
\label{7b2}
H_{{\rm eff}}=\frac{p^2}{2}+J_m(\varepsilon)\cos(x\pm m\omega t)+ Fx 
\end{equation}
(the sign of the Bessel function and the sin or cosine dependence does not
matter). By substituting $x'=x\pm m\omega t$, the Hamiltonian (\ref{7b2})
is transformed to the time-independent form
$H'_{\rm eff}=(p\pm m\omega )^2/2+J_m(\varepsilon)\cos x' +Fx'$. The latter Hamiltonian
can support localized Wannier-Stark states\footnote{Note that 
these states move $m$ lattice periods to the left or right per period of 
the driving frequency.}  $\Psi_{\alpha,l}(x')$ or, alternatively,
extended Wannier-Bloch states $\Phi_{\alpha,\kappa}(x')$. Denoting
by $\mathcal{E}'_\alpha$ the degenerate band of these Wannier-Bloch
states, the dispersion relation for the quasienergy spectrum of the
effective Hamiltonian (\ref{7b2}) reads (up to an additive term) 
\begin{equation}
\label{7b3}
\mathcal{E}_\alpha(\kappa)
=\left\{\left(\frac{(m\omega)^2}{2}+\mathcal{E}'_\alpha \right)
\pm \kappa \,m\hbar\omega\right\}_{{\rm mod:}\hbar\omega} 
=\left\{\left(\frac{(m\omega)^2}{2}+\mathcal{E}'_\alpha \right)
\pm \kappa \,m 2\pi F \,\frac{p}{q}\right\}_{{\rm mod:}\hbar\omega} 
\;.
\end{equation}
It follows from the last equation that the nonlinear resonance index
$m$ can be extracted from the slope of the dispersion lines. 
In particular, one can clearly identify the stability islands with
$m=\pm2$ and the remnant of the stability islands with $m=\pm1$ in
Fig.~\ref{fig7a1}.

To summarize, the quasienergy spectrum of the Wannier-Stark system consists
of two components, associated with the regular and chaotic components
of the classical phase space. The ``chaotic'' component of the spectrum
shows a rather complicated structure. This suggests a statistical
analysis of the spectrum, which will be done in Sec.~\ref{sec7d}. Before 
doing this, however, we shall briefly discuss some results
of random matrix theory.

\section{Random matrix model}
\label{sec7c}

As was mentioned in the introductory part of this chapter, the main
conjecture of random matrix theory of quantum chaos is that the spectral statistics
of a classically chaotic system coincides with those of an appropriate
ensemble of the random matrix. Let us first discuss which ensemble is
``appropriate'' to model the spectral statistics of the system of our interest.

According to the results of Sec.~\ref{sec5c}, the quasienergy resonances
of the Wannier-Stark system are given by the eigenvalues of the
nonunitary matrix (\ref{5c2}), which enters in the definition of the scattering 
matrix (\ref{5c5}). In the random matrix approach it is 
reasonable to keep the same structure of the matrix.
In other words, we model the case of rational $\omega/\omega_B=p/q$ by
the random scattering matrix
\begin{eqnarray}
\label{7c3}
S(E) = e_M
[ B - {\rm e}^{-{\rm i}E}\openone]^{-1}e^M \;,\quad
e_M = \left(\begin{array}{cc} 
  O_{N,M} & \openone_{M,M} \end{array}\right) \;,\quad
e^M = \left(\begin{array}{c} 
  \openone_{M,M}\\[2mm]  O_{N,M} \end{array}\right) \;,
\end{eqnarray}
where {\em nonunitary} matrix $B$ is given by
\begin{equation}
B  = \left(\begin{array}{cc}
        O_{M,N} & O_{M,M} \\[2mm]
              W_{N,N} & O_{N,M} \\
      \end{array}\right) \;, 
\label{7c0}
\end{equation}
and $W_{N,N}$ is a random {\em unitary} matrix of size $N\times N$.
In Eq.~(\ref{7c3}) and Eq.~(\ref{7c0}), the parameter $M$ is identical with
the denominator $q$ in the condition of comensurability.\footnote{
In this section, we use the standard notation of RMT, i.e. $N$ for the matrix
size and $M$ for the number of scattering channels.}
Moreover, we choose $W_{N,N}$ to be a member of Circular Unitary Ensemble
(CUE). The reason for this is that matrix $W_{N,N}$ should model the
unitary matrix $\widetilde{U}^{(\kappa)}$,
\begin{equation}
\label{7c1}
\widetilde{U}^{(\kappa)}= \widehat{\exp}\left(- \frac{{\rm i}}{\hbar}
\int_0^{T} {\rm d}t \, \left[\frac{(p+\hbar\kappa-Ft)^2}{2}+V(x,t)\right]\right)  
\end{equation}
which, excluding the cases $\kappa=0$ and $\kappa=\pm 1/2$, has no
time-reversal symmetry.

\begin{figure}[t]
\begin{center}
\includegraphics[width=7cm,height=7cm]{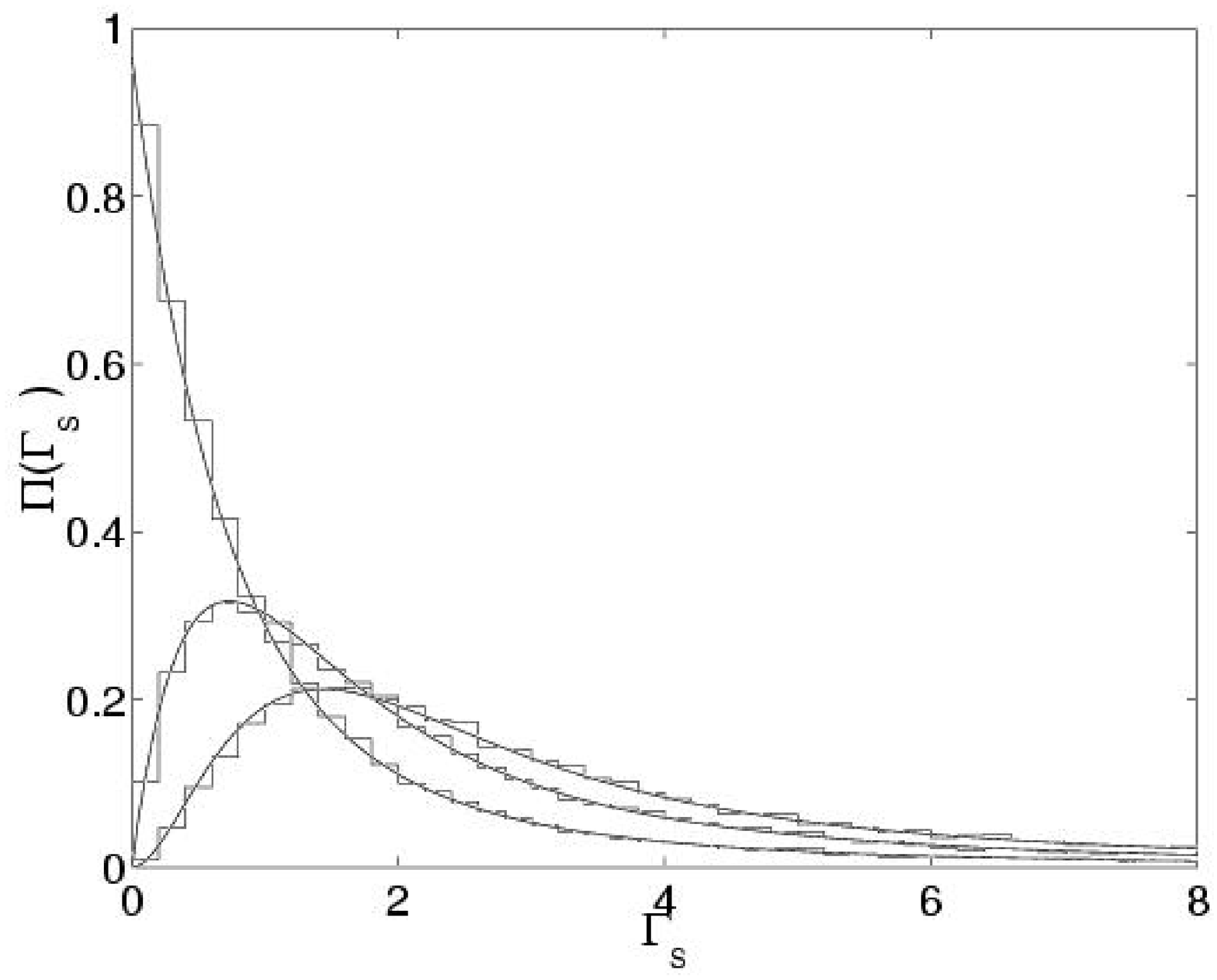}
\includegraphics[width=7cm,height=7cm]{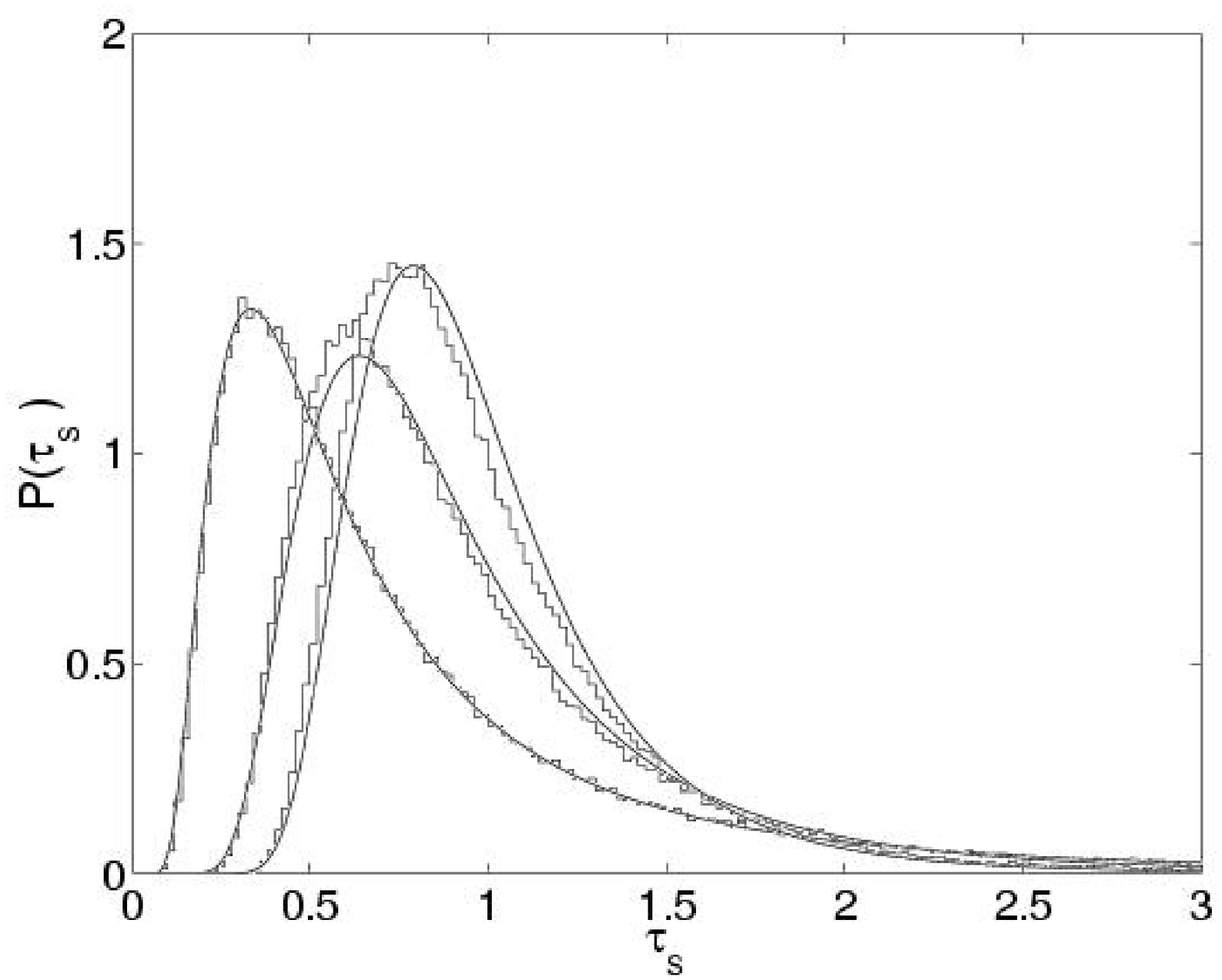}
\caption{\it Distribution of the resonance widths (left panel) 
and distribution of the sum of partial delay time (right panel)
for $M = 1,\, 2,\, 3$ decay channels. Numerical data (histograms) 
are compared with theoretical curves (solid lines). 
With increasing $M$, the maxima of the distributions shift to the right.}
\label{fig7c1}
\end{center}
\end{figure}

Now we discuss the statistics of the resonance widths.
The histograms in the left panel of Fig.~\ref{fig7c1} show the 
distribution of the scaled resonance widths for the random matrix model (\ref{7c3}),
(\ref{7c0}) for $M=1,2,3$. These histograms are obtained in the following way. First, 
we generate a random $40\times40$ GOE (Gaussian Orthogonal Ensemble)
matrix, i.e., a symmetric matrix with Gaussian-distributed random elements.
Then, multiplying the eigenvectors of this matrix (arranged column-wise
in a square matrix) by a random-phase factor, we obtain a member
of CUE \cite{Pozn98}. This CUE matrix is enlarged to a nonunitary matrix
$B$ and diagonalized. After diagonalization, we have $(N-M)$ non-zero
eigenvalues $\lambda=\exp[-{\rm i}(E-{\rm i}\Gamma/2)]$. 
To ensure the convergence in the limit $N\rightarrow\infty$, the 
resonance widths $\Gamma$ are scaled based on the mean level spacing
$\Delta=2\pi/(N-M)$ as $\Gamma_s=N\Gamma/2\approx\pi\Gamma/\Delta$.
Finally, the distribution of the scaled widths is calculated for
an ensemble of 1000 random matrices.

In Fig.~\ref{fig7c1} the distribution of the resonance widths
is compared with the analytical expression
\begin{eqnarray}
\label{7c2}
\Pi(\Gamma_s) =  \frac{(-1)^M}{(M-1)!} \,\Gamma_s^{M-1} 
\frac{{\rm d}^M}{{\rm d}\Gamma_s^M}\left[ 
\frac{1 - \exp(- 2 \Gamma_s)}{2 \Gamma_s} \right]\, , 
\end{eqnarray}
valid in the limit $N\rightarrow\infty$ \cite{Zycz00}. Note that the 
distribution (\ref{7c2}) was originally obtained for a different random
matrix model, which was aimed to model the chaotic scattering of the
ballistic electrons in the mesoscopic cavities \cite{Seba96a}, and
corresponds there to the so-called case of perfect coupling \cite{Fyod97},
which is realized in the case considered here.
The asymptotic behavior of the distribution (\ref{7c2}) is given by
$\Pi(\Gamma_s)\approx M/2\Gamma_s^2$ for $\Gamma_s\gg1$, and
$\Pi(\Gamma_s)\sim\Gamma_s^{M-1}$ for $\Gamma_s\ll1$.
A perfect coincidence between the depicted numerical data and analytical 
results is noticed in all three considered cases.

We proceed with the distribution of the Wigner delay time.
The advantage of the Wigner delay time is that it can be directly
compared to classical delay time (\ref{7a1}). Within the random
matrix approach discussed above, the Wigner delay time can be calculated
by taking the trace of the Smith matrix (\ref{5c7}), where the
random matrix analogue of the scattering matrix (\ref{5c5}) is given in
Eq.~(\ref{7c3}). Alternatively, we can calculate
the Wigner delay time by using an $M$-channel analogue of Eq.(\ref{2c11})
\begin{equation}
\label{7c3a}
\tau(E) =  \frac{1}{M}\, {\rm Tr}\left( e^{M,{\rm t}}
[ B^\dagger - {\rm e}^{ {\rm i}E}\openone]^{-1} 
[ B         - {\rm e}^{-{\rm i}E}\openone]^{-1}e^M \right)\;.
\end{equation}
Note that the scattering matrix (\ref{7c3}) yields only positive delay times 
whereas the Wannier-Stark system
(where the delay time is compared to the ``free'' motion)
also allows for negative values. However,  we can easily take this fact
into account by shifting the delay time (\ref{7c3a}) by $N$ units.

The distributions $P_{\rm qu}(\tau)$ of the Wigner delay times require an 
additional remark. The random matrix theory predicts only the 
distribution of the {\em partial} delay times [see Eq.(\ref{7c4}) below], 
whereas we are interested in the Wigner delay time, which is the sum
of partial delay times divided by the number of channels.
Because the partial delay times are correlated, the exact distribution of the
Wigner delay time is a rather complicated problem in random 
matrix theory \cite{Savi01}. 
However, ``in the first order approximation", the correlation of the partial delay
times may be neglected. According to \cite{Savi01pc}, the correlation
between partial delay times decrease as $1/(M+1)$ with increasing number
of scattering channels.

Then the distribution $P_{\rm qu}(\tau)$
of the Wigner delay time is the $M$-fold convolution of the distribution 
$P(\tau)$ for the partial delay times. According to the results of 
Ref.~\cite{PRE2,Fyod97}, the latter is given by
\begin{eqnarray}
\label{7c4}
 P(\tau_s) = \frac{1}{M!} \, \tau_s^{-M-2} \, e^{-1/\tau_s} \,,
\end{eqnarray}
where $\tau_s=\tau/N\approx\tau\Delta/2\pi$ is the scaled delay time.

In the right panel of Fig.~\ref{fig7c1}, the distributions of the sum
of the partial delay times are compared with the $M$-fold convolution
of the distribution (\ref{7c4}).\footnote{To obtain the distribution for
the Wigner delay time, the displayed histograms should be scaled as 
$P_{\rm qu}(\tau) \rightarrow M P(\tau/M)$.}   
An ensemble contains $1000$ random matrices of the size $40 \times 40$, 
and for each matrix the delay time is calculated at $100$ equally 
spaced values of $E$. In the one channel 
case both results agree perfectly, whereas in the
other cases the curves are slightly shifted. However, even here the
agreement is pretty good. Thus, the assumption of independent partial
delay times really yields a good approximation to the data.

\section{Resonance statistics}
\label{sec7d}

In the previous section we introduced a random matrix model 
of the driven Wannier-Stark system which
yields analytical results for the distribution of the resonance
width and Wigner delay time. In this section we compare the
actual distributions, obtained numerically, to these theoretical
predictions. In our calculation, we construct the statistical ensemble by 
scanning the quasimomentum $\kappa$ with a step $\Delta \kappa$ over 
the first Brillouin zone $-1/2p\le \kappa \le 1/2p$.
To get a good statistics, $\Delta \kappa$ should be as small as possible.
On the other hand, because the widths and the delay times depend 
smoothly on the quasimomentum, there is a characteristic 
value of $\Delta \kappa$ such that a further decrease does not improve 
the statistics. In the following calculations we choose 
$\Delta \kappa = 1/200 p$, i.e.~we average over $200$ spectra. 

The other problem arising in the statistical analysis of the numerical
data is the appropriate rescaling of the
resonance width and the delay time. In fact, the notion of matrix size
$N$ is not directly specified in our approach. However, we can use the
semiclassical estimate (\ref{7b0}) to specify the parameter $N$. 
For the value of the scaled 
Plank constant $\hbar=0.25$ considered below this gives
$N\approx32$. In what follows, however, we use a slightly smaller
value $N=28$ which accounts for the embedded islands of stability.
\begin{figure}[t]
\begin{center}
\includegraphics[width=5.5cm,height=5.5cm]{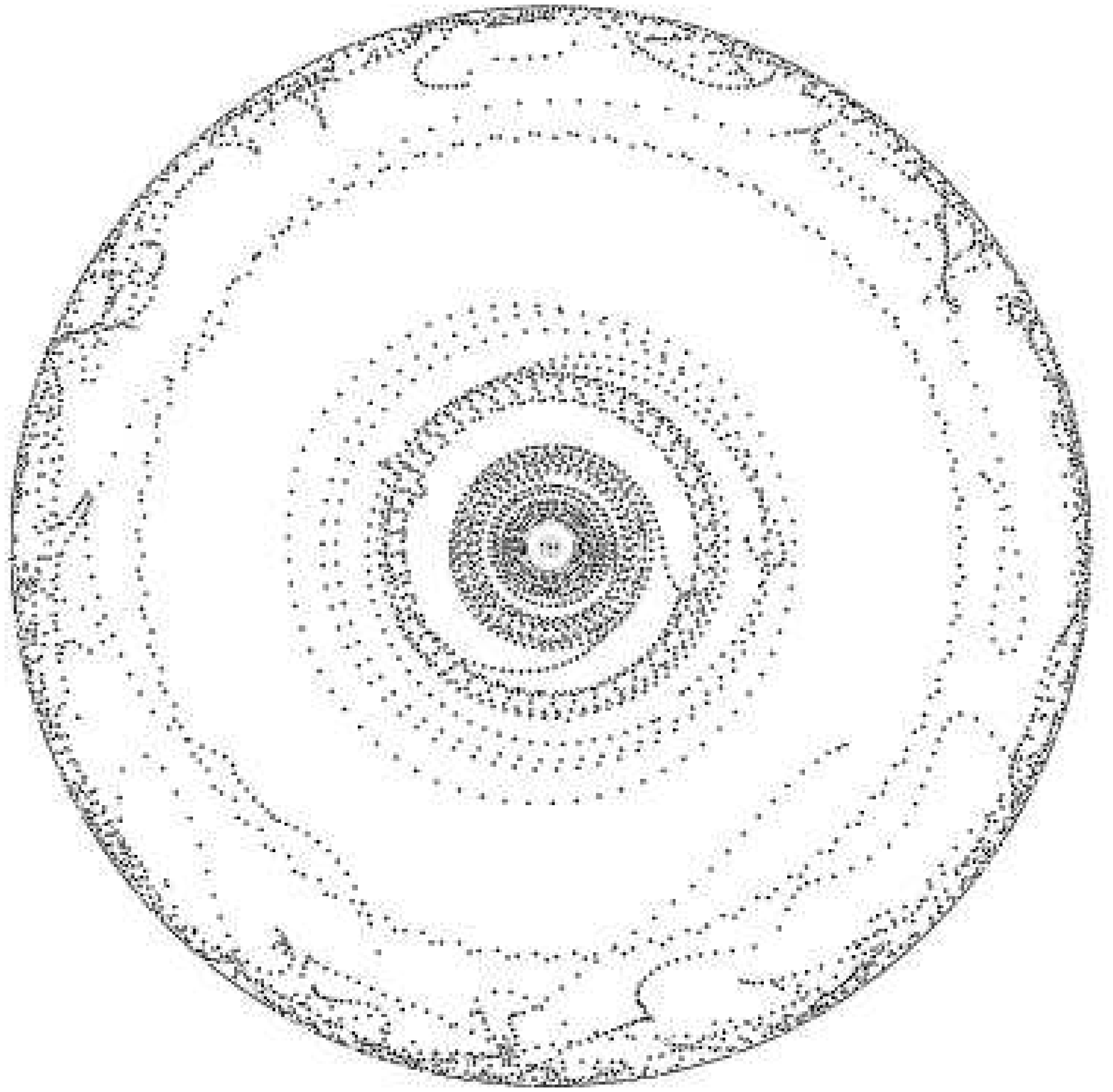}
\includegraphics[width=5.5cm,height=5.5cm]{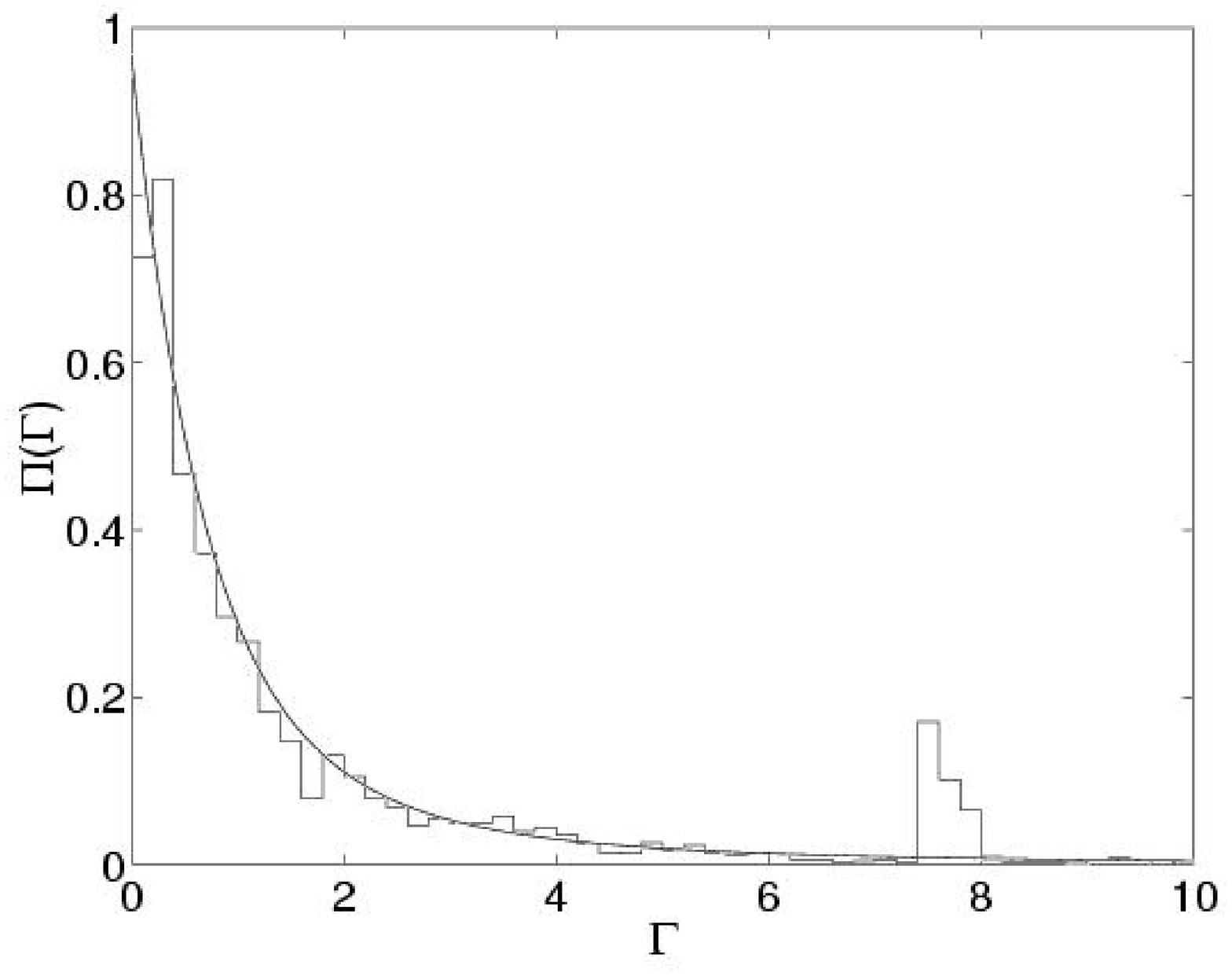}
\includegraphics[width=5.5cm,height=5.5cm]{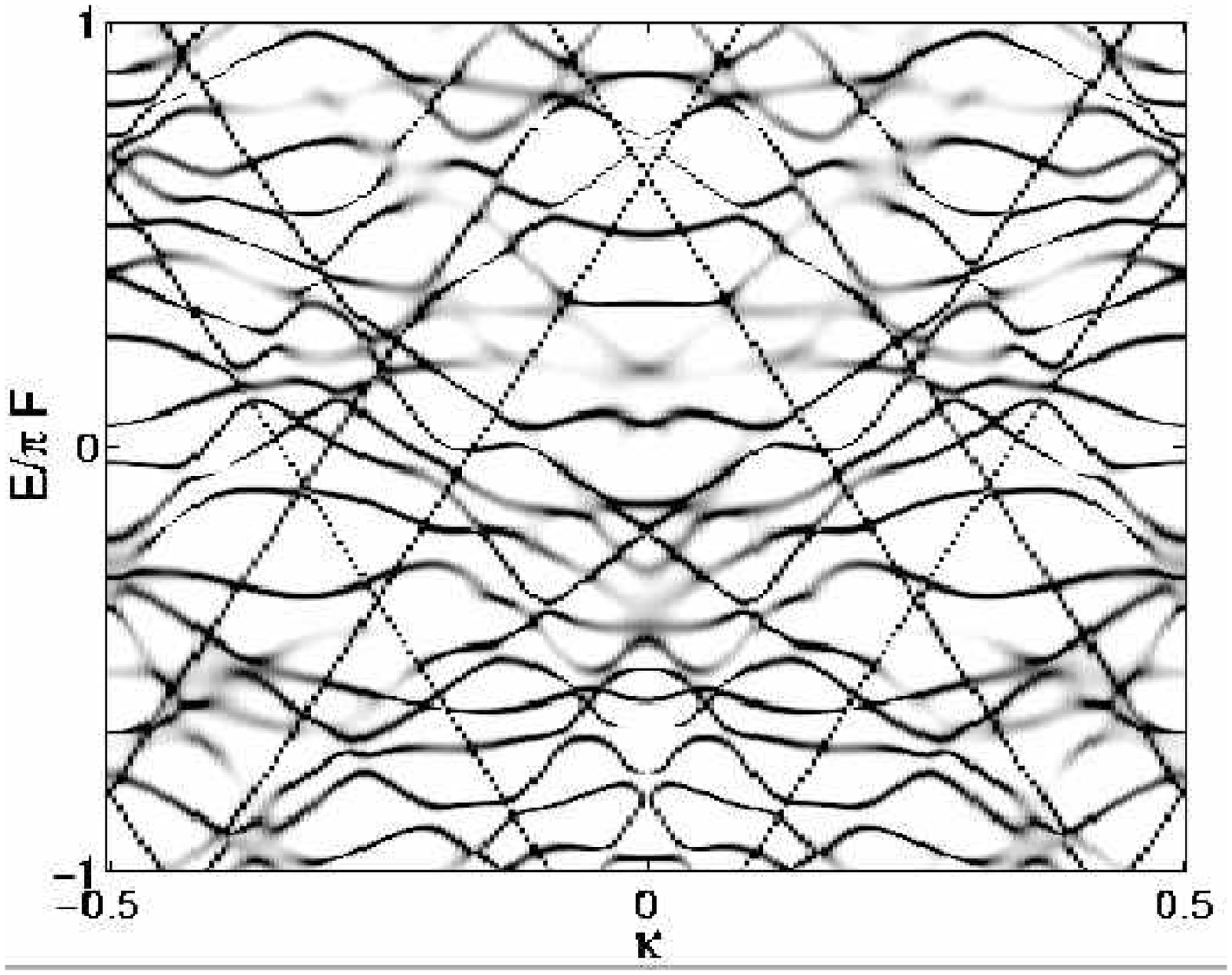}
\includegraphics[width=5.5cm,height=5.5cm]{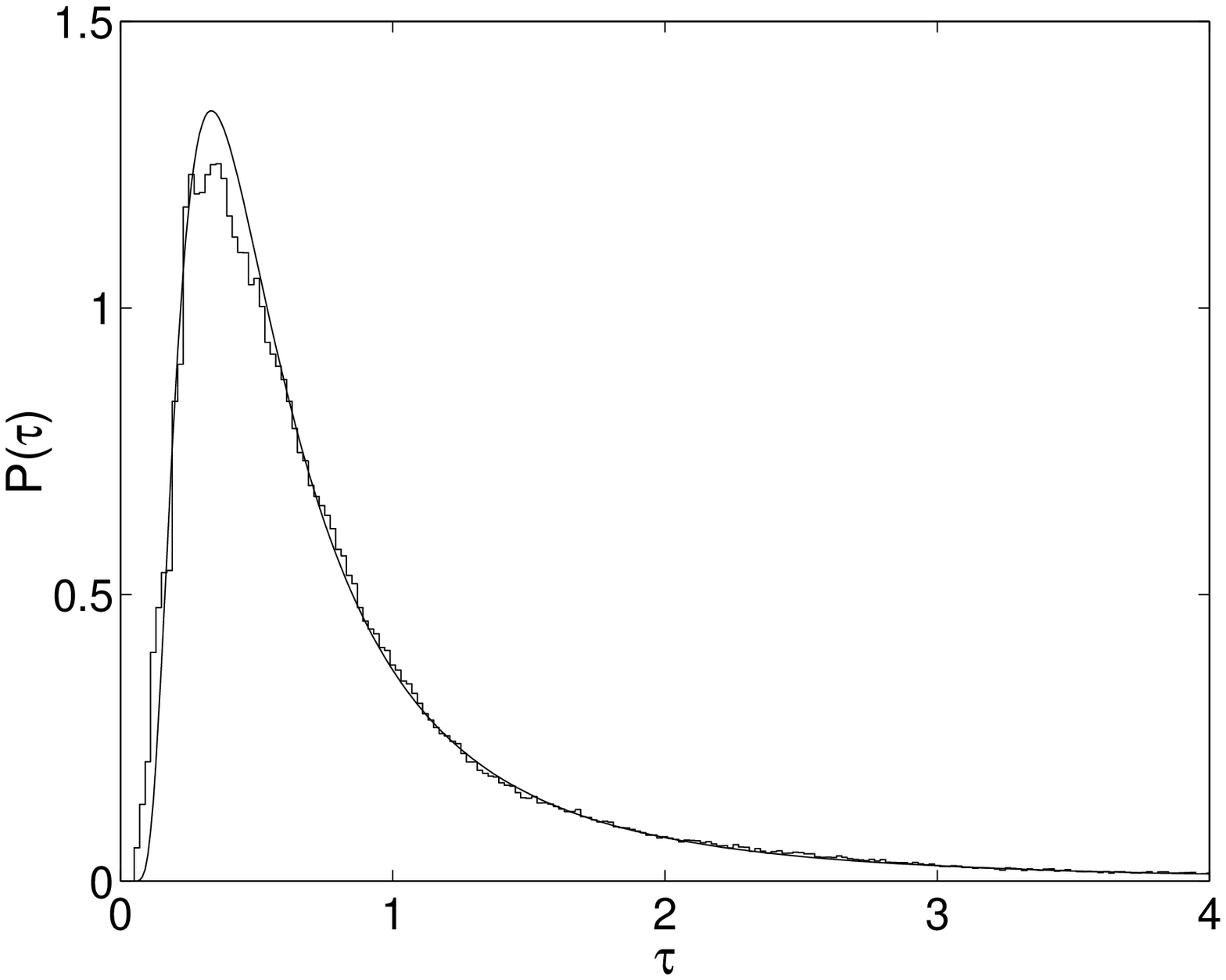}
\caption{\it Complex quasienergies, distribution of the widths,  
delay time and distribution of the delay times of the system (\ref{7})
for the case $p/q=1$ with parameters $\varepsilon = 1.5$, $\omega =10/6$
and $\hbar = 0.25$. In this case the constant force is $F \approx 0.066$.}
\label{fig7d1}
\end{center}
\end{figure}
\begin{figure}[p]
\begin{center}
\includegraphics[width=15cm,height=18cm,angle=270]{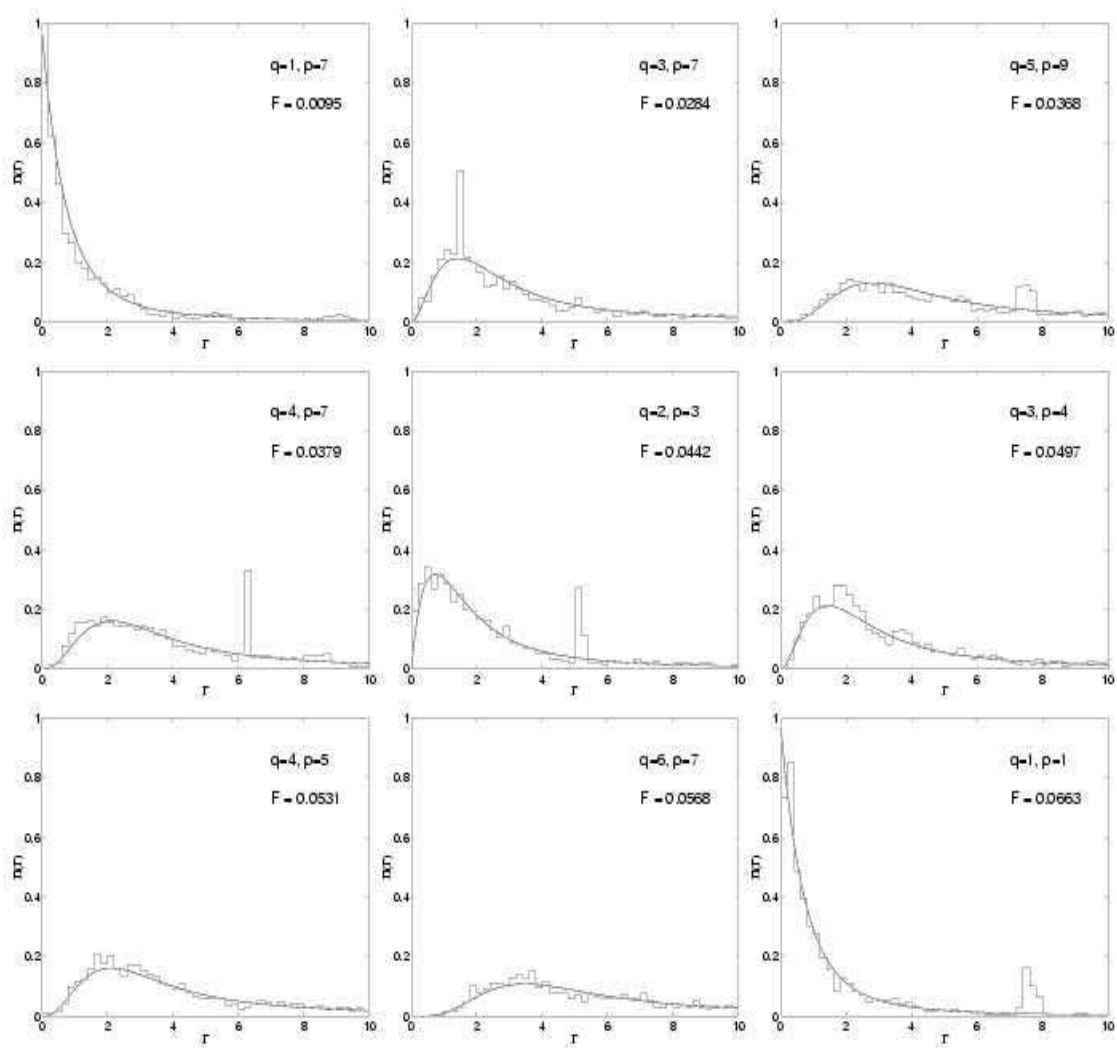}
\caption{\it Distribution of the resonance widths for different field
strengths $F$ with $\varepsilon = 1.5$, $\omega = 10/6$ and $\hbar = 0.25$.
The histogram show the numerical data, the solid lines are 
the random matrix predictions (\ref{7c2}) for the proper number of decay 
channels.}
\label{fig7d2}
\end{center}
\end{figure}

The two upper panels of Fig.~\ref{fig7d1} show the complex quasienergies 
and the distribution of the resonance widths for the most simple single 
channel case $p=q=1$. A good agreement between the random matrix
results and the calculated distribution is noticed. The distribution
has its maximum at $\Gamma_s = 0$, i.e.~the resonances tend to 
be long-lived. The main deviation
is a peak at $\Gamma_s \approx 8$, which is due to states associated 
with stability islands in the classical surface of section.  
One also finds these resonances in the delay time shown in the left lower panel
of Fig.~\ref{fig7d1}. As discussed in Sec.~\ref{sec7b}, resonances corresponding 
to classical stability islands form straight lines in the quasienergy 
spectrum. Indeed, we can see the lines with the slope $\pm4\pi$ and
remnants of two lines with slope $\pm2\pi$. Because
such resonances have approximately the same widths, their signatures 
are easily identified in the distribution of the widths. 

The right lower panel shows the distribution of the delay time
(to facilitate the comparison, the histogram for the scaled delay
time is shifted to the right by one unit.) Here the agreement is pretty 
good, either. The location of the maxima at $\tau_s = 0.33$ and the 
shape of both distributions coincide almost perfectly. 

We proceed with the case $p/q\ne1$, where we restrict ourselves to an analysis
of the resonance widths. The most striking prediction
of the random matrix model of Sec.~\ref{sec7c} is that the statistics
of the resonance widths is solely defined by the
integer $q$. On the other hand, the random matrix model
is supposed to describe the properties of the real system with 
{\em four} parameters. Thus, provided $q$ is the same, the distribution
of the resonance widths should be {\em independent} on the particular
choice of the other system parameters. (Of course, the condition
for chaotic dynamics should be fulfilled.) To check
this prediction we proceed as follows. 

The number $q$ of decay channels is defined by the rationality condition 
$p T_\omega = q T_B$, i.e.~in terms of the system parameters by
$F = q\hbar\omega/p2\pi$. As in the preceding cases, we choose 
$\varepsilon = 1.5$ and $\omega = 10/6$ to ensure that the system is 
classically chaotic, and $\hbar = 0.25$ in order to be in the 
semiclassical regime. Then we calculate the distribution 
of the resonance widths for several combinations of the integers $p$ and $q$, 
which correspond to increasing values of $F$.
Naively one would expect that with increasing $F$ 
the resonances tend to destabilize. Instead the distributions follow
closely the RMT distributions for the $q$-channel case as can be seen 
in Fig.~\ref{fig7d2}. The first and the last picture for the smallest and 
the largest field strength correspond to the one-channel case $q=1$. 
Note that the field strength differs by a factor seven, but the 
distributions are essentially the same. For the intermediate
field strengths the distributions vary according to the number of decay
channels. We should   stress that the only adjusted parameter, 
the number of states $N = 28$ defining the scaled width $\Gamma_s$, 
is constant in all figures.

\section{Fractional stabilization}
\label{sec7f}

In this section we discuss an interesting application of the results
of preceding section, which can be referred to as fractional
stabilization of the Wannier-Stark system. 

Let us discuss again the spectroscopic experiment \cite{Wilk96},
where the survival probability for the cold atoms in the accelerated 
optical lattice was measured as a function of the driving frequency (see
Sec.~\ref{sec4c}).We assume now the following modifications of the experimental 
setup. The value of the scaled Planck constant (which is inversely proportional
to the laser intensity) is small enough to insure the semiclassical
dynamics of the system. The value of the driving  amplitude is large
enough to guarantee the classical chaotic dynamics of the atoms. (Note
that both these condition were satisfied in a different experiment
\cite{Rob95}.) The atomic survival probability is measured as a function
of the acceleration but not as the function of
the driving frequency, i.e. we vary $\omega_B$ instead of varying
$\omega$. (This condition is actually optional.)
\begin{figure}[t]
\begin{center}
\includegraphics[width=15cm]{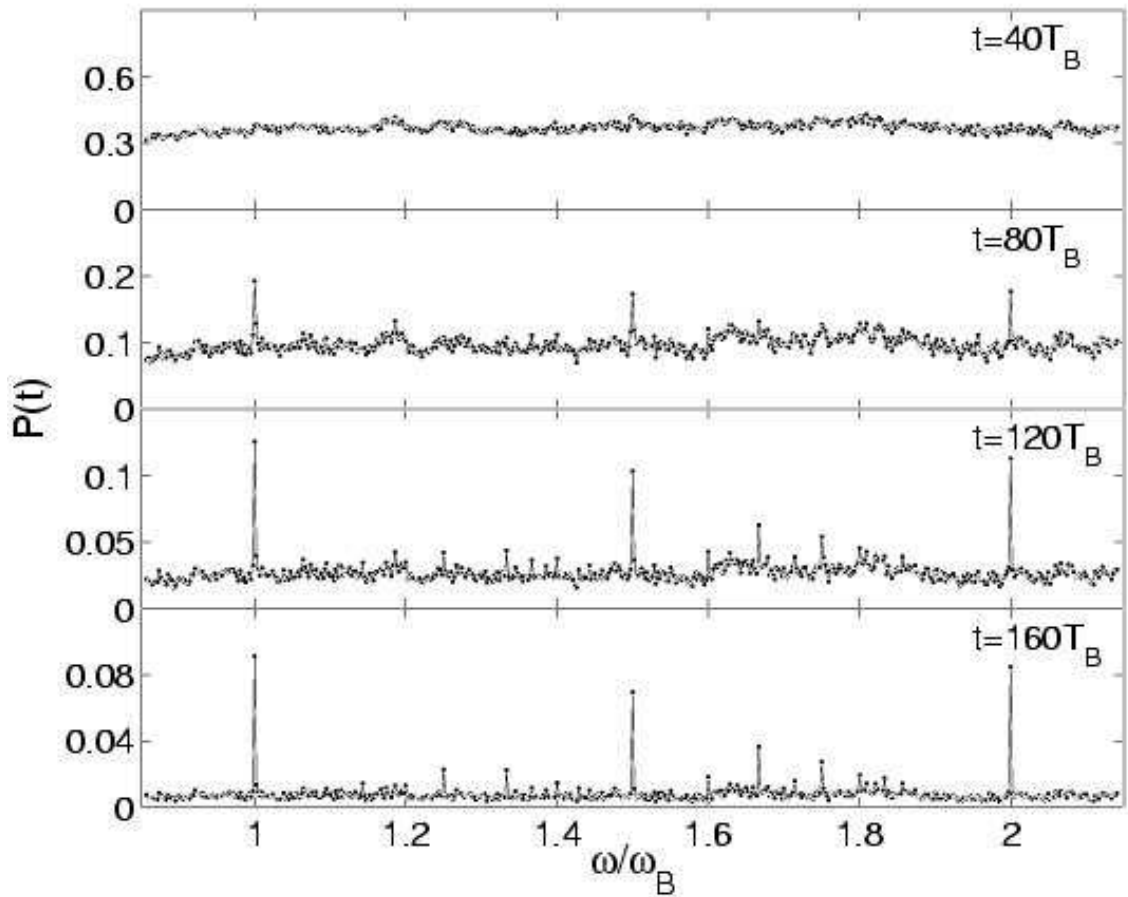}
\caption{\it Survival probability as function of the control parameter
$\gamma = \omega / \omega_B$. The system parameters are $\omega=10/6$,
$\varepsilon=1.5$ and $\hbar=0.25$.}
\label{fig7f1}
\end{center}
\end{figure}

Figure \ref{fig7f1} shows the results of the numerical calculation
of the quantum survival probability $P_{\rm qu}(t)$ based on direct numerical 
simulation of the wave packet dynamics.\footnote{Explicitly, we 
calculate wave function $\psi(p,t)$ in the momentum representation 
with the localized Wannier state as an initial condition. Then the
probability for a quantum particle to stay within the chaotic region
is given by  $P_{\rm qu}(t)=\int_{|p| < p^*} |\psi(p,t)|^2 \, {\rm d}p$,
where $p^*$ is the classical boundary between the chaotic and regular 
components of the classical phase space.} 
The survival probability shows an interesting behavior. For small times,
the curve fluctuates around an approximately constant value.  
When the time is increased, this average value decreases exponentially. 
In addition, however, peaks develop at integer values $\gamma=\omega/\omega_B$ and, 
incrementally, at rational $\gamma=p/q$ with small denominator. 
Thus, the decay is slowed down for rational $\gamma$. 
In what follows we explain this stabilization effect by using RMT 
approach. 

Indeed, the system parameters were chosen to ensure the regime of
chaotic scattering. Then the distribution of the resonance widths
is given by equation (\ref{7c2}). Let us assume that the initial state 
uniformly populates all resonances. If we then neglect the overlap of the 
resonances (this is the so-called diagonal approximation) 
the survival probability is given by the integral \cite{Savi97}
\begin{eqnarray}
\label{7f0}
P_{\rm qu}(t) = \int_0^\infty {\rm d}\Gamma \, \Pi(\Gamma) \, 
{\rm e}^{-\Gamma t/\hbar}\, , 
\end{eqnarray}
where $\Gamma  = 2 \Gamma_s/N$ and $N$ is the number of states in the
interaction region. The long-time asymptotics of this integral is defined
by the behavior of $\Pi(\Gamma)$ at small $\Gamma$, where it increases
as the power law $\Pi(\Gamma) \sim \Gamma^{q-1}$. Consequently, the
survival probability asymptotically follows the inverse power law
$P_{\rm qu}(t) \sim t^{-q}$. Thus, the asymptotics depend on the number of decay
channels and therefore on the denominator of the control parameter
$\gamma=\omega/\omega_B= p/q$.
\begin{figure}[t]
\begin{center}
\includegraphics[width=7.9cm,height=7.5cm]{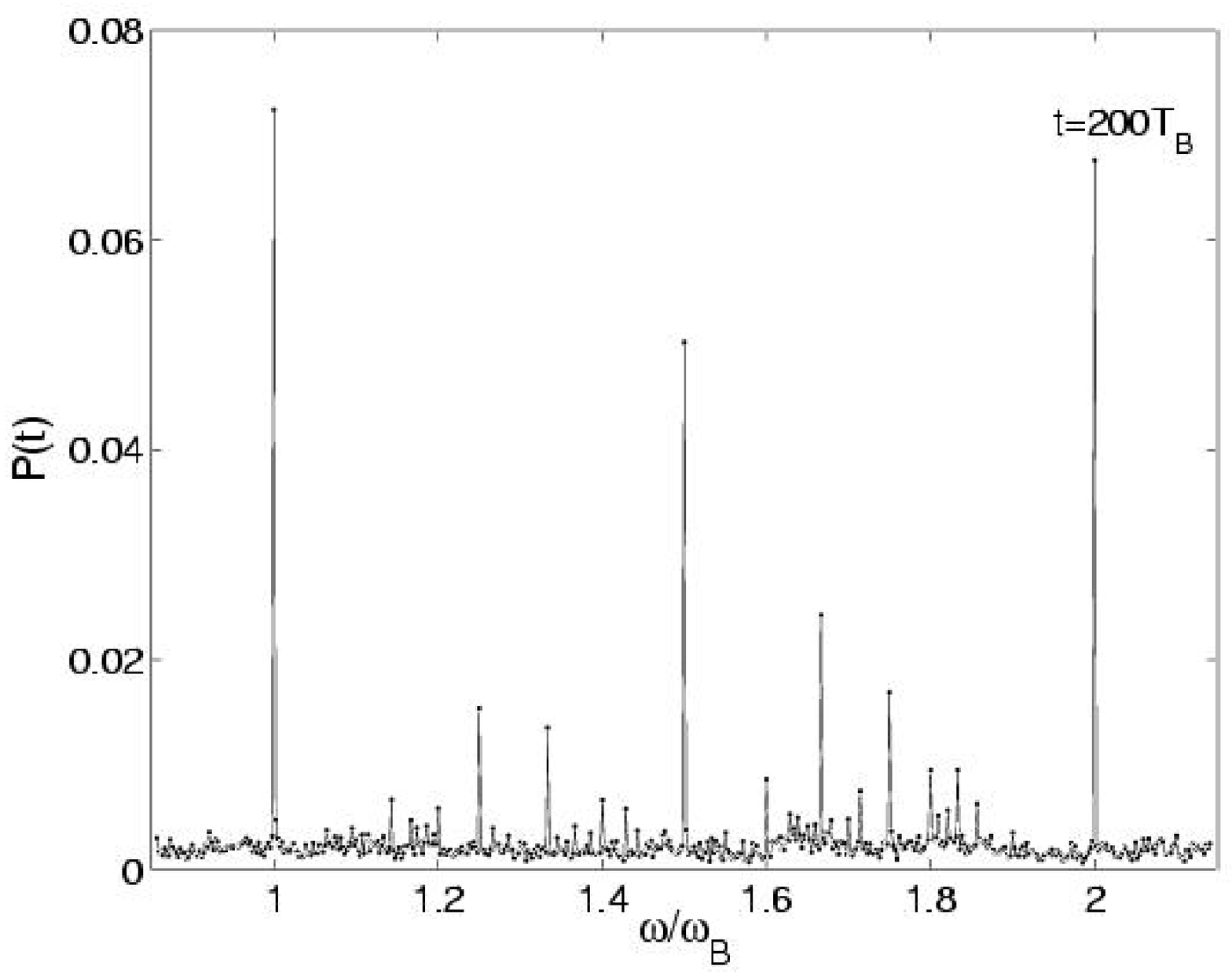}
\includegraphics[width=7.9cm,height=7.5cm]{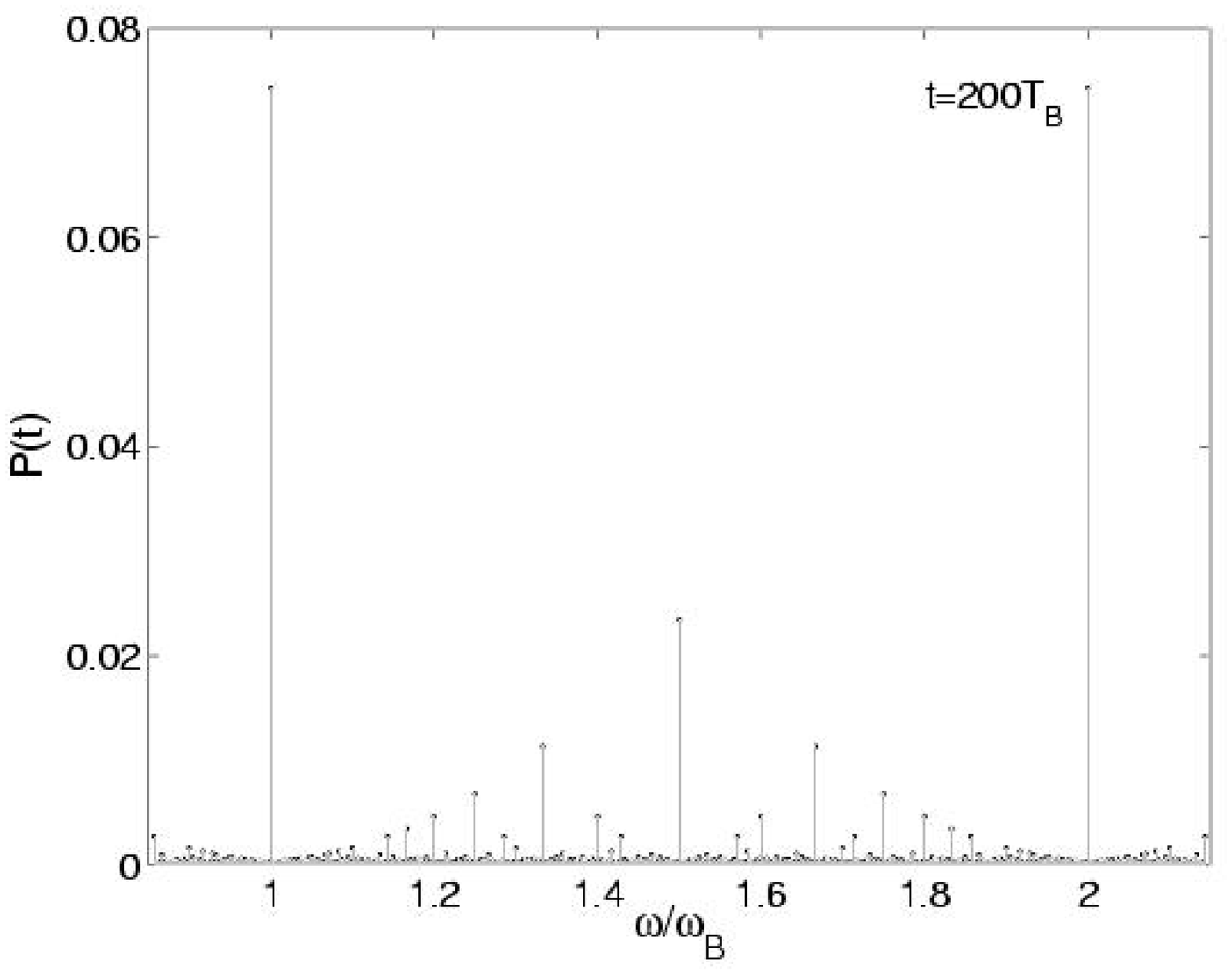}
\caption{\it Survival probability at $t = 200 T_B$. The left part shows
the numerical data, the right part the theoretical curve based on equation
(\ref{7f1}). To stress the discontinuous character of the latter function, 
we slightly changed its graphical representation.}
\label{fig7f2}
\end{center}
\end{figure}

With the help of supersymmetric techniques, $P_{\rm qu}(t)$ can be calculated 
beyond the diagonal approximation. This gives more elaborated result 
\cite{Savi97}
\begin{eqnarray}
\label{7f1}
P_{\rm qu}(t)\approx\left(1 + \frac{\Gamma_W t}{\hbar q}\right)^{-q} \;.
\end{eqnarray}
where $\Gamma_W$ is the so-called Weisskopf width (which is a free
parameter in the abstract random matrix theory). For rational $\gamma$
and large times the decay of the survival probability is algebraic, 
$P_{\rm qu}(t) \sim t^{-q}$, as found in the diagonal approximation.  
The case of irrational $\gamma$ can be approximated by the
limit $q \rightarrow \infty$. Then the system shows the  
exponential decay, $P_{\rm qu}(t) = \exp(- \Gamma_W t/\hbar)$ and its
natural to identify the parameter $\Gamma_W/\hbar$ with the classical
decay coefficient $\nu$. 

The right panel in Fig.~\ref{fig7f2} shows the values of the function (\ref{7f1}) 
for $t=200\,T_B$ and some rational values of $\gamma=\omega/\omega_B$. 
Here we use a slightly different 
graphic presentation of $P_{\rm qu}(t)$ to stress that the
function (\ref{7f1}) is a discontinuous function of $\gamma$ for
any $t$. In contrast, the atomic survival probability  shown in the left
panel is a continuous function of $\gamma$ where its discontinuous structure
develops gradually as $t\rightarrow\infty$. 
In fact, the survival probabilities calculated for two close
rational numbers $\gamma_1$ and $\gamma_2$ follow each other
during a finite ``correspondence" time. (For instance, for
$\gamma_1=1$ and $\gamma_2=999/1000$ the correspondence time
is found to be about $50\,T_B$.) Thus it takes some time to distinguish two
close rationals, although they may have very different 
denominators and, therefore, very different asymptotics.
With this remark reserved, a nice structural (and even semiquantitative) 
correspondence is noticed. 

The described numerical experiment suggest a simple laboratory experiment
with cold atoms in optical lattice, where one can test the statistics
of the resonance width indirectly, by measuring the survival probability
for atoms.

\chapter{Conclusions and outlook}

In this section we review the main results of the work and outline
some problems which are still waiting for their solutions.  In the
overview we shall mainly follow the table of contents.

The approach introduced in Sec.~2 gives us a powerful tool for
analyzing an arbitrary one-dimensional Wannier-Stark system, i.e.
a system with potential energy given by the sum of periodic and
linear terms. The success of the method is ensured by two key
points. First, we inverted the traditional solid state approach, where
the linear term has been treated as ``a perturbation" to the periodic term,
and formulate the problem as scattering of  a quantum particle by a
periodic potential. Second, instead of dealing with the Hamiltonian,
we work with the evolution operator. Although both these points
were discussed earlier, it is only a combination of them, which
provides solution of the Wannier-Stark problem. Let us also note
that the use of the evolution operator provides a way to an analysis of the
Wannier-Stark system affected additionally by a time-periodic
perturbation. The corresponding generalization of the method,
which leads to the notion of the metastable quasienergy Wannier-Stark 
states, is discussed in Sec.~5 of this review.

We apply the developed theory to analyze the Wannier-Stark
ladder of resonances in two particular systems --– undoped
semiconductor superlattices in a static electric field and the system of
cold atoms in optical lattices in an accelerated frame. Both of
these systems mimic the crystal electron in a static electric field
(which was the original formulation of the problem) and have their
own advantages and disadvantages. In particular, the
semiconductor superlattices allow (at least, in principle) to create
an arbitrary periodic potential.  One may think, for example, about
a periodic sequence of double wells, where the interaction of the
Wannier-Stark ladders  (which is essentially the resonance
tunneling effect) should have an especially interesting form. In
Sec.~3 we restricted ourselves by considering the cosine and
square-box shaped potentials. The structure of the Wannier-Stark
states and the interaction of the ladders in periodic potentials of a
different form (like the already mentioned double-well array or
asymmetric ratchet-like potential) is an open problem.

A disadvantage of the semiconductor superlattice is that this is a
more ``dirty'' (in comparison with the optical superlattice) system,
where the effects in question interfere with other effects like
electron-hole Coulomb interaction, scattering by impurities, etc.
Nevertheless, if we want to move further, we should learn how to
deal with these complications. In the first turn, the effect of
Coulomb interaction should be taken into account. We believe that
now this problem can be solved rigorously by extending the
one-particle scattering theory of Sec.~2 to the case of two particles.

We turn to the spectroscopic results of Sec.~4. In this section we
derive an analytic expression for the decay spectrum of the system of
cold atoms in an optical lattice and the absorption spectrum of
semiconductor superlattices. This expression involves complex valued
squared transition matrix elements (non-real squared matrix
elements appear because of the resonance nature of the 
Wannier-Stark states), which lead to a non-Lorentzian shape of the
absorption lines. Although the relation of these results to the famous
Fano theory is obvious, the details of this relation remain
unexplored.

The brief Sec.~6 was inspired by the experiment of Anderson and
Kasevich, where a pulsed output from the periodic array of cold
atoms was observed. We give a proper theoretical description of
this phenomenon which, in fact, is the Bloch oscillations in the
case of a strong static field. In this sense, Sec.~6 is the only section
of the review discussing Bloch oscillations. One might be
interested in other regimes of Bloch oscillations. Evolution of the
theory in this direction is reflected by a recent paper \cite{preprint}.

As already mentioned in the introduction, Sec.~7 deals with the very
different problem of chaotic scattering, which is primary of interest
to the members of quantum chaos community. Nevertheless, from
the formal point of view, the results of Sec.~7 are just  the results
beyond the perturbative approach of Sec.~5. Thus, when the
experimentalists overcome the perturbation limit (the
present state of the art),  Sec.~7 may change its status from of 
``pure theoretical interest" to that of ``practical importance".

To conclude, we would like to highlight one more problem. This
work is devoted entirely to one-dimensional Wannier-Stark
systems. However, practically nothing is known about the
Wannier-Stark states in 3D- or 2D-lattices (a first step in this
direction was taken only recently \cite{PRL3}). An extension of the
present theory to higher dimension is of much theoretical and
practical interest and one may expect on this way a variety of new
phenomena which are absent in the one-dimensional case.

Furthermore, the results presented in this
review will also be relevant in connection with recent new developments
in quantum transport in driven periodic lattices with broken symmetry,
i.e.~quantum hamiltonian ratchets \cite{Flac00,Ditt00,Scha01,Goy98,Goy01}.~  
Such ratchets are usually
studied in the case of vanishing mean potential gradient.~An
interesting situation arises, e.g., for ratchets inclined in the direction 
opposed to the current that would occur in the unbiased case.~
In addition, it should be noted that in the previous studies of
the classical -- quantum correspondence for driven Wannier--Stark
systems as discussed in Sec.~\ref{sec7}, the parameters have been
chosen to guarantee (almost) fully chaotic dynamics in the scattering
region, i.e.~classical stability islands are of minor importance.~Larger 
islands can be observed, however, and will certainly effect the 
decay properties discussed in Sec.~\ref{sec7},  as for
instance by chaos--assisted tunneling, a topic of much interest in
theoretical \cite{Uter94,Toms94,Aver95} and very recently also 
experimental studies \cite{Hens01,Stec01,Goss01}, were this phenomenon
was rediscovered.


\section*{Acknowledgments}

The authors gratefully acknowledge the support of the Deutsche Forschungsgemeinschaft 
via the Schwerpunktprogramm SSP 470 ``Zeitabh\"angige Ph\"anomene und Methoden in 
Quantensystemen der Physik und Chemie'' as well as the Graduiertenkolleg 
``Laser- und Teilchenspektroskopie''. Stimulating discussions with many colleagues 
are gratefully acknowledged, in particular, among others, with
J.~E. Avron, Y.~V.~Fyodorov, M.~Holthaus, K.~Leo, N.~Moiseyev, Q.~Niu, M.~G.~Raizen, 
D.~V.~Savin, V.~V.~Sokolov, and K.~Zyczkowski.
It is also a pleasure to thank the graduate students 
Michael Hankel, Christian Hebell, Frank Keck, Stefan Mossmann, Frank Zimmer
who contributed to our studies of Wannier-Stark systems.
Moreover, we thank P.~H\"anggi for many useful suggestions which 
considerably improved the present review.


\end{document}